\documentclass{aa}  

\usepackage{graphicx}
\usepackage{txfonts}
\usepackage[utf8]{inputenc}
\usepackage{dcolumn}
\usepackage{supertabular}
\usepackage{rotating}
\usepackage{longtable}
\usepackage{lscape}
\usepackage{float}
\usepackage{subfig}
\usepackage{url}
\usepackage{color}
\usepackage{natbib}
\usepackage[dvipsnames]{xcolor}
\usepackage{ulem}
\usepackage{soul}
\usepackage[colorinlistoftodos]{todonotes}
\usepackage[version=4]{mhchem} 
\usepackage{supertabular}  
\usepackage[colorlinks=true,
            linkcolor=blue,
            urlcolor=blue,
            citecolor=blue]{hyperref}
\newcommand{\kms}{\mbox{km s$^{-1}$}}
\newcolumntype{d}[1]{D{.}{\cdot}{#1}}
\newcolumntype{.}{D{.}{.}{-1}}

\newcommand{\hii}{H{\sc ii}}
\usepackage[switch]{lineno}
\begin{document}

   \title{The Cygnus Allscale Survey of Chemistry and Dynamical Environments: CASCADE}

   \subtitle{IV. Unveiling the hidden structures in DR18}
    \titlerunning{Unveiling the hidden structures in DR18}
   \author{W.-J.\,Kim
          \inst{1}\and 
          H. Beuther\inst{2}\and
          F. Wyrowski\inst{3}\and
          K. M. Menten\inst{3}\and
          N. Schneider\inst{1}\and 
          \'A. S\'anchez-Monge\inst{4,5} \and 
          A. Brunthaler\inst{3} \and
          T. Csengeri\inst{6} \and 
          C. Romero\inst{7} \and
          N. Cunningham\inst{8} \and 
          L. Bouscasse\inst{9} \and
          J. M. Winters\inst{9} \and 
          F. Comer\'on\inst{10}\and
          V. S. Veena\inst{3} \and
          A. Ginsburg\inst{11} \and 
          D. Semenov\inst{2} \and 
          C. Gieser\inst{12} \and
          A. Hern\'andez-G\'omez\inst{13}\and 
          S. A. Dzib\inst{3} \and 
          I.-M. Skretas\inst{3} \and 
          I. B. Christensen\inst{3} \and
          P. Schilke\inst{1}
          }

     \institute{I. Physikalisches Institut, Universit\"at zu Köln, Z\"ulpicher Str. 77, 50937 K\"oln, Germany\\
        \email{wonjukim@ph1.uni-koeln.de} 
    \and Max-Planck-Institut f\"ur Astronomie, K\"onigstuhl 17,69117 Heidelberg, Germany
    \and Max-Planck-Institut f\"ur Radioastronomie, Auf dem H\"ugel 69, 53121 Bonn, Germany
    \and Institut de Ci\`encies de l'Espai (ICE, CSIC), Campus UAB, Carrer de Can Magrans s/n, 08193, Bellaterra (Barcelona), Spain
    \and Institut d'Estudis Espacials de Catalunya (IEEC), 08860 Castelldefels (Barcelona), Spain
    \and Laboratoire d'astrophysique de Bordeaux, Univ. Bordeaux, CNRS, B18N, all\'ee Geoffroy Saint-Hilaire, 33615 Pessac, France
    \and Center for Astrophysics | Harvard and Smithsonian, 60 Garden Street, Cambridge, MA 02143, USA
    \and SKA Observatory, Jodrell Bank, Lower Withington, Macclesfield, SK11 9FT, United Kingdom
    \and Institut de Radioastronomie Millim\'etrique (IRAM), 300 rue de la Piscine, Domaine Universitaire, 38406 St. Martin d’H\'eres, France
    \and European Southern Observatory, Karl-Schwarzschild-Str. 2, 85748 Garching bei München, Germany
    \and Department of Astronomy, University of Florida, PO Box 112055, USA
    \and Max-Planck-Institut f\"ur Extraterrestrische Physik, Giessenbachstrasse 1, 85748 Garching, Germany
    \and Tecnologico de Monterrey, Escuela de Ingenier\'ia y Ciencias, Avenida Eugenio Garza Sada 2501, Monterrey 64849, Mexico}

   \date{Received 26 August 2024 / Accepted 19 December 2024 }

 
  \abstract
   {The Cygnus-X complex is a massive (a few $10^6$\,M$_{\odot}$ molecular gas mass), nearby (1.4\,kpc) star-forming region with several OB associations. Of these, Cyg OB2 is the largest,  with at least 169 OB stars. DR18 is the largest globule near the OB2 association, making it a perfect target for investigating the influence of ultraviolet radiation on molecular clouds. }
   {By analyzing emission from different molecular species, we aim to study the molecular gas structures toward DR18 using high angular-resolution molecular line observations.}
   {As part of the Cygnus Allscale Survey of Chemistry and Dynamical Environments (CASCADE) program,  we carried out 3.6\,millimeter (mm) continuum and spectral line high-resolution observations ($\sim$\,3 -- 4$''$) toward DR18, covering several molecular species (e.g., HCN, HNC, \ce{H2CO}, \ce{N2H+}, SiO, \ce{C2H}, deuterated species, etc.) with the Northern Extended Millimeter Array (NOEMA) and the Institut de Radioastronomie Millim\'etrique (IRAM) 30\,m telescope. In addition, multi-wavelength archival datasets from mid-infrared (MIR) to centimeter (cm) wavelengths were used to provide a comprehensive analysis of the region.}
   {The spectral index analysis shows significant contamination of the 3.6\, mm continuum by free-free emission from ionized gas. A comparison of the 3.6\,mm and 6\,cm continuum emission confirms that a B2 star (DR18-05) shapes the cometary \hii\ region in the DR18 cavity, with ionized gas escaping toward the OB2 association. On the other hand, the extended 3.6\,mm and 6\,cm continuum emission are likely to trace photoevaporating ionized gas from ultraviolet radiation from the Cyg OB2 association --  not from DR18-05. To study the feedback of the B2 star and the OB2 association on surrounding molecular regions, we analyzed the \ce{HCO+}, HCN, HNC, \ce{N2H+}, and SiO emission lines. The shell structure around DR18-05 indicates photodissociation regions (PDRs) formed by the expanding \hii\ region and photo-erosion from DR18-05 and OB2 stars. We also identified 18 compact cores with \ce{N2H+} emission, half of which are gravitationally bound (virial parameter, $\alpha_{\rm vir}$, $\lesssim$\,2.0), and mostly located in colder regions ($T_{\rm HCN/HNC}$\,$<$\,30\,K) behind the PDRs. The SiO emission is found only in PDRs, with narrow-line widths ($\sim$\,0.8 -- 2.0 \kms) and lower abundances (X(SiO)\,$\sim~5\times10^{-11} - 1 \times10^{-10}$). Comparing with the UV irradiated shock models, we suggest that the SiO emission partially encompassing the \hii\ region arises from the molecular gas region, marginally compressed by low-velocity shocks with $\sim$ 5\,\kms, irradiated by external UV radiation ($G_{\rm 0}\,\sim\,10^2 - 10^3$), as they traverse through a medium with $n_{\rm H} \sim 10^4$ to 
 $10^5$\,cm$^{-3}$. These shocks can be generated by the initial expansion of the \hii\ region and potentially by stellar winds.}
   {}


   \maketitle
%
\section{Introduction}
In the vicinity of massive young stars ($>$\,8 -- 10\,M$_{\odot}$), molecular clouds are exposed to extreme ultraviolet (UV) radiation and ionized gas pressure causing compression between molecular gas and ionized gas regions \citep{Zinnecker2007_massive_star_review}. The far ultraviolet radiation (FUV, 6\,eV\,<\,$h\nu\,$<\,13.6\,eV) of stars, irradiating the nearby molecular gas creates photodissociation regions (PDRs, \citealt{Tielens1985_pdr, Tielens2013_pdr}). Strong, energetic injections reshape parental clouds, triggering low (and intermediate) and high mass star formation (e.g., \citealt{Hester1996_pillar,Sugitani2002_globules,Schneider2021_globules,Comeron2022_DR18}) and also contributing to the complexity of the PDR chemistry. This makes the immediate neighborhood of OB stars an important laboratory for studying the formation and destruction of molecular species.

\begin{figure*}[!ht]
    \centering
    \includegraphics[width=0.90\textwidth]{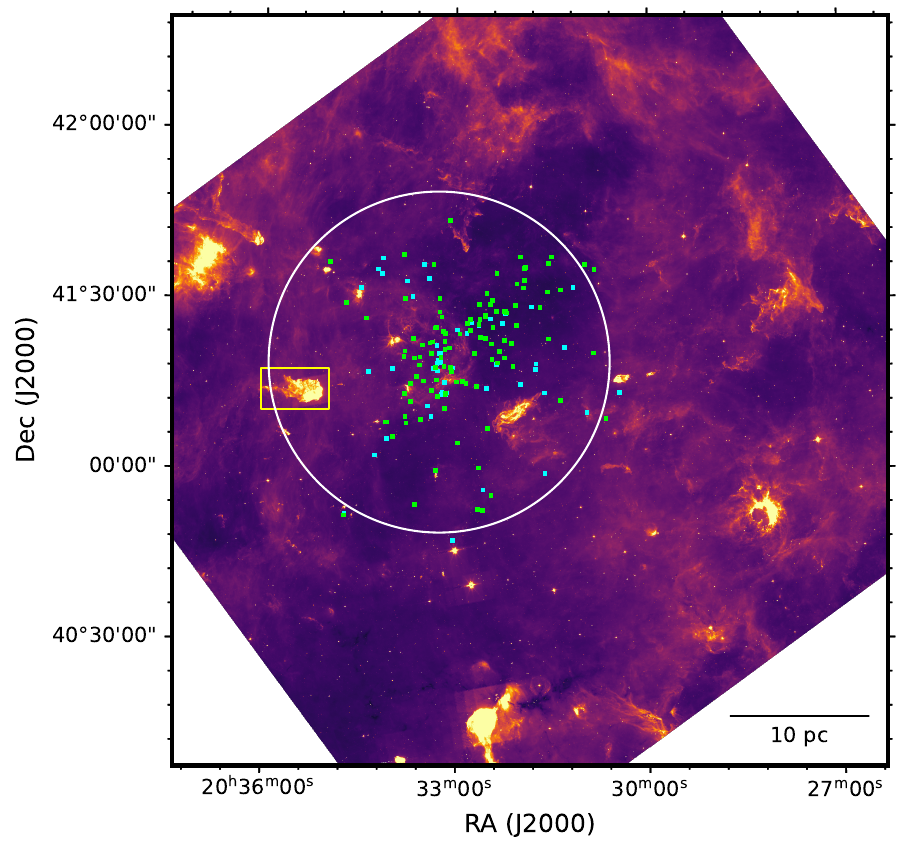}
    \caption{\textit{Spitzer}/IRAC 8\,$\mu$m emission map (in a flux range of 0 -- 98 MJy/sr) of a part of the Cygnus-X complex \citep{Beerer2010_cygnus_x}. The yellow rectangle outlines DR18, which is studied here. The white circle has a radius 12.2\,pc 
    (at a distance of 1.4\,kpc), with its center at the position of the trapezium (Cyg OB2 \#8, R.A. $20^{\rm h}33^{\rm m}16^{\rm s}$ and Dec. $+41^{\circ}18'45''$) of O stars. The cyan and bright green markers are known members of O- and B-type stars, respectively, in the OB2 association \citep{Wright2015}.}
    \label{fig:ob2}
\end{figure*}

The creation of substructures at the interface between an expanding \hii\ region and its native molecular clouds is clear evidence of stellar feedback. Prominent substructures may be comprised of pillars resembling column-like features that are still connected to their natal molecular clouds, with sizes ranging from 0.5\,pc up to a few pc. Unlike pillars, globules are isolated and have a cometary-like shape with a head-tail morphology \citep[e.g.,][]{Schneider2016_herschel_pillars_globules}. These pillars and globules mainly host low-mass star formation \citep[e.g.,][]{Hester1996_pillar, White1999_pillar}, while only a few among them may contain high-mass or intermediate-mass early B-type stars \citep[e.g.,][]{Schneider2012,Djupvik2017,Comeron2022_DR18, Schneider2021_globules}. (Magneto)-hydrodynamic simulations have demonstrated that turbulence in UV-irradiated environments plays a crucial role in the formation of pillars and globules \citep{Gritschneder2010_simulation, Tremblin2012_simulation_turbulence1, Tremblin2012_simulation2}. 

\cite{Schneider2016_herschel_pillars_globules} characterized and classified pillars, globules, and evaporating gaseous globules (EGGs), which are in the form of small globules, along with proplyd-like objects resembling evaporating circumstellar disks, according to their physical properties determined with the \textit{Herschel} far-infrared observations at 70\,$\mu$m -- 500\,$\mu$m toward the Cygnus OB2 (Cyg OB2) association, under the \textit{Herschel} imaging survey of OB Young Stellar objects (HOBYS) program \citep{Motte2010_hobys}. The Cyg OB2 association significantly affects the surrounding medium, especially the nearby molecular clouds, by forming pillars and globules \citep{Schneider2016_herschel_pillars_globules}. DR18 is one of the largest globules in the Cygnus-X region and is located close to the OB2 association region, which contains 169 primary OB stars (age $\sim$ 1--7\,Myr, \citealt{Wright2015}) and it is clearly visible in the \textit{Spitzer}/IRAC 8\,$\mu$m emission shown in Fig.\,\ref{fig:ob2}. 
\cite{Schneider2006_co_observation} also show that DR18 is associated with the Cygnus-X region according to the CO observations and, thus, for the DR18 distance, we adopted a distance of 1.4\,kpc determined by trigonometric parallaxes of 6.7\,GHz \ce{CH3OH} masers \citep{Rygl2012_distance}, which is also consistent with the results of \cite{Dzib2013_dist} measuring the VLBI parallax toward Cyg OB2 \#5. This globule is exposed to the high-UV radiation field ($\sim$1000\,G$_{0}$\footnote{G$_0$ is a measure of the average far-UV (6 -- 13.6 eV) interstellar radiation field; G$_0$ $= 1$ in Habing units corresponds to a flux of $1.86\times10^{-3}$\,erg\,cm$^{-2}$\,s$^{-1}$ obtained by integrating the interstellar radiation field density between 912 and 2400\,\AA\ \citep{Le_Petit_2006_meudon}.}, \citealt{Schneider2016_herschel_pillars_globules}). Despite such strong UV radiation, this region contains several young stellar objects (YSOs) \citep{Wright2015, Comeron2022_DR18}. In addition, its surroundings are filled by low-density ionized gas having filament-like structures (\citealt{Piddington1952_cygnusx,Emig2022_MeerKat}; see \citealt{Reipurth2008_overview_cygnus} for an overview). This makes the DR18 region a perfect laboratory to investigate how stellar feedback, such as turbulence and UV radiation, influences the chemical complexity of a molecular cloud and its evolution. 

\begin{table*}[!ht]
        \centering
        \tiny
        \caption{\label{tab:mol_info} Molecular transitions detected from the NOEMA$+$30~m combined data toward DR18.}
        \begin{tabular}{l c . c . c c c c}
        \hline \hline
    Molecule & Transition & \multicolumn{1}{c}{Rest Frequency} & \multicolumn{1}{c}{$\theta_{\rm maj}\times\theta_{\rm min}$} & \multicolumn{1}{c}{PA}&\multicolumn{1}{c}{$E_{\rm u}/k$} & \multicolumn{1}{c}{$A_{\rm ul}$} & $n_{\rm cr}$ & 1$\sigma_{\rm rms}$\\
        &     & \multicolumn{1}{c}{(GHz)} & \multicolumn{1}{c}{($''\times''$)} & \multicolumn{1}{c}{($^{\circ}$)}& \multicolumn{1}{c}{(K)} & \multicolumn{1}{c}{(s$^{-1}$)} & (cm$^{-3}$)& (mJy beam$^{-1}$)\\
    \hline 
    \ce{HCO+} & $J=1-0$  & 89.189 & $3.33\times2.65$ & 12.55 & 4.28  & $4.19\times 10^{-5}$ & $1.69\times10^5$& 7.416 \\
    HCN   & $\,\,\,J=1-0, F=2-1 ^{\dagger}$  & 88.632 & $3.35\times2.67$ & 12.61 & 4.25  & $2.41\times 10^{-5}$ & $2.19\times10^6$& 6.595 \\
    HNC   & $J=1-0$  & 90.664 & $3.28\times2.65$ & 13.44 & 4.35  & $2.69\times 10^{-5}$ & $2.76\times10^5$& 7.490 \\
    \ce{C2H}& $\,\,N=1-0, J=3/2-1/2, F=2-1 ^{\dagger}$  & 87.317 & $3.40\times2.74$ & 13.50 & 4.19  & $1.53\times 10^{-6}$ & $1.02\times10^5$ & 7.725 \\
    \ce{H2CO} & $J_{K_a, K_c}= 1_{0, 1}- 0_{0, 0}$ & 72.838 & $4.04\times3.19$ & -162.77 & 3.50 & $8.15\times 10^{-6}$ & $1.51\times10^5$ & 9.469 \\ 
    SiO & $J=2-1$ & 86.847 & $3.41\times2.75$ & 13.62 & 6.25 & $2.93\times 10^{-5}$ & $6.38\times10^5$& 6.365 \\ 
    \ce{N2H+} & $J=1-0$  & 93.174 &$3.20\times2.54$ & -165.80 & 4.47 &  $3.63\times 10^{-5}$ & $1.40\times10^5$& 8.512 \\ 
    \ce{HC3N} & $J=8-7$ & 72.784 & $4.04\times3.19$ & -162.77 & 15.72 &  $2.94\times 10^{-5}$ & $1.10\times10^6$& 9.454 \\ 
    \ce{HC3N} & $J=10-9$ & 90.979 & $3.27\times2.64$ & 13.42 & 24.01 &  $5.81\times 10^{-5}$ &$9.48\times10^5$& 6.764 \\ 
    \ce{H^13CO+} & $J=1-0$ & 86.754 & $3.42\times2.75$ & 16.70 & 4.16  & $3.85\times 10^{-5}$ & $1.55\times10^5$& 6.162 \\
    \ce{H^13CN} & $J=1-0, F=2-1$  & 86.340  & $3.43\times2.76$ & 13.66 & 4.14  & $2.23\times 10^{-5}$ & $2.03\times10^6$& 6.836 \\
    \ce{HN^13C} & $J=1-0$  & 87.091  & $3.41\times2.72$ & 12.55 & 4.18  & $1.87\times 10^{-5}$ &$1.93\times10^5$& 6.515  \\
    \ce{^13CS}  & $J=2-1$ & 92.494 & $3.22\times2.56$ & 13.39 & 6.66 &  $1.41\times 10^{-5}$ &$5.26\times10^5$& 7.572 \\
    DCN  & $J=1-0$ & 72.415 & $4.07\times3.23$ & -162.78 & 3.48 &  $1.32\times 10^{-5}$ &$5.38\times10^5$& 9.920 \\
    DNC & $J=1-0$ & 76.306 & $3.89\times3.03$ & -165.94 & 3.66 &  $1.60\times 10^{-5}$ &$1.51\times10^5$& 6.781 \\ 
    \ce{NH2D}  & $J_{Ka,Kc}  = 1_{1, 1}- 1_{0, 1}, F=2-2$  & 85.926 & $3.45\times2.77$ & 13.51 & 20.68 &  $1.76\times 10^{-5}$ &$3.79\times10^6$& 7.757 \\
    \hline
    \end{tabular}
    \tablefoot{For lines with hyperfine structure (hfs), the dagger ($\dagger$) marks the frequency of one of the hfs components. The spectroscopic information on these transitions is taken from Jet Propulsion Laboratory (JPL)\footnote{\url{https://spec.jpl.nasa.gov}} and the Cologne Database for Molecular Spectroscopy (CDMS)\footnote{\url{https://cdms.astro.uni-koeln.de}} catalogs. The 1\,$\sigma_{\rm rms}$ values are determined with a spectral resolution of 0.8\,\kms. The critical densities ($n_{\rm cr}$) are estimated with the collisional rates at a temperature of 10\,K taken from the Leiden Atomic and Molecular Database (LAMDA) \citep{lamda2020}. The collisional rates of isotopologues and deuterated species are assumed to be the same as their main isotopologue. $\theta_{\rm maj}$ and $\theta_{\rm min}$ are the major and minor FWHM sizes of a beam, respectively. PA is the position angle (east of north) of a beam.  Finally, $E_{\rm u}$ and $A_{\rm ul}$ are the energy of the upper state for a chosen transition and the Einstein coefficient, respectively.} 
\end{table*}

Previous observations of Cygnus-X have mostly focused on continuum emission from submillimeter (submm) \citep[e.g.,][]{Motte2007_mambo_cygnusx,Schneider2016_herschel_pillars_globules, Cao2019_core_surveys} to centimeter (cm) wavelengths \citep[e.g.,][]{Brunthaler2021, Emig2022_MeerKat}. Most of the molecular line observations were done with a low angular resolution ($>$\,15$''$) for this environment, mainly from CO surveys with single dish telescopes \citep{Schneider2006_co_observation,Gottschalk2012_CO}, except for the [OI] observations with a better resolution of 6$''$ \citep{Schneider2021_globules}. Several high angular resolution interferometric observations targeting molecular outflows \citep{Duarte_Cabral2013_outflow,Duarte-Cabral2014_sio_outflows, Skretas2022_outflow},  6.7 GHz \ce{CH3OH} masers \citep{Ortiz-Leon2021_6p7maser}, and \ce{H2CO} absorption line \citep{Gong2023_h2co}, the latter two studies as part of the GLObal view on STAR formation in the Milky Way (GLOSTAR) survey \citep{Brunthaler2021}. Thus, more research needs to be carried out on the dense, cold gas components of DR18 and Cygnus-X as a whole with high angular resolution to fully understand the influence of the stellar feedback and UV radiation from the OB2 association. 

This work makes use of the results of observations made in the framework of the Cygnus Allscale Survey of Chemistry and Dynamical Environments (CASCADE; see \citealt{Beuther2022_cascade_overview} for an overview), which provides a unique view of molecular gas components toward the Cygnus-X region, with high angular resolution millimeter (mm) observations of different molecular gas tracers. The observations were carried out at wavelengths of 3--4 mm, centered at $\approx 3.6$~mm (or frequencies of approximately 70.5 -- 78.2\,GHz and 85.2 -- 93.9\,GHz) with unprecedentedly high angular resolutions (3$''$--4$''$) over an area of 5.1$'\times5.1'$. The spatial resolution is approximately 0.02 -- 0.03\,pc (or $\sim$ 4000--6000\,au) at an adopted distance of 1.4\,kpc. In Sect.\,\ref{sec:obs}, we describe the observations obtained within the CASCADE program and aspects related to the data reduction process. Supplementary data sets for comparisons with our 3.6\,mm data and a description of DR18 are presented in Sects.\,\ref{sec:ancillary}. In Section \ref{sec:result_analysis}, we compare the millimeter and centimeter (mm and cm) continuum emission datasets and present molecular gas structures traced by \ce{HCO+}, \ce{N2H+}, SiO, and \ce{NH2D}. In that section, we describe compact core identification and the physical properties of the identified cores. In Section\,\ref{sec:discussion}, we discuss the origin of the free-free radiation contribution to the 3.6 mm continuum emission to understand the ionized gas structures toward DR18 and the stellar feedback to the surrounding dense molecular gas. Furthermore, we investigate the association between SiO emission and the PDRs of DR18. Lastly, we summarize our findings and emphasize the main conclusions in Section \ref{sec:summary}.

\section{Observation and data reduction}\label{sec:obs}
\subsection{IRAM 30 m and NOEMA}
The observations of the DR18 region were done with the Northern Extended Millimeter Array (NOEMA) for high angular resolution imaging and the Institut de Radioastronomie Millim\'etrique (IRAM) 30~m telescope for lower angular resolution imaging. The NOEMA observations comprised one mosaic tile consisting of 78 pointings in the C and D configurations between December 2019 and April 2020. The center coordinates of the covered field of view (FoV) of 5.1$'\times5.1'$ are R.A. 20$^{\rm h}$35$^{\rm m}$08.370$^{\rm s}$ and Dec. $+$41$^{\circ}$13$'$30.40$''$ ($l$, $b = $ 080.3622, $+$00.4479). With the goal to restore missing fluxes of extended molecular line resolved out by the interferometer, we used the IRAM 30~m telescope to cover the 5.1$'\times5.1'$ area, in on-the-fly (OTF) mode, which resulted in typical root mean square (rms) noise values of $\sim 0.1$\,K on the $T_{\rm mb}$ scale. A total bandwidth of 16\,GHz is covered in dual polarization, the lower sideband covering 70.4\,GHz to 78.2 GHz, and the upper sideband from 85.8\,GHz to 93.6\,GHz. The 3.6~mm continuum emission is obtained over the whole bandwidth with a spectral resolution of 2.0\,MHz, equivalent to a velocity resolution of $\sim$7.3\,\kms\ after masking and discarding portions of the frequency range with emission or absorption lines. In the NOEMA observations, a number of individual spectral lines targeted for the CASCADE program were covered with the higher spectral resolution of 62.5\,kHz, yielding velocity resolutions between 0.26 and 0.19\,\kms, while for the IRAM 30~m observations, the spectral resolution is consistently resampled to $\sim 0.8$\,\kms. Since this study uses the NOEMA plus IRAM 30~m combined data, we note that all analyses for spectral lines are limited to a velocity channel width of 0.8\,\kms. Table\,\ref{tab:mol_info} lists all the molecular lines detected toward DR18 and the spectroscopic parameters of the lines, along with information on the synthesized beams, as well as noise levels of 1\,$\sigma_{rms}$. All the data reductions, including calibration and imaging, for the NOEMA and 30~m telescope were performed with the Grenoble Image and Line Data Analysis Software (GILDAS) package \citep{Pety2005_gildas}. The further details of the observations with NOEMA and the IRAM 30~m telescope and data reduction procedures for the CASCADE program are described in the CASCADE overview paper (see \citealt{Beuther2022_cascade_overview}).
\begin{table*}
    \centering
    \small
    \caption{Continuum and line emission ancillary data}\label{tab:cont_ancillary}
    \begin{tabular}{l c c c c }
    \hline\hline
   Data type & Frequency/wavelength & Reference & Telescope & Beam [PA] or pixel scale \\
    \hline
Continuum &  90 GHz$^{\dagger}$  & CASCADE & GBT 100~m \& NOEMA (C + D configurations) & $3.32\times2.66''$ [$-$165\,$^{\circ}$] \\
  Continuum &  4 -- 8 GHz & GLOSTAR & JVLA (B + D configurations) & $4''\times4''$ [0\,$^{\circ}$] \\
  Continuum \& line$^*$ &  8\,$\mu$m & Cygnus-X legacy & \textit{Spitzer}/IRAC &  0.6$''$ \\
  Continuum \& line$^*$ &  4.5\,$\mu$m & Cygnus-X legacy  & \textit{Spitzer}/IRAC &  0.6$''$ \\
Continuum \& line$^*$ &  3.6\,$\mu$m & Cygnus-X legacy  & \textit{Spitzer}/IRAC &  0.6$''$ \\
  Br$\gamma$ line &  2.166\,$\mu$m & \cite{Comeron2022_DR18} & PANIC Calar Alto 2.2~m  &  0.45$''$ \\
  \ce{H2} (1--0) $S$(1) & 2.122\,$\mu$m & \cite{Comeron2022_DR18} & PANIC Calar Alto 2.2~m &  0.45$''$ \\
    \hline
    \end{tabular}\
    \tablefoot{The dagger ($^{\dagger}$) marks the central frequency used for combining the GBT MUSTANG-2 and NOEMA USB datasets. References for the GLOSTAR survey and Cygnus-X legacy program are \cite{Brunthaler2021} and \cite{Beerer2010_cygnus_x}, respectively. The \textit{Spitzer}/IRAC data of the Cygnus-X legacy program used in this study is re-processed by the Galactic Legacy Infrared Midplane Survey Extraordinaire (GLIMPSE) team\footnote{\url{https://irsa.ipac.caltech.edu/data/SPITZER/GLIMPSE/overview.html}}, and thus the pixel scale (0.6$''$ pixel scale) of the used data is smaller than the initial data (0.86$''$ pixel scale) from the Cygnus-X legacy program. }
\end{table*}

\begin{figure*}
\centering
    \includegraphics[width=0.48\textwidth]{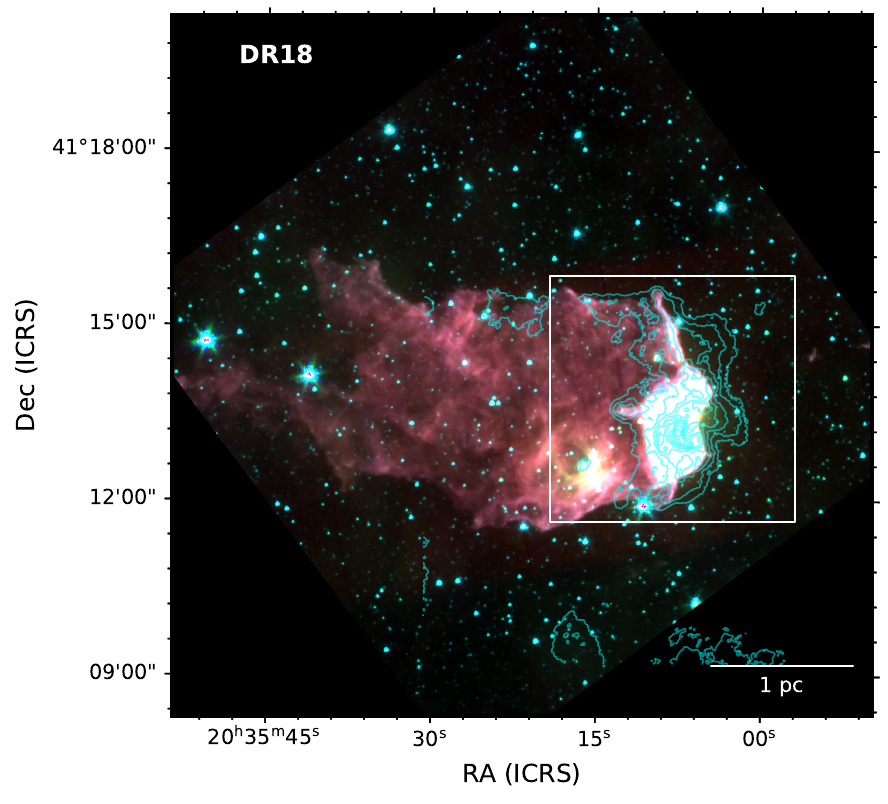}
    \includegraphics[width=0.48\textwidth]{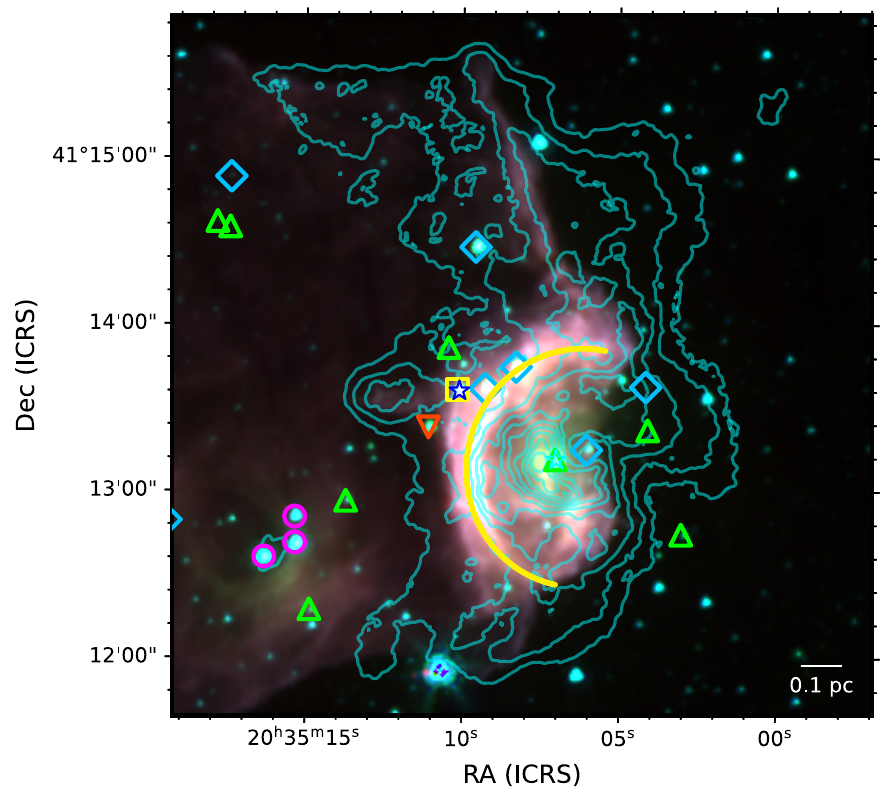}
    \caption{Three-color composite images of DR18 (8\,$\mu$m in red, 4.5\,$\mu$m in green, and 3.6\,$\mu$m in blue) with different maximum flux levels for three \textit{Spitzer}/IRAC bands to show the extended emission in a larger FoV (left panel) and the emission details, which are saturated in the left panel, in a zoom-in view (right panel) covering our 5.1$'\times5.1'$ target region (outlined by a white square box in the large FoV image). In both images, cyan contour lines represent the combined VLA B+D array 6\,cm radio continuum emission from the GLOSTAR survey by levels of 0.35, 0.70, 1.05, 1.40, 1.75, 2.10, 2.45, 2.80, 3.15, and 3.50 mJy/beam. In the zoom-in image, the different color symbols indicate young stellar objects identified by infrared colors \citep{Comeron2022_DR18}: yellow squares mark Class I YSOs, light blue diamonds for Class II YSOs, red upside-down triangle for a flat spectrum YSO, purple circles for Class III YSOs, and bright green upside triangles represent transition-disk YSOs. The yellow half-arc line indicates the open shell or the ear-like feature in the infrared emission image. }
    \label{fig:rgb_image}
\end{figure*}

\subsection{GBT MUSTANG-2}
To complement also the mm continuum data with short-spacing information, we used the MUSTANG-2 array on the Green Bank Telescope (GBT). The MUSTANG-2 instrument at GBT has a $4.2'$ instantaneous FoV comprised of a 215-element array of feedhorn-coupled TES bolometers \citep{Dicker2014_mustang2}. It has a 30 GHz bandpass with an effective bandpass center near 90 GHz 
\citep{Ginsburg2020_mustang2}. Observations for this work were performed under project GBT22A-280 on 2022 February 20 and March 14, with 3.7 hours spent on source (spread over multiple pointings). The data were calibrated and reduced with the MUSTANG
IDL Data Analysis System (MIDAS), with improvements in the iterative mapmaking pipeline \citep[e.g., see Appendix C in][for details of the pipeline]{Romero2020_MIDAS}. The primary improvements are the reduction of ringing, which is inherent in the reduction pipeline, as well as the improved recovery of larger scales.

\subsection{Merging millimeter continuum emission data obtained with  NOEMA and MUSTANG-2}
The bandpass of the MUSTANG-2 camera on the GBT  roughly covers the frequency range between 75 and 109\,GHz. With 90\,GHz being almost exactly at the center of that, we use for the combination of the data only the NOEMA USB data that are centered around 90\,GHz as well. We acknowledge that the bandpass widths of MUSTANG-2 and NOEMA are different (30 versus 8\,GHz), but that the difference is not accounted for in the combination. Having the same central frequency should make the two datasets compatible in good approximation. To combine NOEMA and MUSTANG-2 data, we tried different approaches employing both the Common Astronomy Software Applications (CASA) and GILDAS packages. The results obtained by the two approaches agreed very well. Therefore, in the following, we describe and use the data combined in GILDAS. The MUSTANG-2 data have a beam size of 9$''$ and a pixel size of 1$''$. We first reprojected the MUSTANG-2 data to the phase center of the NOEMA data and updated a few MUSTANG header parameters to the GILDAS format. For example, the rest frequency or beam size have to be explicitly given in the MUSTANG-2 data header for proper data combination. The final data combination was then conducted with the GILDAS task uv\_short using a $uv$ truncation radius of the MUSTANG-2 data of 15\,m. The synthesized beam of the combined data is 
$3\rlap{.}''32\,\times\,2\rlap{.}''66$ with a position angle of $-165^{\circ}$ (E of N). The 1$\sigma_{\rm rms}$ noise of the combined data is 0.07\,mJy\,beam$^{-1}$.

\section{Ancillary and archival data}\label{sec:ancillary}
To better understand the environment of the DR18 region, we utilized ancillary continuum data from various surveys carried out at different wavelengths. To trace the ionized gas, we used the high angular resolution radio continuum emission at 6\,cm wavelength from the GLOSTAR survey\footnote{https://glostar.mpifr-bonn.mpg.de/glostar/} \citep{Medina_glostar_2019,Brunthaler2021}. The 6\,cm radio continuum data were generated by combining the Karl G. Jansky Very Large Array (VLA) D- and B-configuration data, which yields a synthetic beam size, $\theta$, of 4$''$ and a sensitivity of $\sim$0.1\,mJy\,beam$^{-1}$. 

In addition, to investigate the properties of emission from dust and polycyclic aromatic hydrocarbons (PAHs) tracing PDRs, we compared the \textit{Spitzer} Legacy Survey of the Cygnus-X complex (Cygnus-X; \citealt{Hora2007_cygnus_x_irac, Beerer2010_cygnus_x}), which mapped a $6^{\circ}\times6^{\circ}$ area with \textit{Spitzer}/IRAC four bands (3.6, 4.5, 5.8, and 8.0\,$\mu$m; \citealt{Papovich2016_cygnusX}). We also used \ce{H2} column density and dust temperature maps obtained from \cite{Bonne2023_dust_nh2_data}, adopting the methods described in \cite{Palmeirim2013_dust_temperature}. The \ce{H2} column density is determined with the difference maps of the convolved maps from 250\,$\mu$m to 500\,$\mu$m from the HOBYS program \citep{Motte2010_hobys} and the dust temperature map is generated with the 160\,$\mu$m/250\,$\mu$m flux ratio. Lastly, we utilized 2.166\,$\mu$m Br$\gamma$ line and 2.122\,$\mu$m \ce{H2}\,(1--0)\,S(1) vibrational line obtained from \cite{Comeron2022_DR18} to identify PDRs in DR18 by comparing our NOEMA data and other ancillary datasets. Table\,\ref{tab:cont_ancillary} summarizes the continuum emission datasets and line ancillary data.

\section{Results and analysis}\label{sec:result_analysis}
The head of the DR18 globule observed in the CASCADE survey is visually striking in infrared emission, particularly in the \textit{Spitzer}/IRAC 8\,$\mu$m wavelength band, which is dominated by radiation from PAHs. The head faces the main Cyg OB2 population (see Fig.\,\ref{fig:ob2}) and exhibits excellent spatial alignment between the highly saturated 8\,$\mu$m emission and strong 6\,cm radio continuum emission obtained from the GLOSTAR survey (as shown in the larger FoV image in the left panel of Fig.\,\ref{fig:rgb_image}). Also, this mid-infrared emission image reveals an apparent hollow feature, such as an ear-like shape (delineated by a yellow curved line). In addition, from the distinct appearance of the infrared emission, several YSOs have been identified and classified by \citet{Comeron2022_DR18}, indicating active star formation inside of the globule. 
 
A group of Class II and transition-disk YSOs is clearly situated in the \hii\ region traced by the 6\,cm radio continuum emission. In the vicinity of the curved illuminated 8\,$\mu$m emission (yellow curve line), a Class I YSO (yellow square) and a flat spectrum YSO (red upside-down triangle) are found, which are younger than the other YSOs in this region. Further out toward the tail of the globule, only more evolved YSOs have been found (e.g., Class III YSOs, purple circles). A recent near-infrared spectral study conducted by \cite{Comeron2022_DR18} found the spectral type of the transition disk YSO (DR18-05) close to the 6\,cm radio continuum emission peak to be a B2 star, which is thought to have created the cavity in the head of DR18 \citep{Comeron1999_DR18,Comeron2022_DR18}. 
According to \cite{Comeron2022_DR18}, the B star could have required approximately $3\times10^4$ years to carve the arc-shape with a radius of 0.22\,pc on the tip of DR18.

\begin{figure*}
    \centering
    \includegraphics[width=0.84\textwidth]{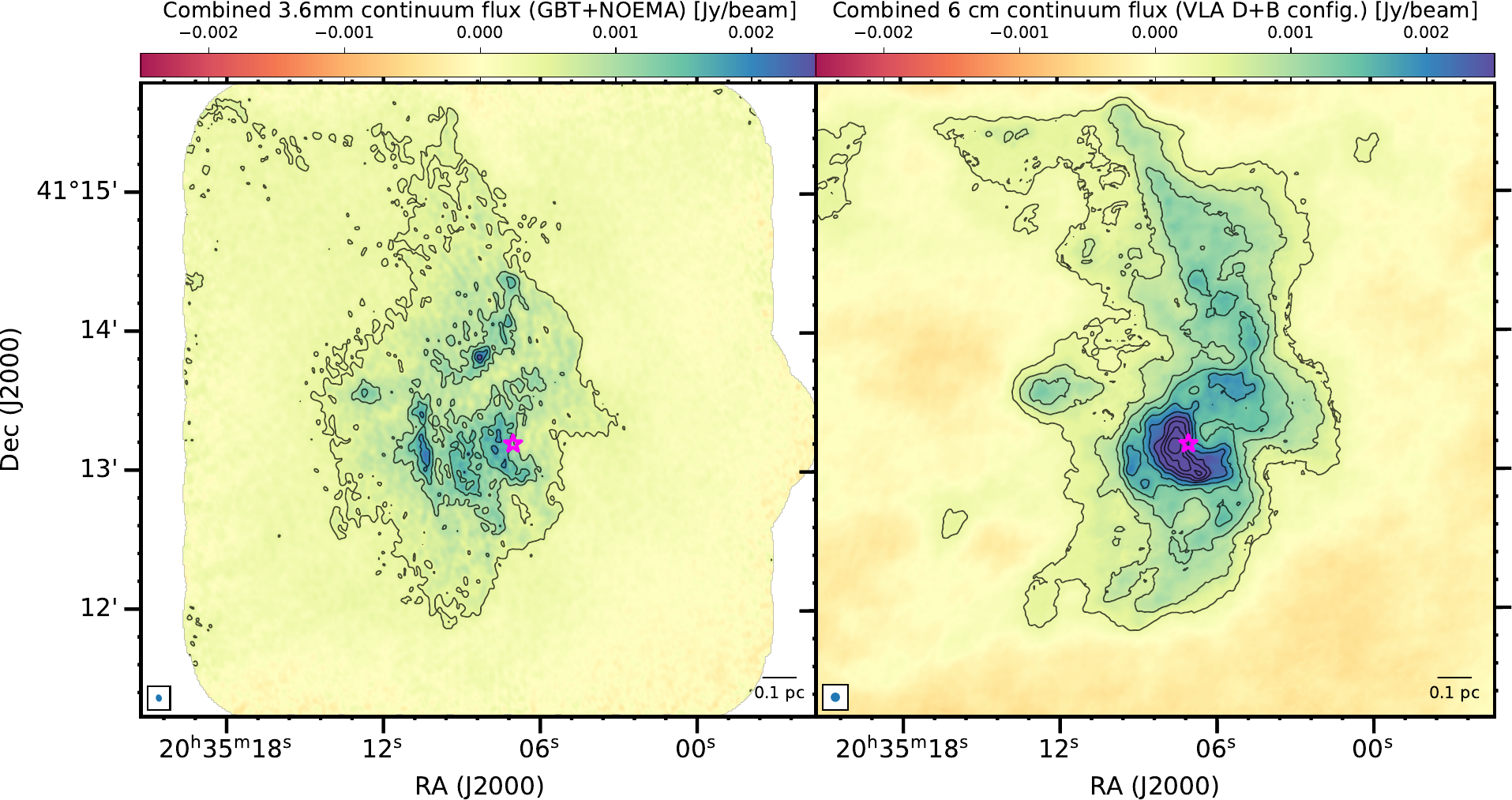}
    \caption{Millimeter (left) and centimeter (right) continuum emission maps. The mm continuum emission map represents the combination of the GBT MUSTANG2 and NOEMA data sets, whereas the cm continuum emission map is generated by combining the VLA D and B configuration observations. The FWHM synthetic beams of these maps are displayed in the lower left corners. The contour levels of the mm continuum emission are 0.46, 0.96, 1.38, 1.80, 4.14, 5.33, and 6.51 mJy~beam$^{-1}$, whereas for the cm emission, the levels are 0.35, 0.70, 1.05, 1.40, 1.75, 2.10, 2.45, 2.80, 3.15, and 3.50 mJy~beam$^{-1}$. The star symbol indicates the position of the B2 star.}
    \label{fig:mm_cm_images}
\end{figure*}
\begin{figure*}[!ht]
\centering
    \includegraphics[width=0.84\textwidth]{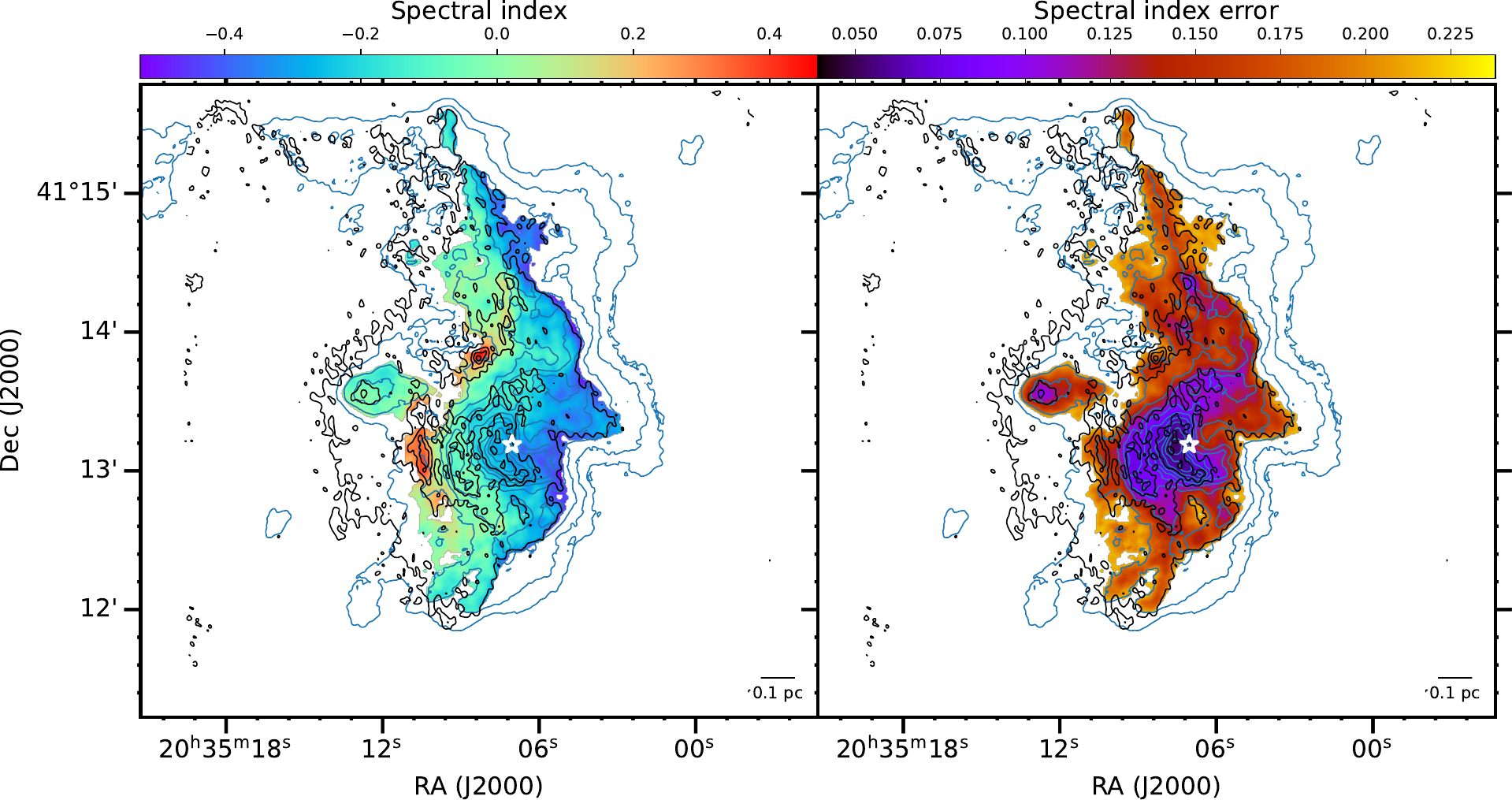}
    \caption{90\,GHz -- 5.8\,GHz spectral index map in the left panel and its spectral index error map in the right panel. The higher frequency data are smoothed to the beam size of the lower frequency data to achieve a common resolution of 4$''$. The star symbol indicates the position of the B2 star. The black contours are the 3.6\,mm continuum flux, and their levels are the same as in the left panel of Fig.\,\ref{fig:mm_cm_images}, and the blue contours are the 6\,cm continuum emission with the same contour levels shown in the right panel. }
    \label{fig:gbt-glostar}
\end{figure*}

\subsection{Dust thermal and free-free emission contributions to 3.6\,mm continuum emission}
\begin{figure*}[!ht]
    \centering
    \includegraphics[width=0.44\textwidth]{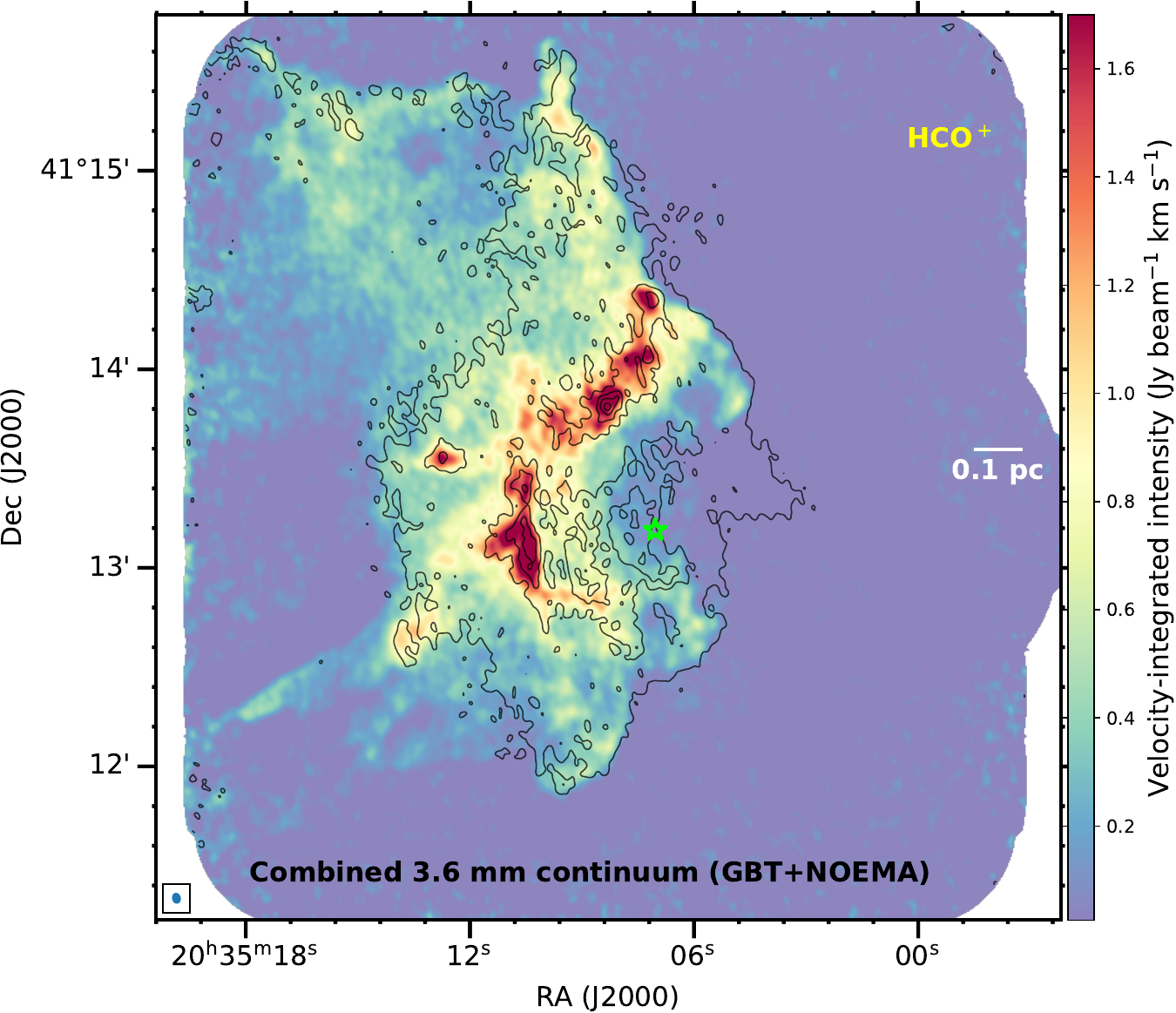}
    \includegraphics[width=0.44\textwidth]{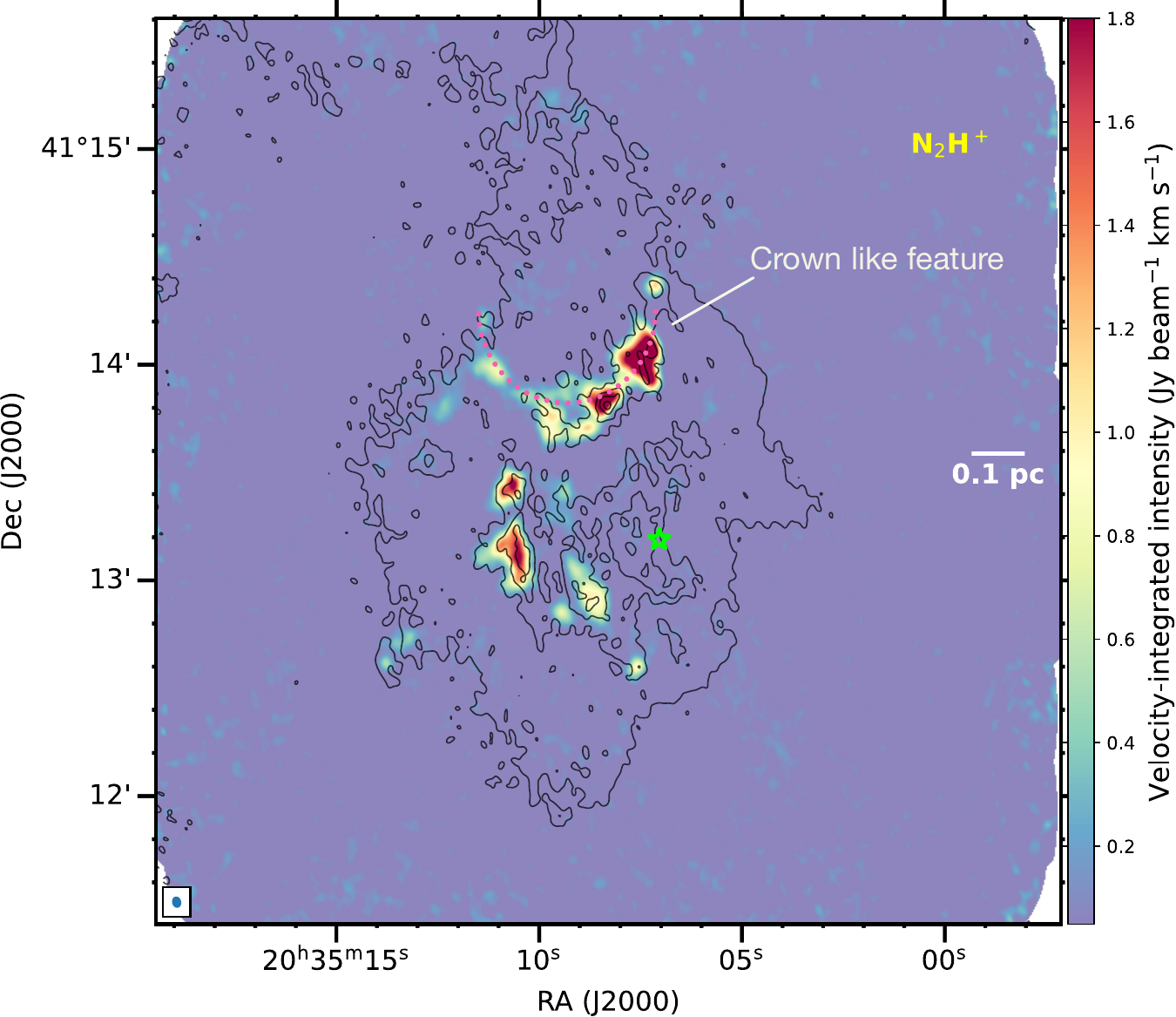}
    \includegraphics[width=0.44\textwidth]{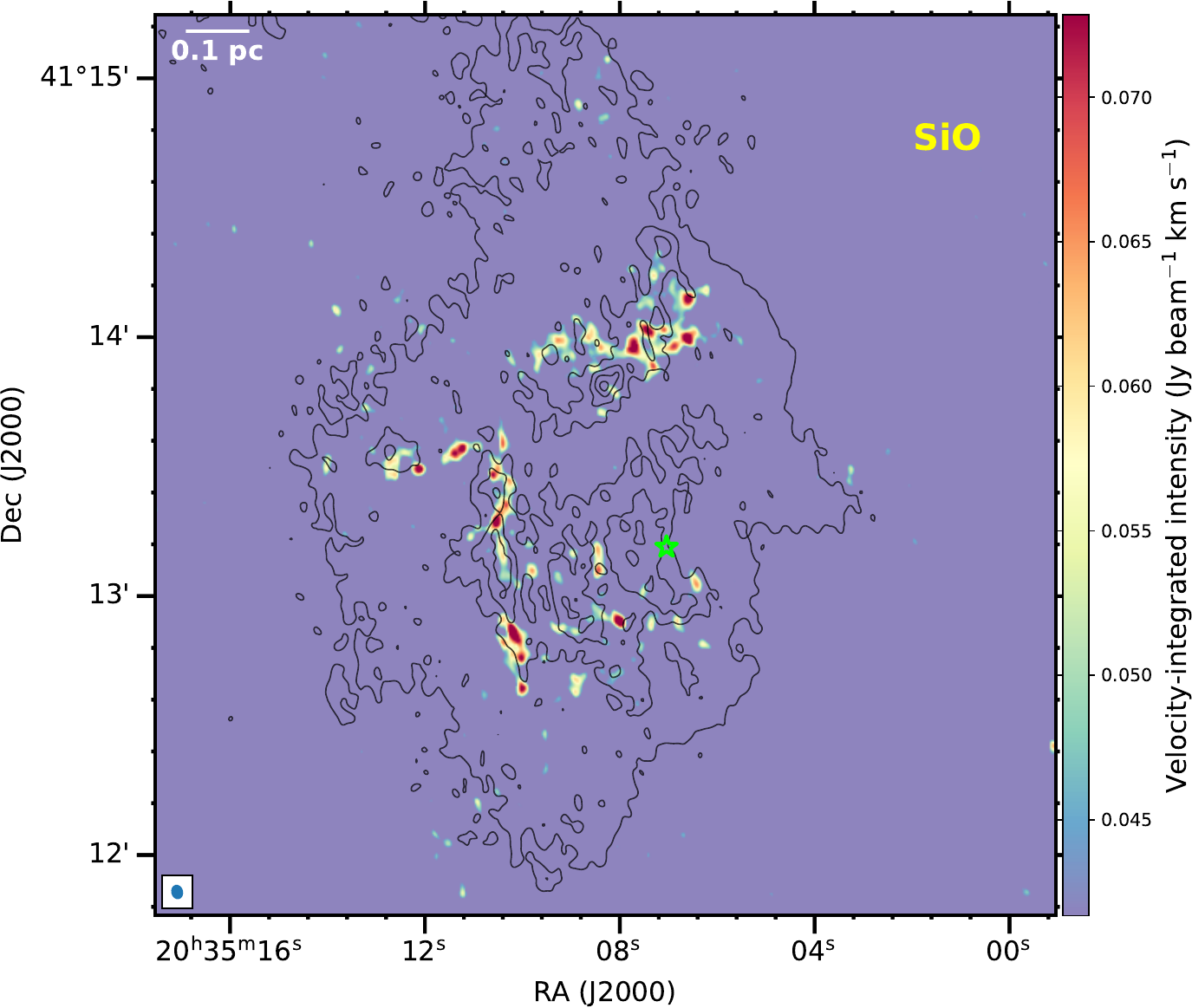}
    \includegraphics[width=0.44\textwidth]{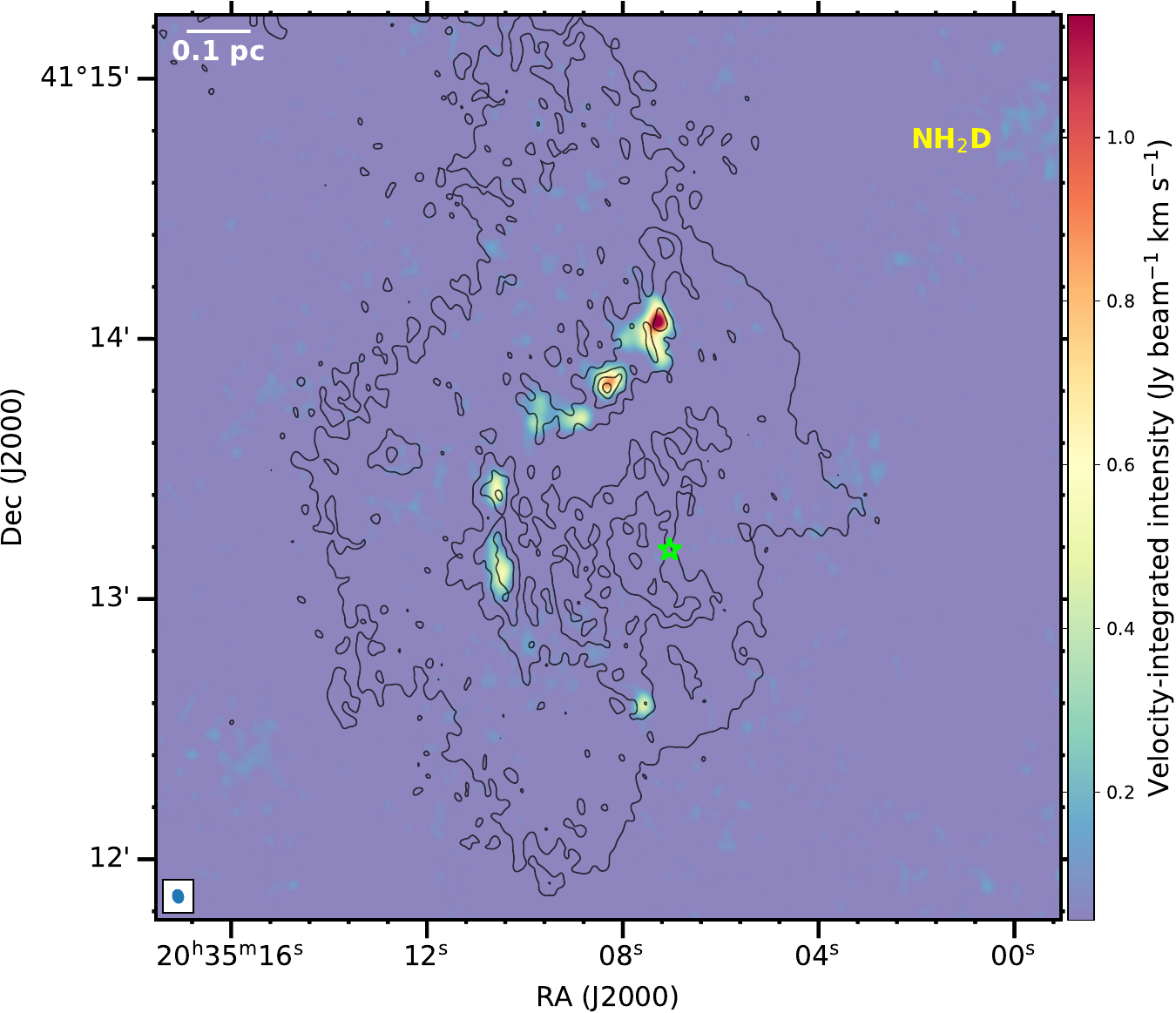}  
    \caption{Velocity-integrated intensity maps of \ce{HCO+} (6\,\kms\ to 15.6\,\kms), \ce{N2H+} ($-$1.2\,\kms\ to 17.2\,\kms\ spanning all the hfs transitions) in the upper row and SiO (8\,\kms\ to 10\,\kms) and \ce{NH2D} (0\,\kms\ to 15.2\,\kms\ covering all the hfs transitions) in the lower row. The black contours represent the GBT+NOEMA combined 3.6\,mm continuum emission, and the bright green marker indicates the position of the ionizing source, DR18-05. The contour levels of the mm continuum are the same as shown in Fig.\,\ref{fig:mm_cm_images}. The horizontal white scale bar has a size of 0.1\,pc at a given distance of 1.4\,kpc. The beam of the maps of each species is displayed on the left corner of each image panel. }
    \label{fig:mom0_maps}
\end{figure*}

 Figure\,\ref{fig:mm_cm_images} shows the 3.6\,mm continuum and 6\,cm continuum emission maps from this study and the GLOSTAR survey (from left to right). While the 3.6\ mm and the 6\,cm continuum emission are similarly extended, it is notable that the 6\,cm emission extends more to the west compared to the 3.6\,mm emission.  On the other hand, many of the bright, compact features on the 3.6\,mm continuum emission do not have significant 6\,cm emission counterparts. The overall emission distributions at these two wavelengths imply that the 3.6\, mm continuum emission potentially has significant contributions from free-free emission originating in the ionized gas, apart from the dust thermal emission. 

To examine the contribution of free-free radiation emission to the 3.6\,mm continuum emission, we analyzed spectral indices between  6\,cm and 3.6\,mm wavelengths using the 6\,cm GLOSTAR combined the D$+$B VLA array continuum image and the 3.6\,mm MUSTANG-2$+$NOEMA combined continuum image. To compare both datasets, we re-convolved the MUSTANG-2$+$NOEMA combined data ($3\rlap{.}''32\,\times\,2\rlap{.}''66$ with a position angle of $-165^{\circ}$) to the exact angular resolution (4$''$) of the VLA D$+$B combined data, and then resampled the reconvolved MUSTANG-2$+$NOEMA data to match the pixel size of the VLA data. This spectral index measurement consists of fitting a simple slope to the continuum fluxes at the central frequencies  of the two datasets (5.8\,GHz for the GLOSTAR survey and 90\,GHz for the CASCADE survey) in logarithmic scale, given by the formula 
\begin{equation}
    \alpha_{\rm 90\,GHz - 5.8\,GHz} = \frac{{\rm log}(S_{\rm 90\,GHz}/{S_{\rm 5.8\,GHz})}}{\rm log(90\,GHz/5.8\,GHz)}.
\end{equation}
The $S_{\rm 90\,GHz}$ and $S_{\rm 5.8\,GHz}$ are the continuum flux densities per pixel. The spectral index uncertainty is the propagating error estimated as 
\begin{equation}
    \sigma_{\rm 90\,GHz - 5.8\,GHz} = \frac{\sqrt{(3\sigma_{S_{\rm 90\,GHz, rms}}/S_{\rm 90\,GHz})^2+(3\sigma_{S_{\rm 5.8\,GHz, rms}}/{S_{\rm 5.8\,GHz})^2}}}{{\rm log(90\,GHz/5.8\,GHz)}}.
\end{equation} 
The $\sigma_{S_{\rm 90\,GHz}} $ and $\sigma_{S_{\rm 5.8\,GHz}}$ are the continuum flux uncertainties corresponding to the frequencies. Instead of using 1\,$\sigma_{\rm rms}$ observational uncertainty for the spectral index uncertainty, we applied 3\,$\sigma_{\rm rms}$ threshold because the GLOSTAR continuum emission data used for the spectral index does not include the total-power emission data, which is more sensitive to extended emission structures. This concerns missing fluxes for faint, extended emission in the outer regions of the cm continuum emission image. To mitigate this, we restricted our analysis to regions with emission above 3\,$\sigma_{\rm rms}$ at 5.8 and 90\,GHz, respectively. 

The determined spectral index in large parts is consistent with free-free emission close to a spectral index of $\alpha=-0.1$ with several compact regions that show a clear dust thermal contribution with positive spectral indices. While the western edge shows slightly more negative values than those values expected from dust thermal emission and free-free emission regions, these parts also have larger uncertainties approaching 0.2. Furthermore, these values must be taken cautiously because of uncertainties introduced by different UV coverages of both observations. In general, the spectral index is between $-0.1$ and 2.0 for typical \hii\ regions and in a range of $2.0\,<\alpha<4.0$ for sources with pure dust thermal emission (e.g., \citealt{sanchez_monge2017_spectral_index}). We note that the compact regions with thermal emission are found exclusively east of the B2 star, while the western part of the emission, which faces the OB2 association, is dominated by free-free emission.

\subsection{Molecular line profiles and emission
distributions}\label{sec:molecular_lines}
The CASCADE observations have provided, for the first time, a high angular resolution view of spectral line emission in the DR18 region, allowing for detailed investigations of its molecular gas content to be undertaken on physical scales of $\sim$ 0.02 -- 0.03\,pc. Figure\,\ref{fig:mom0_maps} shows the velocity-integrated maps of \ce{HCO+}, \ce{N2H+}, SiO, and \ce{NH2D} emission lines with the overlaid mm continuum emission in black contours. The intensity maps of the other species detected in this region are presented in Figs.\,\ref{appendix:mom0_extend_gas} and \ref{appendix:mom0_dense_gas}. The \ce{HCO+} intensity integrated over a velocity range from 6\,\kms\ to 15.6\,\kms\ shows an extended distribution and exhibits the arc-shape morphology resembling the bright 8\,$\mu$m emission feature indicated by a yellow arc line in the right panel of Fig.\,\ref{fig:rgb_image}. In addition,  weaker \ce{HCO+} emission extends toward the tail of the DR18 visible in the infrared emission. Such spatially widespread emission distributions are also found in the intensity maps of CCH, \ce{H2CO}, HCN, and HNC exhibited in Fig.\,\ref{appendix:mom0_extend_gas}. The outline of \ce{HCO+} line emission follows the morphology of the extended 3.6\,mm and 6\,cm continuum emission regions. Such spatial coincidence features between molecular gas and free-free emission are also found toward the pillars in other OB association regions, such as the elephant trunks in M16 \citep{Sofue2020}, as well as toward an ultracompact \hii\ region Mon R2\,\citep{Trevino-Morales2016_MonR2}. The bright \ce{HCO+} emission peaks spatially match with the 3.6\,mm compact continuum emission features, which also coincide significantly with bright features in the \ce{N2H+} and \ce{NH2D} maps. 

\begin{figure}[!ht]
    \centering
    \includegraphics[width=0.5\textwidth]{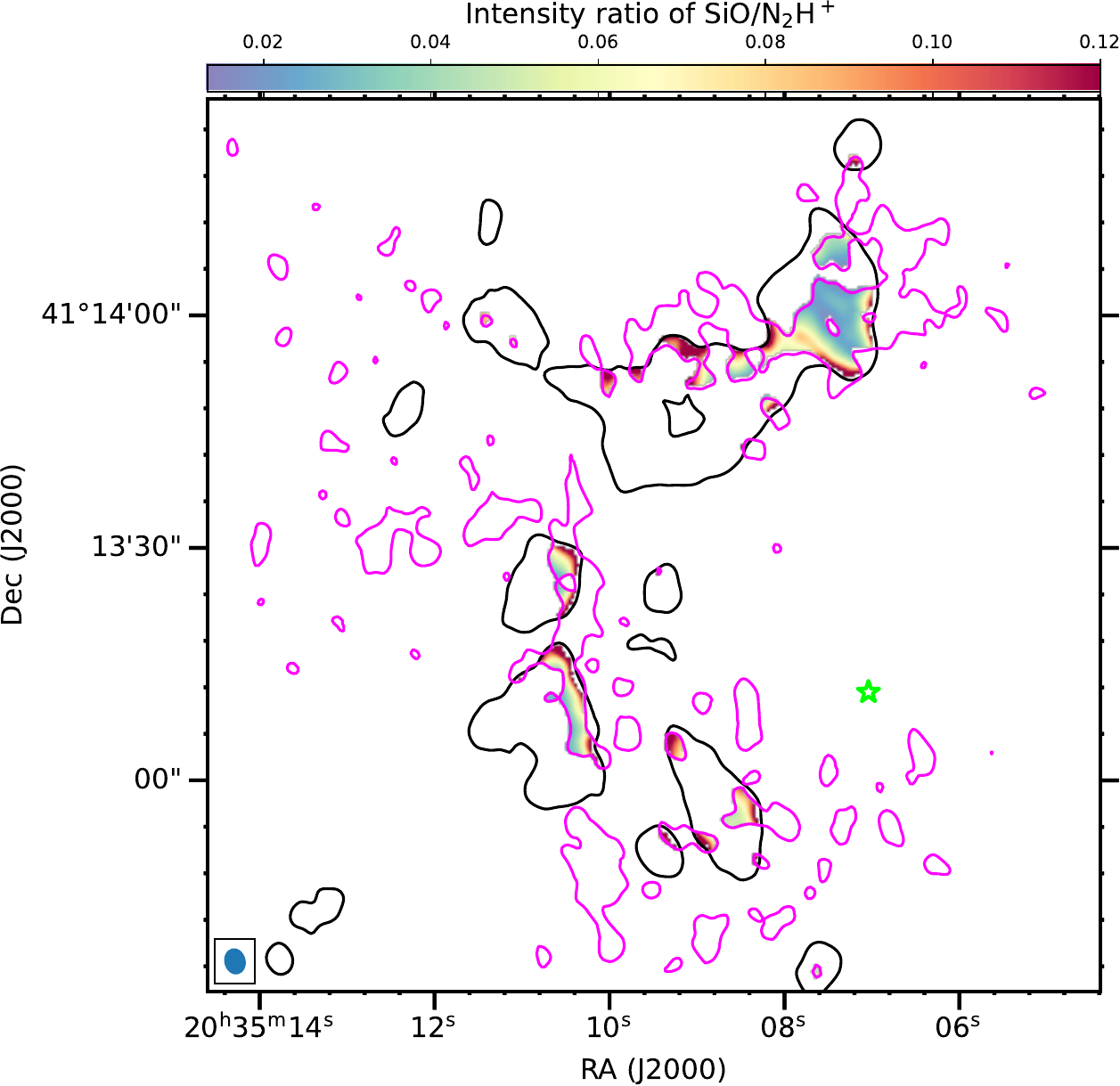}
    \caption{Intensity ratio map of SiO/\ce{N2H+} in color scale for the overlapping areas of both molecular species. The black and pink contours represent the 5\,$\sigma$ level of \ce{N2H+} velocity-integrated intensity and a 3\,$\sigma$ level of SiO velocity-integrated intensity, respectively. The bright green star indicates DR18-05, the ionizing source in DR18.}
    \label{fig:sio_n2hp_ratio}
\end{figure}


In contrast, \ce{N2H+} line emission better coincides with the compact mm continuum emission (see the upper right panel in Fig.\,\ref{fig:mom0_maps}) rather than with pure free-free emission found at cm wavelengths. This supports that the continuum emission features traced by dense molecular tracers ($n_{\rm cr}>10^5$\,cm$^{-3}$) represent dust thermal emission rather than free-free emission. These dense gas regions also have counterparts in the emission from other molecular lines (as shown in Fig.\,\ref{appendix:mom0_dense_gas}), including lines from deuterated species; for example, from \ce{NH2D} in the right bottom panel in Fig.\,\ref{fig:mom0_maps}. Interestingly, some of the compact emission components are aligned as a ``crown-like" feature, indicated by the magenta-dotted curve opening toward the northern direction in the \ce{N2H+} intensity map. The feature is also pronounced in the HNC, \ce{HN^{13}}, \ce{HC3N}, and \ce{H^{13}CO+} emission maps. The eastern portion of the crown-like feature does not show any continuum emission counterpart and only appears in those species. \ce{NH2D}, DCN, and DNC shown in Figs.\,\ref{fig:mom0_maps} and \ref{appendix:mom0_dense_gas} are also detected toward the dense molecular gas regions. However, DCN shows weaker emission and is spatially less matched with DNC and \ce{NH2D} emission. 


\begin{figure}[!ht]
    \includegraphics[width=0.44\textwidth]{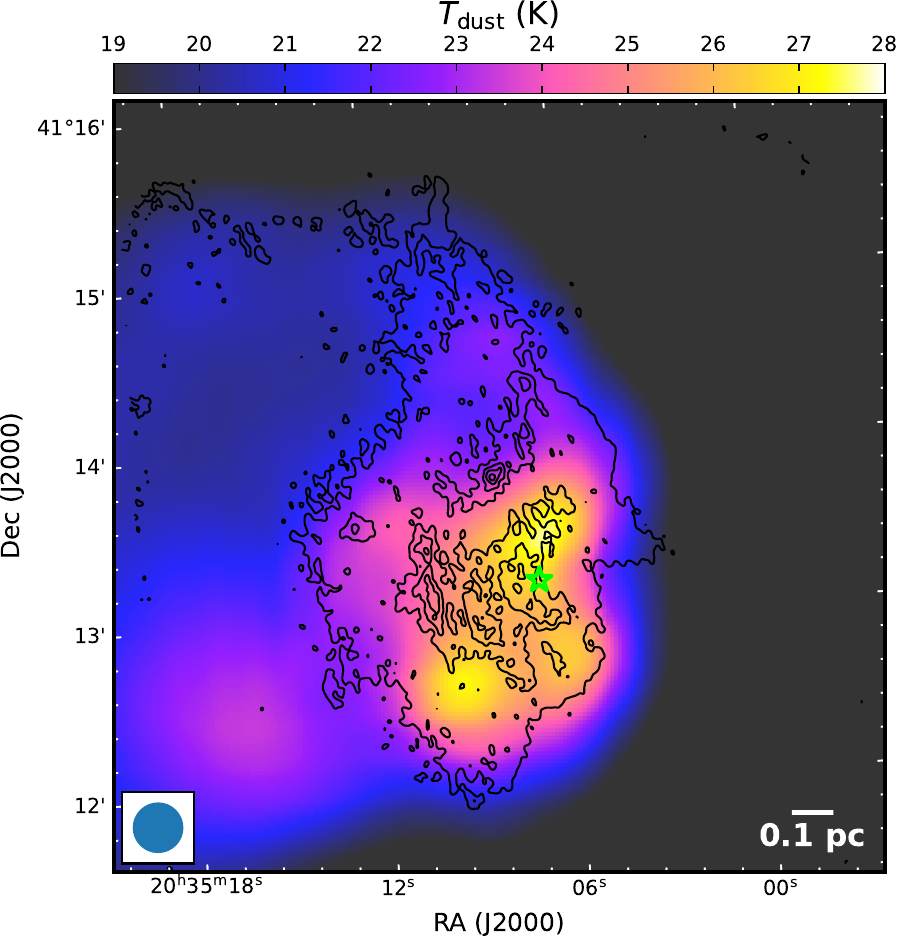}
    \vskip 0.3cm
    \includegraphics[width=0.44\textwidth]{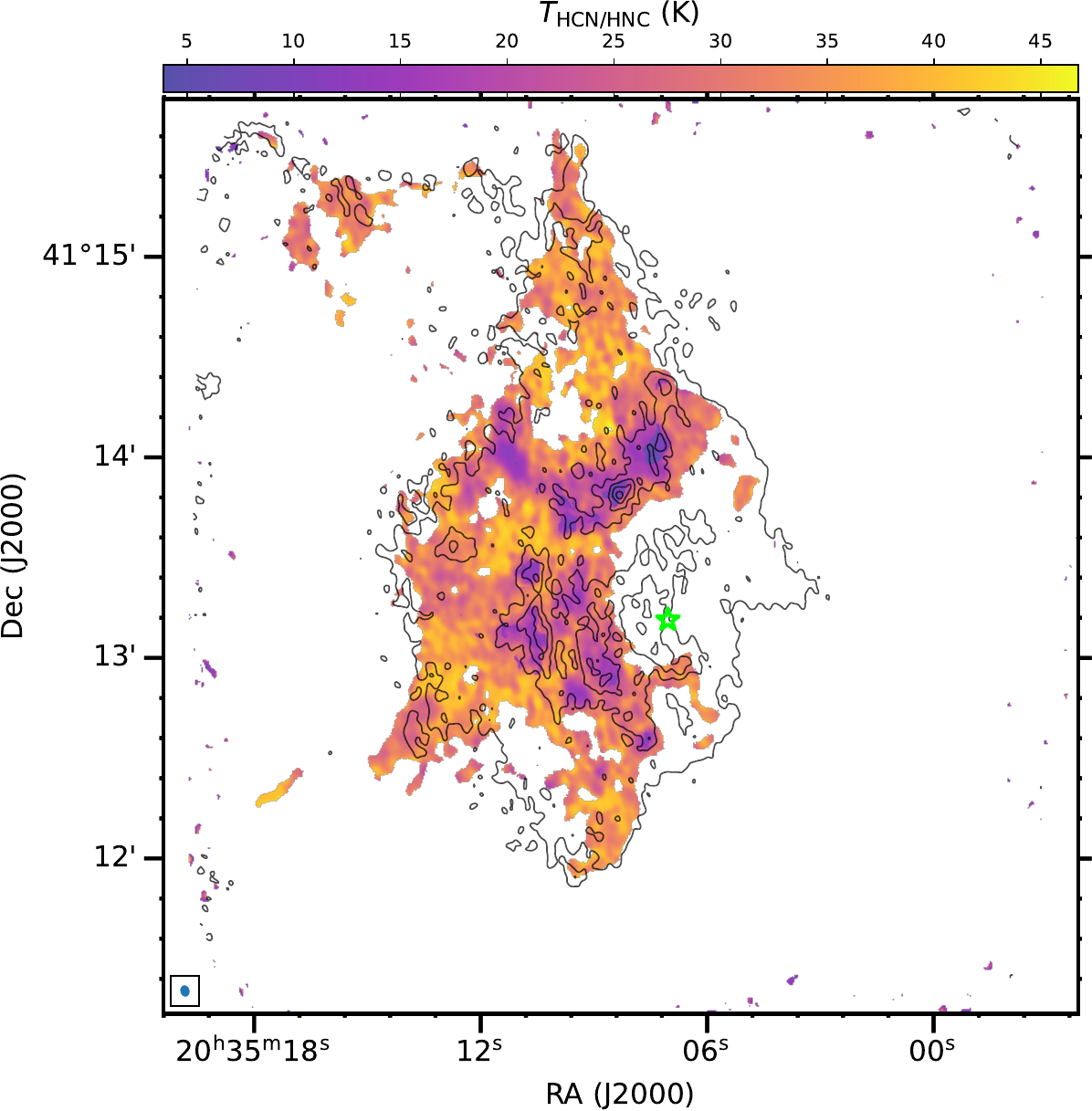}
    \caption{Dust and gas temperature maps. Top: Dust temperature map obtained from \cite{Bonne2023_dust_nh2_data} generated with the 160\,$\mu$m/250$\mu$m flux density ratio as described in the work by \cite{Palmeirim2013_dust_temperature}. Bottom: Gas kinetic temperature map combined temperature maps derived from the HCN/HNC and \ce{H^13CN}/\ce{HN^13C} emission line ratios. The black contours represent the 3.6\,mm continuum emission. The bright green star indicates the position of the B2-type star D18-05. The beams for the $T_{\rm dust}$ and $T_{\rm HCN/HNC}$ maps are 18$''$ and $3.43\times2.76$ with a position angle of $13.66$\,$^{\circ}$ displayed in the lower left corner. }
    \label{fig:T_hcn_hnc}
\end{figure}


The left bottom panel of Fig.\,\ref{fig:mom0_maps} displays the SiO emission integrated over a velocity range from 8\,\kms\ to 10\,\kms. The SiO emission has a remarkably narrow linewidth, considering that SiO emission is often used as a tracer of shocked gas and this type of emission has very broad line wings \citep[e.g.,][]{Martin-Pintado1992_SiO_outflow,Bachiller1997_SiO_outflow,Codella1999_SiO_outflow,Csengeri2016_atlasgal_sio}. Not only does SiO have a narrow linewidth, but its emission shows clumpy distributions concentrated in spatially compact regions around the \hii\ region of D18-05. Compared to other molecules, the distribution of SiO emission is distinct and spatially separated from that of other dense gas tracers. The SiO emission peaks do not coincide with 3.6\,mm continuum emission peaks or the bright \ce{N2H+} emission components and appear in regions without dense gas, unlike other dense gas-tracing molecular lines, which are preferably detected toward the mm continuum emission peaks. There are only a few weak SiO detections toward the compact mm features. The north of the \hii\ region shows slightly more extended SiO emission. Figure\,\ref{fig:sio_n2hp_ratio} shows the intensity ratio map of SiO over \ce{N2H+} lines. It clearly shows that the intensity ratios increase outwards of the \ce{N2H+} emission regions (close to the black contours), indicating the relatively weaker \ce{N2H+} and brighter SiO in the regions. On the other hand, in the inward parts of the \ce{N2H+} emission regions, the ratios decrease. This pattern appears consistently in all the areas having detections of both species. We explore the origin of this narrow line SiO emission in Section\,\ref{sec:disscusion_sio}. 

\subsection{Gas temperature distribution}

Based on the dust temperature map (shown in the upper panel of Fig.\,\ref{fig:T_hcn_hnc}) taken from \cite{Bonne2023_dust_nh2_data}, we know that toward the interior \hii\ region of DR18 the dust temperatures are higher than 23\,K and can reach up to 28\,K. However, due to the larger angular resolution of 18$''$ \citep{Bonne2023_dust_nh2_data} than the NOEMA synthesized beams, it is difficult to investigate the temperature structure toward the \ce{N2H+} cores. However, it is still evident that the cores are relatively colder than the interior of the \hii\ region, given the large spatial resolution. As these cores are expected to be denser, colder parts in the DR18 region, we expect that they have lower gas temperatures than those obtained from the dust temperature map. To determine the temperature distributions of extended and dense molecular gas components, we used the line emission ratios of HCN (1--0) and HNC (1--0) and their isotopologues, as suggested by \cite{Hacar2020_Tgas}. The authors show that the ratio of these ground state transitions of the two species for values from 1 to 4 (corresponding to temperatures between 10 and 40 K) provides a good approximation of gas kinetic temperature, as derived from the \ce{NH3} inversion line data toward the Orion molecular cloud. Unlike other molecular transitions, the  HCN and HNC lines have similar distributions as the \ce{HCO+} emission and ridge structures in the $N$(\ce{H2}) map, with $N$(\ce{H2}) $>$ a few $\times 10^{22}$\,cm$^{-2}$. To generate HCN/HNC 1--0 ratio maps, we smoothed HCN, HNC, and \ce{HN^{13}C} maps with the beam size of \ce{H^{13}CN}, which has the largest beam size among these lines. After the spatial smoothing, we sequentially performed regridding all the smoothed data to the same pixel size and coordinates of \ce{H^{13}CN}. From the HCN/HNC and \ce{H^{13}CN}/\ce{HN^13C} maps, we generated gas temperature maps following equations presented in \cite{Hacar2020_Tgas}:
\begin{equation}
    T_{\rm HCN/HNC}~{\rm (K)}=10\times \left[ \frac{I(\rm HCN)}{I(\rm HNC)}\right]~~{\rm if}~~ \frac{I(\rm HCN)}{I(\rm HNC)} \leq 4
,\end{equation}

\begin{equation}
    T_{\rm HCN/HNC}~{\rm (K)}=3\times \left[ \frac{I(\rm HCN)}{I(\rm HNC)} - 4\right] + 40~~{\rm if}~~ \frac{I(\rm HCN)}{I(\rm HNC)} > 4.
\end{equation}

The bottom panel of Fig.\,\ref{fig:T_hcn_hnc} displays a map of the gas temperatures in high angular resolution ($\sim 3\,''$) obtained by combining both gas kinetic temperature maps (see Fig.\,\ref{fig:appendix_gas_temp}) determined with the HCN/HNC and \ce{H^13CN}/\ce{HN^13C} line ratios. The combination was performed by substituting the temperatures of HCN/HNC with those of \ce{H^13CN}/\ce{HN^13C}. The study by \cite{Hacar2020_Tgas} mentions that the expected uncertainty with this method is $\Delta T_{\rm HCN/HNC} \approx 5$\,K for temperatures ranging from $10 < T_{\rm HCN/HNC} < 40$\,K and $\Delta T_{\rm HCN/HNC}\approx 10$\,K for temperatures above 40\,K. The effect of optical depth for HCN and HNC can be ignored because the highest column density of hydrogen molecules toward the DR18 region is not above $4.2\times10^{22}$\,cm$^{-2}$, which corresponds to a visual extinction,  $A_{\rm V}$, of 42\,mag calculated using the $A_{\rm V} = N$(\ce{H2})$/9.4\times10^{20}$ cm$^{-2}$. The study by \cite{Hacar2020_Tgas} states that for regions with up to $A_{\rm v}\approx100$, their method provides HCN/HNC line ratios that are still acceptable. Nevertheless, toward higher \ce{H2} column density regions, we estimated the gas kinetic temperatures using  \ce{H^13CN} and \ce{HN^13C} emission lines, whose optical depths are smaller than 0.2--0.3.

Figure\,\ref{fig:T_hcn_hnc} shows the kinetic temperature map determined from HCN/HNC and \ce{H^13CN}/\ce{HN^13C} ratios. In the $T_{\rm HCN/HNC}$ map, the east of the \hii\ region and the crown-like structure appearing in the \ce{N2H+} intensity map (Fig.\,\ref{fig:mom0_maps}) show colder temperatures 5$\,{\rm K} \lesssim T_{\rm HCN/HNC} \lesssim 25$\,K, whereas the $T_{\rm dust}$ map shows slightly higher temperatures of 25 -- 27\,K. In contrast to the dense regions, the extended molecular gas regions show higher temperatures as high as above 40\,K in the $T_{\rm HCN/HNC}$ map compared with those $\sim 22$\,K -- 26\,K in the $T_{\rm dust}$ map. This temperature difference occurs because the far-infrared wavelengths between 160\,$\mu$m and 500\,$\mu$m used for determining the dust temperature map are not sensitive to warmer gas above 40\,K. In addition, higher gas temperatures have been observed toward the extended features and between the \ce{N2H+} cores, which are considered to be the lower gas density regions and highly UV-illuminated regions as seen in Fig.\,\ref{fig:rgb_image}. Thus, in such regions, dust and gas could be decoupled as observed in other \hii\ and photodissociation regions \citep[e.g.,][]{Koumpia2015_s140,Salgado2016_orion}. However, as mentioned above, in these regions with $T_{\rm HCN/HNC}>$ 40\,K, the uncertainties are higher about 10\,K. Nevertheless, the high-temperature values in the $T_{\rm HCN/HNC}$ map could be unlikely to be artifacts and might imply an interaction between the ionized gas and molecular gas, with careful considerations of the absolute gas temperature values. 

\begin{table*}[!ht]
\centering
\tiny
    \caption{\label{tab:cores}Summary of the extracted \ce{N2H+} cores and physical properties.}
    \begin{tabular}{c c c c c c c c c c c c c c}
    \hline \hline
    Cores & R.A. (J2000) & Dec. (J2000) &  $\theta_{\rm core}$ & $R_{\rm core}$ & $\upsilon_{\rm \ce{N2H+}}$ & $\Delta\upsilon_{\rm \ce{N2H+}}$ & $T_{\rm ex}$ & $N$(\ce{N2H+})$^a$& $T_{\rm HCN/HNC}$ & $N$(\ce{N2H+})$^b$ & $M_{\rm core}$ & $M_{\rm vir}$ & $\alpha_{\rm vir}$\\  
      & (hh:mm:ss) & (dd:mm:ss)& ($''$) & (pc) & (\kms) & (\kms) & (K) & ($10^{13}$\,cm$^{-2}$) & (K) & ($10^{13}$\,cm$^{-2}$) & (M$_{\odot}$) & (M$_{\odot}$) &  \\
    \hline
        C1 &  20:35:07.60  & +41:12:35.76 & 5.98  & 0.020 & 7.4 & 0.9 & 26 & 2.45 & 19 & 2.73 &1.72&3.40&2.0 \\
        C2 &  20:35:13.77  & +41:12:36.90 & 3.66  & 0.012 & 10.7 & 0.8 & 15 & 1.16 & 27 & 2.11 &0.29&1.61&5.5 \\
        C3 &  20:35:13.33  & +41:12:43.25 & 5.59  & 0.019 & 10.4 & 0.9 & 25 & 1.12 & 29 & 1.60 &0.71&3.23&4.5 \\
        C4 &  20:35:09.42  & +41:12:50.77 & 6.12  & 0.021 & 8.4 & 0.9 & 26 & 1.87 & 17 & 1.76 &1.45&3.57& 2.5\\
        C5 &  20:35:08.67  & +41:12:54.95 & 8.51  & 0.029 & 8.3 & 0.9 & 28 & 3.15 & 18 & 3.44 &4.66&4.93&1.1 \\
        C6 &  20:35:10.66  & +41:13:06.86 & 16.41 & 0.056 & 8.5 & 1.1 & 19 & 2.56 & 22 & 3.62 &14.12&14.23&1.0 \\
        C7 &  20:35:09.08  & +41:13:02.40 & 3.10  & 0.011 & 8.2 & 1.1 & 21 & 2.23 & 22 & 2.99 &0.47&2.8&5.9 \\
        C8 &  20:35:10.74  & +41:13:25.48 & 10.73 & 0.036 & 8.8 & 1.1 & 20 & 2.69 & 23 & 3.81 &6.13&9.15&1.5\\
        C9 &  20:35:09.39  & +41:13:24.59 & 5.45  & 0.018 & 8.0 & 1.1 & 22 & 1.23 & 21 & 1.52 &0.70&4.57&6.5\\
        C10 & 20:35:09.71  & +41:13:44.15 & 8.36  & 0.028 & 8.6 & 1.0 & 20 & 2.64 & 17 & 3.04 &3.64&5.88&1.6\\
        C11 & 20:35:08.89  & +41:13:42.15 & 5.17  & 0.018 & 8.8 & 0.9 & 23 & 3.16 & 19 & 3.56 &1.80&3.06&1.7\\
        C12 & 20:35:12.34  & +41:13:47.83 & 5.43  & 0.018 & 7.7 & 1.0 & 24 & 0.96 & 24 & 1.23 &0.55&3.78&6.9\\
        C13 & 20:35:08.42  & +41:13:50.27 & 9.73  & 0.033 & 8.6 & 1.1 & 20 & 4.93 & 16 & 6.07 &9.44&8.39&0.9\\
        C14 & 20:35:09.34  & +41:13:53.28 & 4.27  & 0.015 & 8.5 & 0.9 & 27 & 1.53 & 23 & 1.66 &0.61&2.55&4.2\\
        C15 & 20:35:07.47  & +41:14:01.69 & 13.51 & 0.046 & 8.5 & 1.2 & 17 & 5.90 & 15 & 7.95 &21.96&13.91&0.6\\
        C16 & 20:35:11.16  & +41:13:58.54 & 9.58  & 0.032 & 8.4 & 0.8 & 30 & 1.99 & 17 & 1.61 &3.58&4.3& 1.2\\
        C17 & 20:35:11.37  & +41:14:12.11 & 4.06  & 0.014 & 8.7 & 0.8 & 24 & 1.28 & 19 & 1.30 &0.44&1.88&4.3\\
        C18 & 20:35:07.15  & +41:14:21.94 & 6.04  & 0.021 & 9.6 & 0.9 & 20 & 1.98 & 25 & 2.95 &1.54&3.57&2.3\\
        \hline
    \end{tabular}
    \tablefoot{$\theta_{\rm core}$ is the angular diameter of a core and estimated from the area of the \texttt{astrodendro} leaf. Heree, $N$(\ce{N2H+})$^a$ is the \ce{N2H+} column density derived with the fitted \ce{N2H+} excitation temperature, while $N$(\ce{N2H+})$^b$ is obtained with the fixed temperature using $T_{\rm HCN/HNC}$. }
\end{table*}
\begin{figure}[ht!]
        \includegraphics[width=0.49\textwidth]{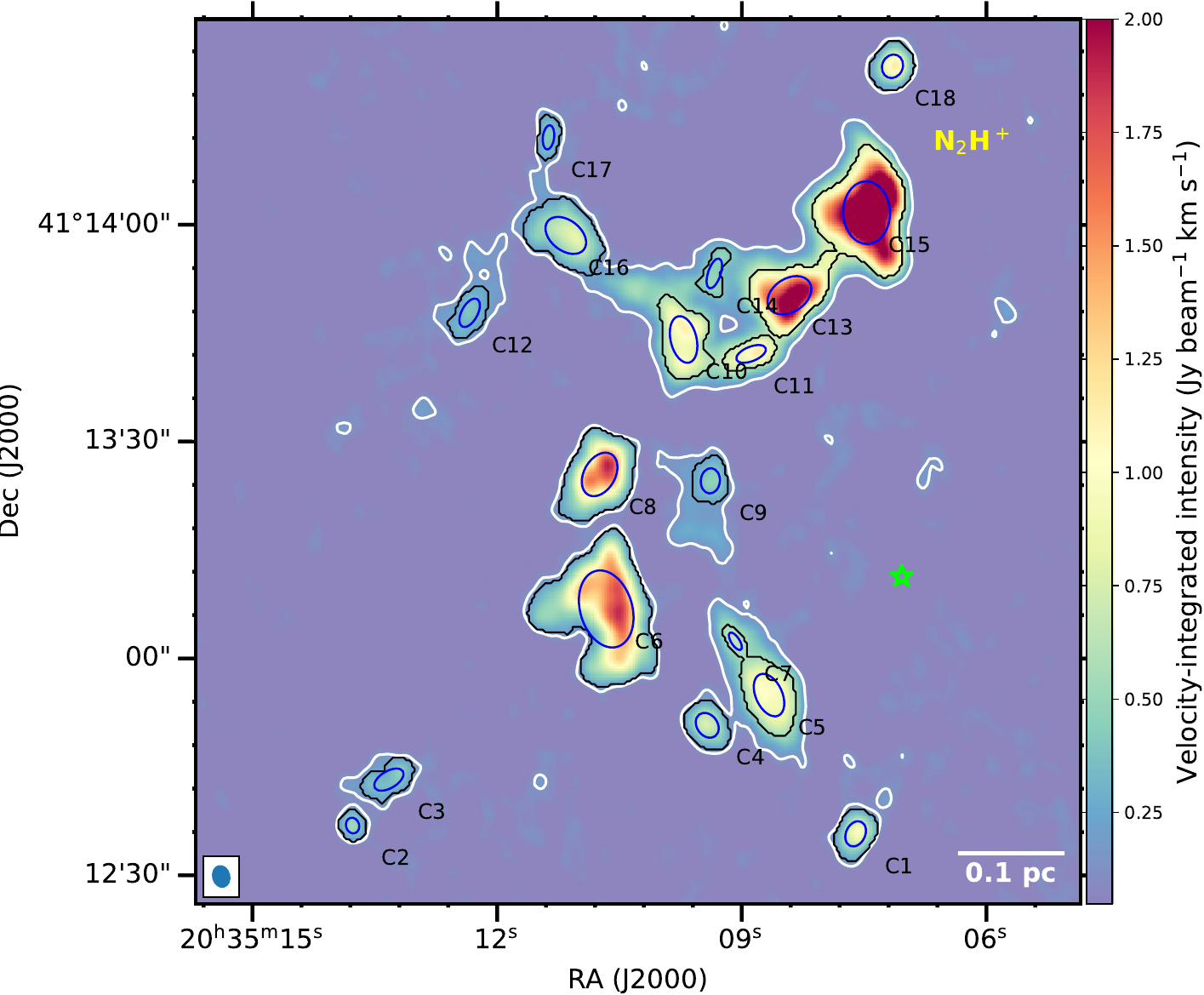}
        \caption{\ce{N2H+} velocity-integrated intensity over all the hfs transitions. The leaf structures identified by the \texttt{Dendrogram} algorithm are shown in black contours, and schematic structures of the extracted cores computed from the intensity-weighted second moment (e.g., variance) of intensities in the plane of the sky are marked as blue ellipses. The white contours are 3$\sigma_{\rm rms}$ (0.146\,Jy beam$^{-1}$\,\kms) of the intensity map integrated over a velocity range of 18\,\kms.  All the identified cores are labeled the same as in Table\,\ref{tab:cores}. The synthesized beam size for \ce{N2H+} emission is displayed in the lower left corner.}
        \label{fig:sou_extraction}
\end{figure}
\begin{figure}[ht!]
    \centering
    \includegraphics[width=0.33\textwidth]{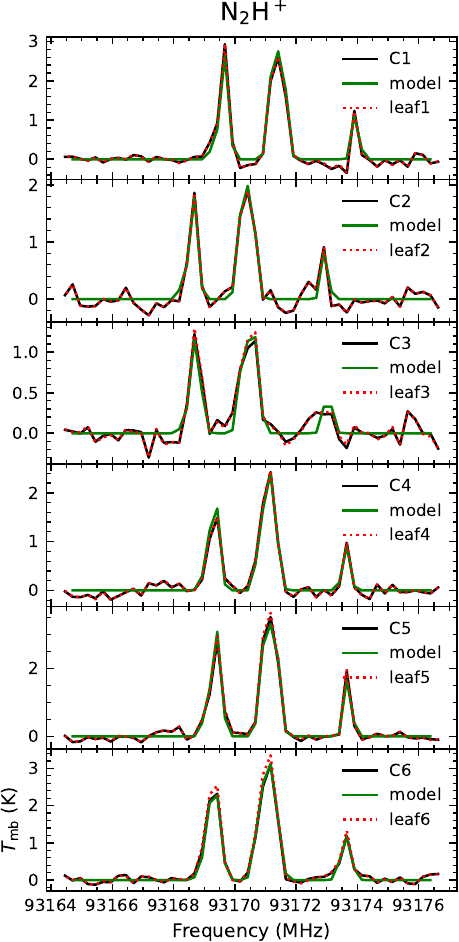}
    \caption{\ce{N2H+} spectral lines (black curves) extracted over the circular regions of the compact cores, overlapped with \ce{N2H+} spectral lines (red dotted curves) extracted over the leaf structures from the \texttt{astrodendro} and the XCLASS modeled spectra (green curves) in a $T_{\rm mb}$ scale.}
    \label{fig:n2hp_xclass_fit}
\end{figure}
\subsection{Compact core identification}\label{sec:sou_extraction}
We utilized the \ce{N2H+} intensity map, integrated over a velocity range encompassing all the hyperfine structure (hfs) transitions, to extract compact cores that trace the dense gas regions. To identify substructures (e.g., cores) in DR18, we used the \texttt{astrodendro} package \footnote{https://dendrograms.readthedocs.io/en/stable} which applies the Dendrogram algorithm \citep{Rosolowsky2008_dendrogram}. Among the substructures classified by \texttt{astrodendro}, "leaves" are the smallest substructures referred to here as compact cores. Three input parameters were required for the \texttt{astrodendro} package: \textit{min\_value}, \textit{min\_delta}, and \textit{min\_npix}. We applied a 5$\sigma$ threshold value to \textit{min\_value}, where the $1\sigma_{\rm rms}$ noise (0.049\,Jy\,beam$^{-1}$\,\kms) is measured from the velocity-integrated intensity map. Additionally, we chose \textit{min\_delta} for the significance of a leaf compared with its neighboring leaf or branch features to be equal to 1\,$\sigma_{\rm rms}$. Lastly, we used \textit{min\_npix} as the minimum number of pixels to identify a leaf, set as 45 pixels, the number of pixels covered by the beam for \ce{N2H+} emission. The leaf structures are defined as brighter than the five \,$\sigma_{\rm rms}$ level and larger than the beam size. With these input parameters, we identify 18 leaf substructures that we hereafter consider as cores. In Fig.\,\ref{fig:sou_extraction}, the black contours indicate the substructures, and the blue ellipses show their schematic structures, which are not extracted source sizes. The central coordinates of the structures are their mean positions in the x and y directions and are listed in the second and third columns of Table\,\ref{tab:cores}.

\begin{figure*}[!ht]
    \centering
    \includegraphics[width=0.97\textwidth]{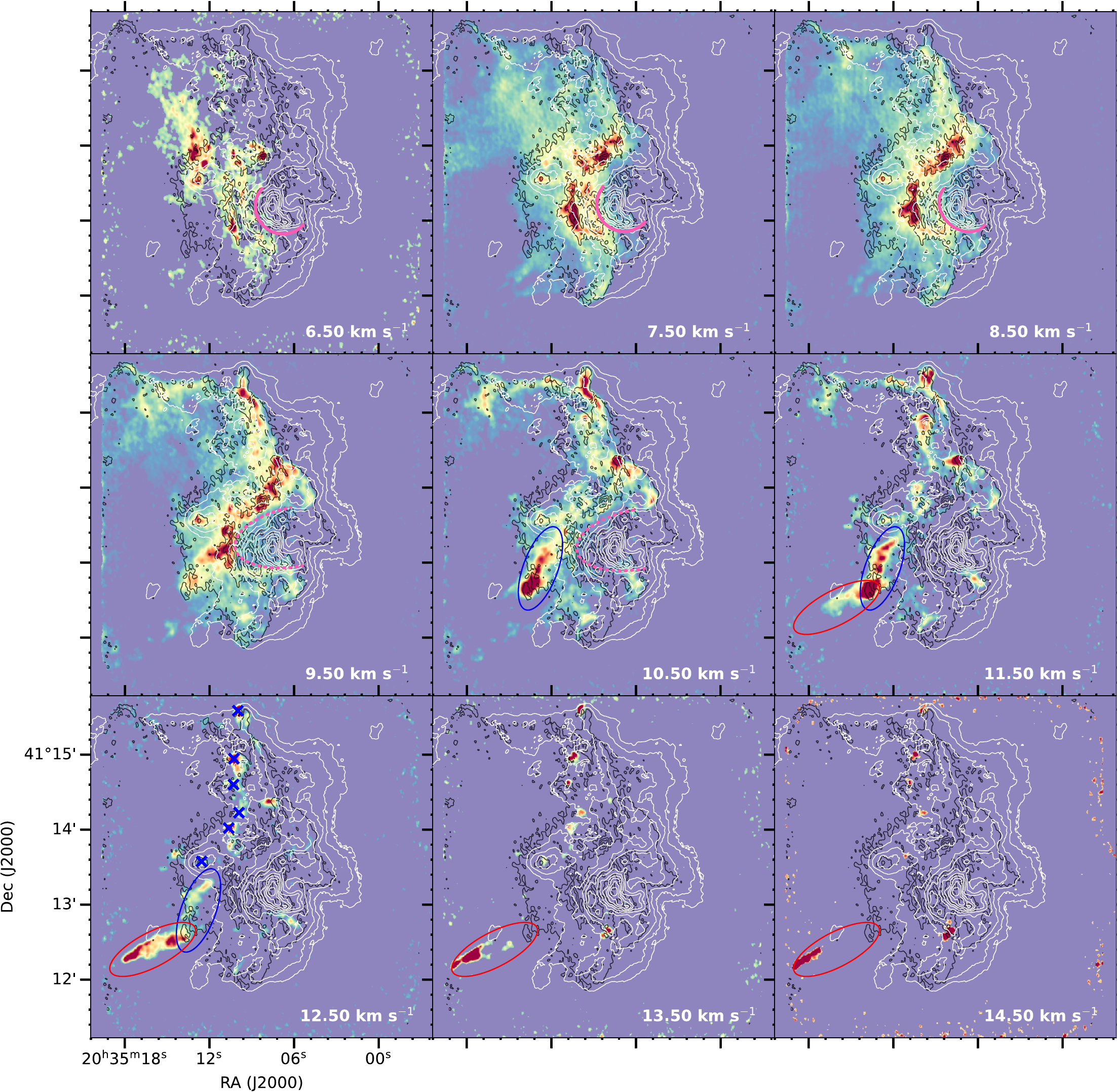}
    \caption{Velocity channel maps of \ce{HCO+} from 6.5\,\kms\ until 14.5\,\kms. The velocity difference between the channel maps is 1\,\kms. Each channel map has different color scales and shows emission stronger than a 10\,$\sigma_{\rm rms}$. The flux scale per velocity channel map is varied to better highlight the individual emission features. The black and white contours are the mm and cm continuum emission, respectively, and the contour levels are the same as in Fig.\,\ref{fig:mm_cm_images}. The pink sold lines and dashed lines at velocities from 6.5\,\kms\ to 10.5\,\kms\ show the outlines of the cavity created by the \hii\ region of DR18-05. The blue ellipses marked at velocities from 10.5\,\kms\ to 12.5\,\kms\ and the red ellipses marked at velocities of 11.5\,\kms\ -- 14.5\,\kms\ are indicated to blue and red-shifted lobes of the potential outflow (or jet) candidate, respectively. In the 12.5\,\kms\ channel map, the blue cross symbols mark the positions of the extracted \ce{HCO+} and \ce{H^13CO+} spectral lines (shown in Fig.\,\ref{fig:spectra_knots}) toward compact features.}
    \label{fig:hcop_channel}
\end{figure*}
\subsection{Derivation of the core physical parameters}
Here, we determine the physical quantities of the \ce{N2H+} cores identified in the previous section. The effective core radius is derived as $R_{\rm core} = \sqrt{A/\pi}$, where $A$ is the area of the leaves on the plane of the sky and adopting a distance of 1.4\,kpc. The average radius of the cores is 0.025\,pc. The obtained core radius in pc and its diameter in arcseconds are listed in Table\,\ref{tab:cores}. With the given core radius, we extracted the \ce{N2H+} spectra toward the 18 compact cores to derive \ce{N2H+} column densities and to determine line parameters (e.g., linewidths).

To deconvolve the hyperfine structure of the \ce{N2H+} 1--0 line, we utilized the eXtended CASA Line Analysis Software Suite (XCLASS\footnote{https://xclass.astro.uni-koeln.de}, \citealt{moeller2017}) by solving the one-dimensional (1D) radiative transfer equation. Here, we assume an isothermal source under the assumption of local thermodynamic equilibrium (LTE). To model and generate synthetic spectra for fitting spectral lines extracted toward the cores and for creating a velocity field map, the \texttt{myXCLASSFit} and \texttt{myXCLASSMapFit} functions were used, respectively. These functions compute a synthetic spectrum for each pixel or an input spectrum by fitting an observed spectrum with Gaussian profiles, assuring that opacity effects on the line shapes are appropriately taken into account. The optimization package \texttt{MAGIX} \citep{moeller2013} was used to perform the fit. This software acts as an interface between the input codes and an iterating engine. The aim is to minimize the deviation ($\chi^2$ values) between the modeled outputs and the observational data. An embedded SQLite database was used to acquire molecular properties, such as Einstein $A$ coefficients and partition functions. This database contains entries from the Cologne Database for Molecular Spectroscopy (CDMS, \citealt{mueller2005}) in its Virtual Atomic and Molecular Data Center (VAMDC, \citealt{endres2016}) implementation. Additionally, an extended set of partition function calculations is also available. The initial parameter set for fitting each emission line component comprises the excitation temperature ($T_{\rm ex}$), the total column density ($N_{\rm tot}$), the line width ($\Delta \varv$), and the velocity offset ($\varv_{\rm off}$) from the systemic velocity. The column density is related to the optical depth, $\tau$, by 
\begin{equation}
\begin{aligned}
\label{eq:optically_thin}
N = \frac{8\pi\nu^{3}}{c^{3}} \frac{Q(T_{\rm ex})}{g_{u}A_{ul}} \frac{{\rm exp} ({E_{u}}/{kT_{\rm ex}} )}{{\rm exp}(h\nu/{kT_{\rm ex}}) -1} \tau ~d\varv ~{\rm cm^{-2}},
\end{aligned}
\end{equation}
where $E_{\rm u}$ and $g_{\rm u}$ are the energy of the upper state for the selected transition and its degeneracy, respectively; $\nu$ is the rest frequency of a selected molecular transition, $c$ is the speed of light, and $h$ is Planck constant. The quantity $A_{\rm ul}$ is the Einstein coefficient for spontaneous emission, and $Q$($T_{\rm ex}$) and $k$ are the rotational partition function and the Boltzmann constant, respectively. For the excitation temperature of the \ce{N2H+} components, we set a prior with a minimum $T_{\rm ex}$ of 5\,K and a maximum $T_{\rm ex}$ of 30\,K. To acquire the best-fit output parameters by minimizing the $\chi^2$ value, we used the Levenberg-Marquardt (LM) algorithm. In the fitting procedure, all the hfs transitions ($F_1 = 1-1$, $F_1 = 2-1$, and $F_1 = 0 -1$) of \ce{N2H+} are considered, and thus the output parameters of the fitting are the hfs deconvolved values. To improve the signal-to-noise level, we separately fitted spectral lines extracted from the areas of cores. Thus, we fixed the core size to the core diameter to fit the individual spectra. Figures\,\ref{fig:n2hp_xclass_fit} and \ref{appendix:n2hp_xclass} display the XCLASS modeled spectra and the input spectra, extracted over the core size, in a main beam temperature scale ($T_{\rm mb}$). The average values of the decomposed line widths and the column density of \ce{N2H+} are 1\,\kms\ and $2.4\times10^{13}$\,cm$^{-2}$, respectively. In addition, we also estimate $N$(\ce{N2H+}) with a fixed temperature using the $T_{\rm HCN/HNC}$ averaged over an individual core size. The derived column densities are consistent with the results obtained with  $T_{\rm ex} (\ce{N2H+})$ and, thus, we use the $N$(\ce{N2H+}) values obtained with $T_{\rm ex} (\ce{N2H+})$.

With the fitted line parameters and the \texttt{astrodendro} output results, we can examine the physical properties of the identified cores and derive the core mass as: 
\begin{equation}
    M_{\rm core} = \mu_{\rm \ce{H2}}m_{\rm H} \pi N{\rm (\ce{H2})} R_{\rm core}^2~{\rm M_{\odot}},
\end{equation} by integrating $N$(\ce{H2}) over the core size, where $\mu_{\rm \ce{H2}} = 2.8$ is the mean molecular weight, and $m_{\rm H}$ is the mass of a hydrogen atom. Here, $N$(\ce{H2}) is the \ce{H2} column density and it is determined by adopting the relative \ce{N2H+} abundance to be $4\times10^{-10}$ to \ce{H2} because this is the average value between the abundances of $3.0\times10^{-10}$ in low-mass molecular cloud cores \citep{Caselli2002_n2hp_lowmass} and $5.2\times10^{-10}$ in high-mass molecular cloud cores \citep{Pirogov2003_n2hp_highmass}. This abundance assumption is reasonable at the given \ce{N2H+} core sizes with the average radius of 0.025\,pc, based on the 1D prestellar core model presented in Fig. 5 of \cite{Redaelli2019_xn2hp_model} showing $X$(\ce{N2H+}) between 10$^{-10}$ and 10$^{-9}$ for core radius of 0.02 -- 0.03\,pc. The median and maximum \ce{H2} column densities are $5.27
\times10^{22}$\,cm$^{-2}$ and $1.48\times10^{23}$\,cm$^{-2}$, respectively.The mean and median values of $M_{\rm core}$ are 4.10\,M$_{\odot}$ and 1.63\,M$_{\odot}$. With the determined core masses, to address whether the cores are gravitationally bound or unbound, we calculate the virial mass ($M_{\rm vir}$) and the virial parameter ($\alpha_{\rm vir}$) using the \ce{N2H+} line widths. The virial mass of an equivalent homogeneous density sphere \citep{Caselli2002_n2hp_lowmass} is derived as 
\begin{equation}
    M_{\rm vir} = 210\left( \frac{R_{\rm core}}{\rm pc}\right)\left(\frac{\Delta\upsilon}{\rm \kms} \right)^{2}~{\rm M_{\odot}}.
\end{equation}
Thus, the mean and median values of $M_{\rm vir}$ are 5.27\,M$_{\odot}$ and 3.68\,M$_{\odot}$.
The virial parameter \citep{Bertoldi1992_alpha_vir} is defined as the ratio of the virial mass and the actual core mass, 
\begin{equation}
    \alpha_{\rm vir} = \frac{M_{\rm vir}}{M_{\rm core}},  
\end{equation}
and can be used to evaluate the stability of a core against gravitational collapse. This simplified measurement assumes the core is isolated and does not interchange mass, momentum, or energy with its surrounding medium. \cite{Kauffmann2013_vir} shows that $\alpha_{\rm vir} \lesssim 2$ means that a core is gravitationally bound, while if $\alpha_{\rm vir} \gtrsim 2$, a core is unbound as random turbulent gas motions create significant support against collapse. In addition, $\alpha_{\rm vir} =1 $ indicates that a core is 
in virial equilibrium, and for the case with $\alpha_{\rm vir}\sim\,2$, a core is marginally gravitationally bound. Based on the determined virial parameters, 50\,\% (9/18) of the cores are unbound, while the other half of the cores are bound. This virial parameter could change with the abundance of \ce{N2H+}, which can vary in different physical environments. If the \ce{N2H+} abundance were higher, $\sim4\times10^{-9}$  \citep[as found toward star-forming
clumps; e.g.,][]{Sanhueza2012_irdc_line_survey} than the assumption ($\sim4\times10^{-10}$), we would obtain smaller core mass as $N(\ce{H2})$ becomes smaller ($\sim~5\times10^{21}$\,cm$^{-2}$). Consequently the virial parameters of all the cores are derived to be much larger, 6 -- 69, indicating a gravitationally unbound nature. According to the 1D prestellar core model presented by \cite{Redaelli2019_xn2hp_model}, the \ce{N2H+} abundances are lower, about $10^{-10}$, in the inner core regions where a core radius is smaller than 0.03\,pc; while the abundance peaks to $10^{-8}$ at a core radius of 0.05\,pc. Thus, the virial parameters obtained with $\sim4\times10^{-10}$ are reasonable as the \ce{N2H+} abundances toward the clumps \citep[e.g.,][]{Sanhueza2012_irdc_line_survey} were derived from larger physical scales compared to the cores analyzed in this study. We will discuss the distribution of these bound and unbound cores in connection with stellar feedback in Sect.\,\ref{sec:discussion}.

\begin{figure}
    \centering
    \includegraphics[width=0.325\textwidth]{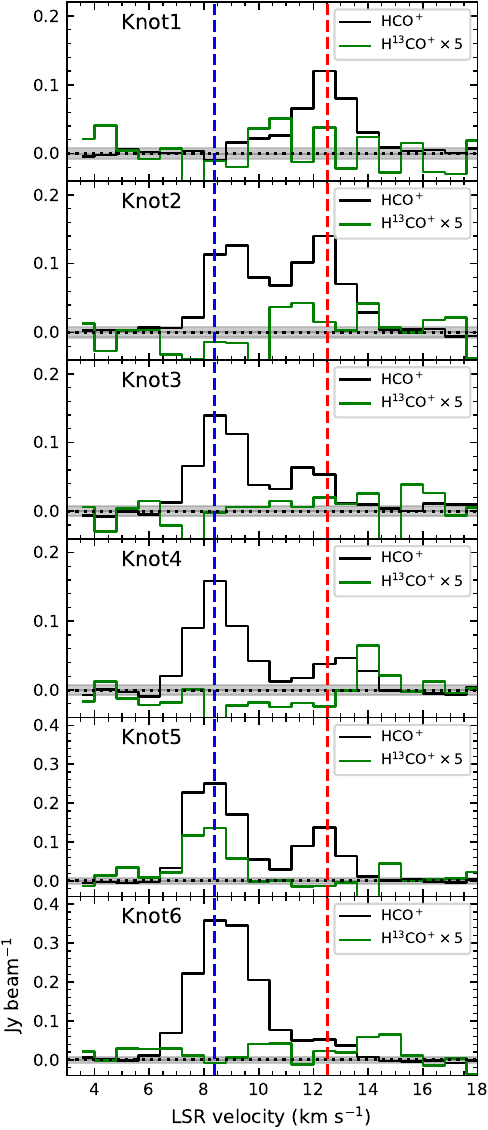}
    \caption{\ce{HCO+} (black) and \ce{H^{13}CO+} (green) spectral lines extracted over an area with a diameter of $2\times\theta_{beam}$ toward pronounced bright knot features marked on the velocity channel map at 12.5\,\kms\ in Fig.\,\ref{fig:hcop_channel}. The blue dashed-vertical lines indicate the systemic velocity of DR18, and the red dashed lines have a velocity of 12.5\,\kms. The gray filled areas indicate $\pm$1\,$\sigma_{\rm rms}$ ($=$~7.15\,mJy\,beam$^{-1}$) for \ce{HCO+}.} \label{fig:spectra_knots}
\end{figure}

\subsection{Gas kinematics toward DR18}
To investigate the structures and kinematics of the molecular gas, we generated velocity-channel maps of the \ce{HCO+} line for the velocity range from 6.5\,\kms\ to 14.5\,\kms, with a channel spacing, $\Delta \upsilon$, of 0.8\,\kms. As shown in Fig.\,\ref{fig:hcop_channel}, we detected both compact and extended emission distributions. The peak velocity of the \ce{H^{13}CO+} spectrum over the observed entire area peaks at the systemic velocity, 8.4\,\kms, of DR18 and this velocity component gas is widespread in the Cygnus-X region \citep[e.g.,][]{Schneider2023_CII, Schneider2016}. At velocities from 7.5 -- 9.5\,\kms, close to the systemic velocity, the compact substructures appear, and they show clear emission counterparts in the mm continuum emission and \ce{N2H+} integrated emission maps. The emission morphologies at velocities ranging from 6.5\,\kms\ to 8.5\,\kms\ spatially overlap with the cometary head of the radio continuum emission, indicated by the magenta solid arc lines. The redshifted emission components, which are visible at velocities between 10.5 and 11.5 \kms, are consistently aligned with the extended mm and cm continuum emission and more concave against the inner cometary \hii\ region, as indicated by the dotted magenta curves. In addition, there are several bright emission knots that prevail even at higher redshifted velocities, reaching up to 14.5 \kms. 

In the channel maps from 10.5\,\kms\ to 14.5\,\kms, we notice spatially elongated features as indicated by blue and red ellipses. Unlike other red-shifted features, the locations of these emission components appear to be devoid of cm radio continuum emission; only the northern one (blue ellipse) shows the weak mm continuum. In particular, the southeastern feature clearly shows bullet-like shapes, as are often found in protostellar jets (e.g., \citealt{Tafalla2017_jets}). The jet-like red-shifted features seem to be originating from the region where C2 and C3 are located, but not from YSO counterparts, which were identified by \cite{Comeron2022_DR18}. On the other hand, their possible blue-shifted counterparts might be the features elongated toward the northwest direction marked with blue ellipses at velocities of 10.5\,\kms\ -- 12.5\,\kms. Their bent shapes could be due to an interaction with the ambient gas components surrounding the \hii\ region. A detailed discussion of these features is beyond the scope of this work.

\begin{figure*}[!ht]
    \centering
    \includegraphics[width=0.495\textwidth]{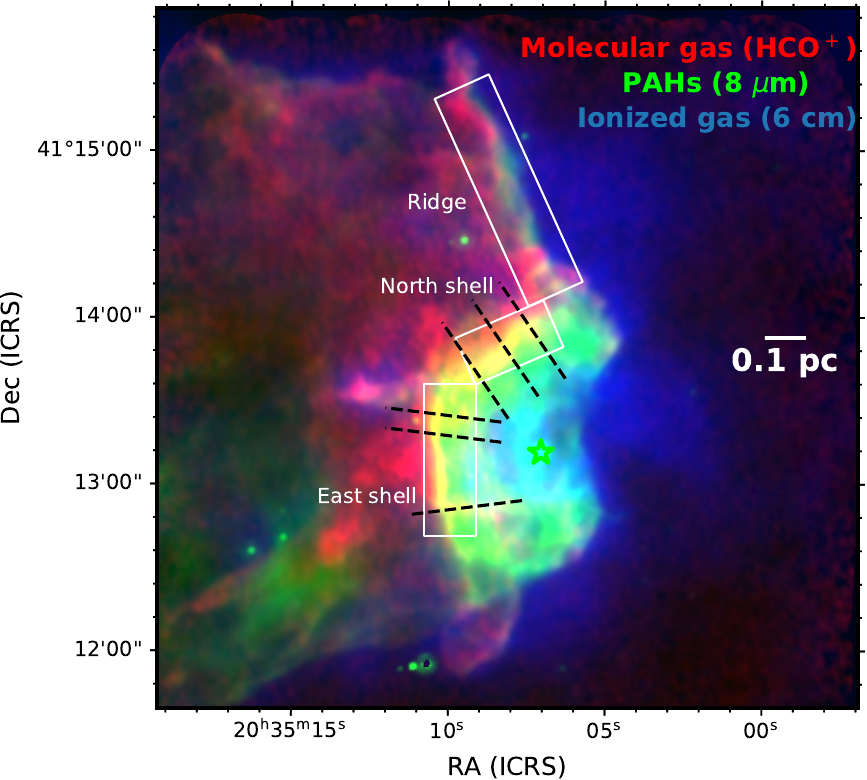}
    \includegraphics[width=0.475\textwidth]{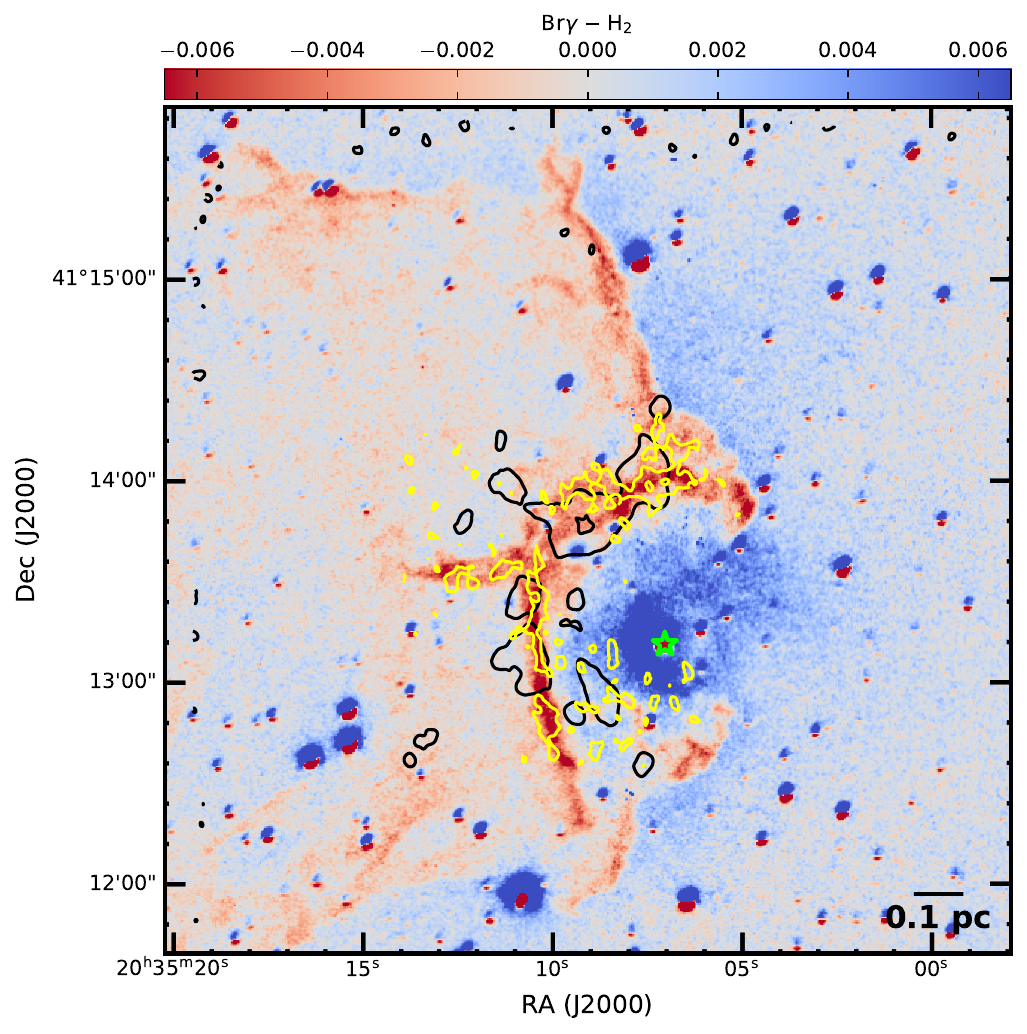}
    \caption{Different emission distributions toward DR18. \textit{Left}: Three-color composite image of DR18; \ce{HCO+} emission in red, 8\,$\mu$m emission in green, and the GLOSTAR 6\,cm radio continuum emission in blue. The 6\,cm emission represents a combination VLA D- and B-configuration data from the GLOSTAR survey. The synthesized beam size of the GLOSTAR data is $4''$,  comparable to that of NOEMA data ($\sim 3-4''$). The pixel size,  of the \textit{Spitzer}/IRAC 8\,$\mu$m data is $0\rlap{.}''6$. The black dashed lines indicate the cuts for intensity profiles of different emission tracers, presented in Fig.\,\ref{fig:intensity_profile}. The white rectangles represent the areas of the East and North shells, as well as the Ridge. \textit{Right}: Flux difference image of Br$\gamma$ and \ce{H2} 1--0 $S$(1) emission lines obtained from \cite{Comeron2022_DR18}. The black and yellow contours show the 5$\sigma$ level of \ce{N2H+} velocity-integrated intensity and a 3.0$\sigma$ level of SiO velocity-integrated intensity, respectively. In both images, the bright green star indicates DR18-05, the ionizing source in DR18. }
    \label{fig:dr18_pdr_image}
\end{figure*}

Apart from the jet-like features, we also identified knot-like features with red-shifted velocities in a range of 11.5 -- 14.5\,\kms, which are distributed along the northeast edges of the extended radio continuum emission. These knot-like features are certainly compact but not associated with any mm continuum emission features or  \ce{N2H+} emission. Also, these features are not bright compared with the \ce{N2H+} cores in the \ce{HCO+} intensity map (Fig.\,\ref{fig:mom0_maps}). To investigate the knots seen in the \ce{HCO+} emission, we extracted \ce{HCO+} and \ce{H^13CO+} spectra over a twice larger area than the average beam size at the locations of six conspicuous knot-like features selected by visual inspection on the channel maps. In the 12.5\,\kms\ channel map, the blue crosses indicate the locations of the extracted \ce{HCO+} and \ce{H^{13}CO+} spectra. The spectra shown in Fig.\,\ref{fig:spectra_knots} correspond to the positions of the markers, numbered 1 to 6, with decreasing declination.  Most of the knots display double-peak profiles, except for Knot 1, which only shows the red-shifted component. The lower velocity component is at the systemic velocity of 8.4\,\kms\ (shown by the blue dashed line), while the other component is red-shifted (approximately 12.5\,\kms) relative to the systemic velocity. We also compared the \ce{HCO+} spectra with \ce{H^{13}CO+} spectra shown in green (scaled up by a factor of 5). The gray filled area indicates the rms noise level (not scaled up). \ce{H^{13}CO+} emission is only detected toward knot 5, at the systemic velocity. On the other hand, the other positions are not detected at our noise level.  Thus, the double-peak profiles of the main-isotopologue spectra cannot be attributed to self-absorption as no optical thin emission is seen at their peak velocities. We observe two kinematic components toward these features, likely associated with the extended free-free radiation detected in the mm and cm continuum emission based on the morphologies of the continuum emission as well as the systemic gas.

\section{Discussion}\label{sec:discussion}

\subsection{Ionized gas and molecular gas emission in DR18}\label{sec:structure}
In the 3.6\,mm and 6\,cm radio continuum emission maps (Fig.\,\ref{fig:mm_cm_images}), the morphology of the continuum emission around the DR18-05 star shows a cometary head pointing toward the dense gas regions. On the other hand,  the extended continuum emission has a cometary shape heading toward the Cyg OB2 association, which is the opposite direction of the arc-like dense molecular gas distribution. To investigate whether the continuum emission at 6\,cm and 3.6\,mm wavelengths is primarily due to DR18-05 or whether there is an additional contribution from the external OB2 association, we estimated its Lyman continuum photon flux ($N_i$), namely, the number of ionizing photons emitted by a star, to define the spectral type of an ionizing source, with the following equation (e.g., \citealt{Carpenter1990_lyman}),
\begin{equation}
    \left(\frac{N_i}{\rm photon s^{-1}}\right) = 9\times10^{43}~\left(\frac{\int S_{\nu}}{{\rm mJy}} \right)~\left(\frac{D}{\rm kpc} \right)^{2}~\left(\frac{\nu}{\rm 5\,GHz} \right)^{0.1}.
\end{equation}
Here, $\int{S_{\nu}}$ is the integrated radio flux density estimated at a frequency of $\nu$, while $D$ is the distance to DR18. The equation assumes that the \hii\ region is optically thin. This is consistent with our spectral index analysis discussed above and is to be expected for the DR18 \hii\ region as it has an extended morphology and is not deeply embedded in molecular clouds, for instance, ultra-compact or hyper-compact \hii\ regions \citep[see, e.g.,][]{Kurtz1994}. 

Considering the flux over the entire 6\,cm radio continuum emission obtained from the GLOSTAR, we obtain a Lyman photon continuum flux of 2.4$\times10^{49}$\,s$^{-1}$, corresponding to the spectral type of an O5.5 star by following the relation between Lyman continuum photons and zero-age main sequence (ZAMS) spectral type presented in \cite{Panagia1973_spectral-type}. Hence, the estimated Lyman continuum photon rate is nearly five orders of magnitude larger than that of a typical B2-type star (i.e., 4.5$\times10^{44}$\,s$^{-1}$). However, if we only consider the continuum flux within the interior \hii\ region exhibited as blue-green color in the three-color composite image in the left panel of Fig.\,\ref{fig:dr18_pdr_image}, we obtained ZAMS spectral types of B0 -- B0.5 (6.9$\times10^{46}$\,s$^{-1}$), which are reasonably consistent with the one (B2 type) from the optical spectral analysis carried out by \cite{Comeron2022_DR18}. In addition, using the 3.6\,mm continuum emission toward the \hii\ region by assuming that the mm continuum emission from the \hii\ region is mainly free-free emission, we obtained a very similar Lyman photon continuum flux (i.e., $7.4\times10^{46}$\,s$^{-1}$). The ionized gas along the ridge cannot be formed by DR18-05 because the molecular gas seems to block the UV radiation emitted from the B2 star as the PDR layer traced by the PAH 8\,$\mu$m and \ce{H2} (1--0) $S$(1) emission lines are located between the ridge and the DR18-05 \hii\ region. Thus, the ionization of the gas on the ridge must have another origin. This suggests that DR18-05 cannot itself excite the entire volume of ionized gas associated with the globule. Thus, the ionization by the external OB stars of the OB2 association likely contributes to the free-free emission observed at 6\,cm and 3.6\,mm to a significant extent, as discovered in another globule in the Cygnus-X region \citep{Schneider2021_globules}.

In the left panel of Fig.\,\ref{fig:dr18_pdr_image}, the three-color composite image of \ce{HCO+}, 8\,$\mu$m, and 6\,cm radio continuum emission shows that strong radio continuum emission is extended toward the west within the eroded cavity delineated by the PAH and molecular emission. This feature is visible in the flux difference map of Br$\gamma$ and \ce{H2} (1--0) $S$(1) emission lines, as shown in the right panel of Figure \ref{fig:dr18_pdr_image}. The 2.12\,$\mu$m \ce{H2} (1--0) $S$(1) emission line is emitted in a cascade after absorbing UV photons (10\,eV\,$<$\,$h\nu$\,$<$\,13.6\,eV) and it is thus an excellent tracer of the photodissociation interfaces, where FUV radiation (6\,eV $<$\,$h\nu$\,$<$\,13.6\,eV) is absorbed in the cold molecular cloud \citep{Black1987_h2_pdr_excitation}. In addition, the 2.17\,$\mu$m Br$\gamma$ emission line is a recombination line of hydrogen coming from an \hii\ region or ionized plasma region (e.g., \citealt{Djupvik2017, Carlsten2018}). In the difference map,
negative values indicate strong \ce{H2} emission, implying the FUV absorbing surface of the molecular cloud, while positive values indicate an ionized gas region. In the image, as expected, the bright Br$\gamma$ emission closely resembles the 6\,cm radio continuum emission (blue color in the three-color image). 

Except for the ionized gas originating from the \hii\ region of DR18-05, we notice widespread ionized gas emission along the rim-brightened edges of DR18, which stretches toward the north of the \hii\ region cavity. The intensities of 6\,cm radio continuum, 8\,$\mu$m, and the \ce{H2} (1--0) $S$(1) emission are lower toward the rim-brightened ridge evident in the three-color image, compared to values close to the \hii\ region. However, the distributions of 6\,cm, 8\,$\mu$m, \ce{H2} (1--0) $S$(1), and \ce{HCO+} toward the ridge show the well-aligned paralleled geometric structure of an ionization front (traced by the 6\,cm continuum), a photodissociation layer (in the 8\,$\mu$m emission), and a molecular gas layer (indicated by the \ce{HCO+} emission), from west to east. As shown in Fig.\,\ref{fig:ob2}, the surface of DR18 is irradiated directly by external O/B star members of the OB2 association, ionizing the low-density gas on its surface. This suggests that the ionized gas in the ridge is likely photoevaporating, which is consistent with the analysis with 3.6\,mm and 6\,cm continuum emission. Such photoevaporating ionized gas surfaces are found in intense radiation regions in the Cygnus-X complex \citep{Emig2022_MeerKat, Carlsten2018} and the M16 region \citep{Sofue2020}. Such cases showing both an internal PDR caused by embedded stars and external PDRs due to OB associations have been found in infrared spectroscopic observations of Br$\gamma$ and \ce{H2} (1--0) $S$(1) lines (e.g., \citealt{Djupvik2017}) and PDR tracers, such as the $158~\mu$m [CII] line (e.g., \citealt{Schneider2012}).

To investigate whether ionizing radiation from the \hii\ region is eroding the globule and affecting the surrounding ambient molecular gas, we must estimate its physical properties. First, we estimated the propagation velocity, $\upsilon_{\rm exp.}$, of the eroding front, using the following equation for its propagation velocity \citep{Comeron2005_v_exp}, assuming that the ionized gas from the molecular cloud is free to flow away from the cloud:  
\begin{equation}
    \upsilon_{\rm exp} = \frac{N_{\rm i}}{4\pi(n_0)r^2}~{\rm km~s}^{-1}, 
\end{equation}
where $n_0$ is 2$n_{\rm \ce{H2}}$, the initial molecular gas density. This assumption is acceptable as the ionized gas is traced by the 6\,cm radio continuum, and the Br$\gamma$ line emission is extended outward from DR18-05. The derived range of $\upsilon_{\rm exp}$  is $\simeq$\,19.8 -- \,24.2\,\kms, by adopting $N_{\rm i}$ of 6.9$\times10^{46}$\,s$^{-1}$ obtained from the 6\,cm radio continuum emission from the volume of the shell, a molecular gas density of $2\times10^{3}$\,cm$^{-3}$ toward DR18, derived from  \textit{Herschel} far-infrared observations from 70\,$\mu$m to 500\,$\mu$m \citep{Schneider2016_herschel_pillars_globules}, and a distance from DR18-05 of $\sim$\,0.24 -- 0.27\,pc. As mentioned by \cite{Comeron2005_v_exp}, the velocities derived from the above formula stand as upper limits because the ionizing flux impinging on the molecular gas is diminished by ionization of the region between the molecular gas and the ionizing star and also because the real distance from the ionizing star DR18-05 can be larger than the projected distance. The propagation velocities determined here are typical velocities found toward expanding \hii\ regions from radio recombination lines (e.g., \citealt{Kim2017_rrl}). The structure of the radio continuum emission displays a cometary head pointing toward the molecular gas regions and a tail extending toward the OB2 association. This could be explained by the motion of the DR18-05 star and the expansion of ionized gas in a champagne flow \citep{Churchwell1999_champagne_flow}. This explanation is also consistent with the results of \cite{Comeron1999_DR18}. 
However, we also notice a bow shock-like feature around the cometary head. Massive stars such as O/B type stars develop strong stellar winds, injecting their mechanical energy into the ambient gas. Thus, winds can play a significant role in the dynamical evolution of an ionized gas region along with the uniform expansion and the moving star (e.g., hybrid model; \citealt{Cyganowski2003_hydrid_model, Veena2017_hydrid}). The shell radius is described by the following equation \citep{Castor1975_stellar_wind},

\begin{equation}
    R_{\rm sh} = 0.042\left(\frac{L_{\rm w}}{10^{36}\,{\rm ergs\,s^{-1}}}\right)^{1/5}\,\left( \frac{n_0}{10^{5}\,{\rm cm^{-3}}}\right)^{-1/5}\left( \frac{t}{10^3\,{\rm yr}}\right)^{3/5}\,{\rm pc},
    \label{eq:rsh}
\end{equation}
where $L_{\rm w}$ is the mechanical luminosity of the wind. The shell expansion velocity is 
\begin{equation}
    V_{\rm sh} = 24.7\left( \frac{L_{\rm w}}{10^{36}\,{\rm ergs\,s^{-1}}}\right)^{1/5}\,\left( \frac{n_0}{10^{5}\,{\rm cm^{-3}}}\right)^{-1/5}\left( \frac{t}{10^3\,{\rm yr}}\right)^{-2/5}\,{\rm \kms}.
    \label{eq:vsh}
\end{equation}

We adopted a distance of 0.2\,pc between the ionizing star and the shell, a velocity of 5\,\kms\ for the shell velocity due to the stellar wind, and an ambient gas density of $2\times10^{3}$\,cm$^{-3}$ \citep{Schneider2016_herschel_pillars_globules}. The shell velocity is defined using the \ce{HCO+} emission (Fig.\,\ref{fig:hcop_channel}). The determined mechanical luminosity and the shell expansion time are $3.9\times10^{33}$\,ergs\,s$^{-1}$ and $2.4\times10^4$\,yr, respectively. Since DR18 is nearly in the plane of the sky, we also estimated a case with a shell velocity of 10\,\kms, and obtained $L_{\rm w}$ of $3.1\times10^{34}$\,ergs\,s$^{-1}$ and a shell expansion time of 1.2$\times10^{4}$\,yr. To investigate whether the wind is dynamically more important for forming the shell than the expansion driven by the difference in pressure between the ionized and surrounding gas, we compare the relation between the mechanical luminosity, $N_{\rm i}$, and $n_0$ \citep{Shull1980_wind_lyman,Garay1994_wind_lyman}, for the case that the wind is more important than the expansion. For this to apply, the following relation must be satisfied,
\begin{equation}
    \left( \frac{L_{\rm w}}{10^{36}\,{\rm ergs\,s^{-1}}}\right)\,>\,0.33\left( \frac{N_{\rm i}}{10^{49}\,{\rm s^{-1}}}\right)^{2/3}\left(\frac{n_0}{10^5\,{\rm cm^{-3}}} \right)^{-1/3}.
\end{equation}
The derived wind mechanical luminosity (for a case with a shell velocity of 5\,\kms, $3.9\times10^{33}$\,erg\,s$^{-1}$, and, for a case with 10\,\kms, $3.1\times10^{34}$\,erg\,s$^{-1}$) is at best comparable to (and possibly only a small fraction of) the luminosity of the ionized gas, $4.4\times10^{34}$\,ergs\,s$^{-1}$. This implies that the stellar wind alone cannot produce the shell. Hence, it is likely that the cavity of DR18 is shaped by the ionized gas produced by DR18-05, with a minor contribution from its stellar wind and also by the influence of the OB2 association. The cavity is open toward the OB2 association and is photoionized by its combined UV radiation. In addition, along with DR18-05, other YSOs located in DR18 could contribute to forming the shell as well. Located around the shell, there are four compact \ce{N2H+} cores (C1, C4, C7, and C9) close to the \hii\ region and, interestingly, these cores are gravitationally unbound, with $\alpha_{\rm vir} \gtrsim 2.0$. This may imply that they are affected by the expansion of the \hii\ region toward ambient gas associated with these unbound cores. The majority of the gravitationally bound cores showing $\alpha_{\rm vir} \lesssim 2.0$ (6 out of 8 cores; C6, C8, C10, C11, C13, and C15) have compact \ce{NH3} core counterparts \citep{Zhang2024_NH3}, which were observed with a comparable spatial resolution ($\sim$ 0.02\,pc) with to CASCADE observations ($\sim$ 0.02 -- 0.03\,pc). In addition, these bound cores partially overlap with the 8\,$\mu$m and \ce{H2} emission, as depicted in Fig.\,\ref{fig:dr18_pdr_image}. This suggests that these bound cores are located in cold molecular gas regions behind the PDR and the core fragmentation might be influenced by the \hii\ region.

\begin{figure}
    \centering
    \includegraphics[width=0.33\textwidth]{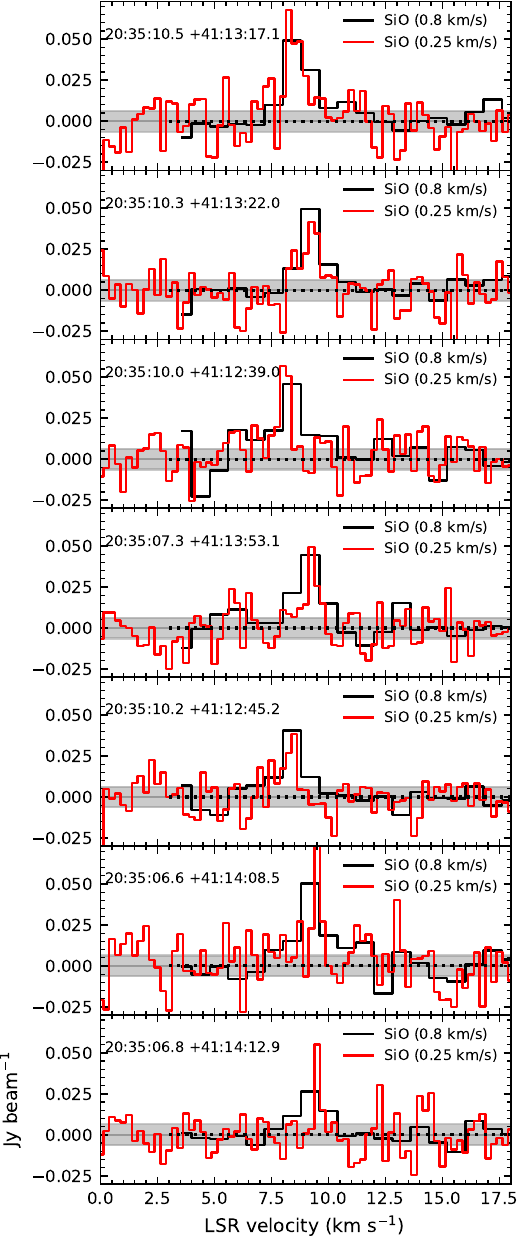}
    \caption{SiO spectral lines extracted from the NOEMA$+$30~m combined data (black curves) and the NOEAM-alone data (red curves). The combined data has a channel width of 0.8\,\kms\ while the NOEMA-only data provides a narrower channel width of 0.25\,\kms. The coordinates of each extracted position on the SiO intensity map are individually displayed on each panel.}
    \label{fig:sio_spectra}
\end{figure}
\begin{figure}
    \centering
    \includegraphics[width=0.48\textwidth]{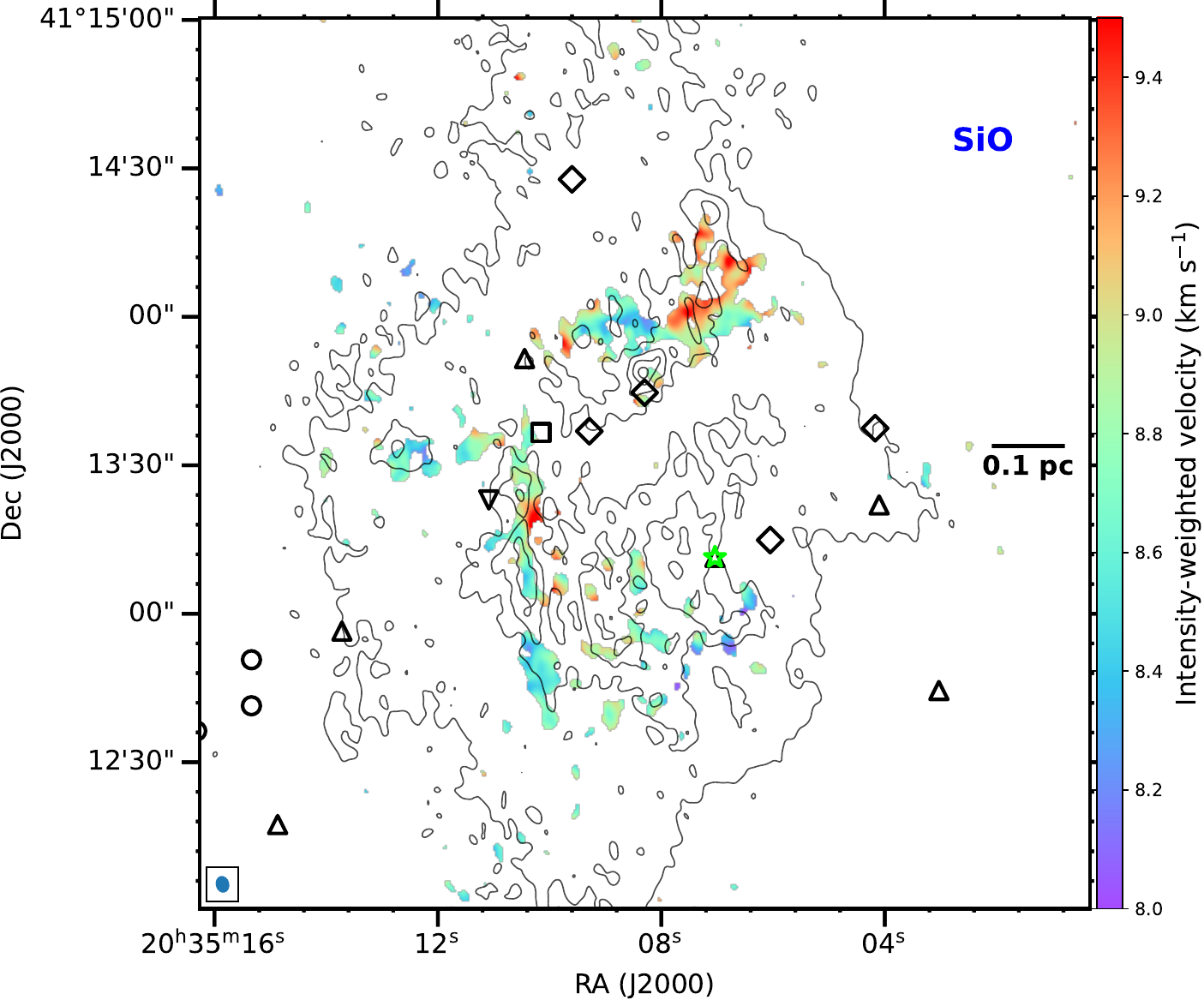}
    \includegraphics[width=0.49\textwidth]{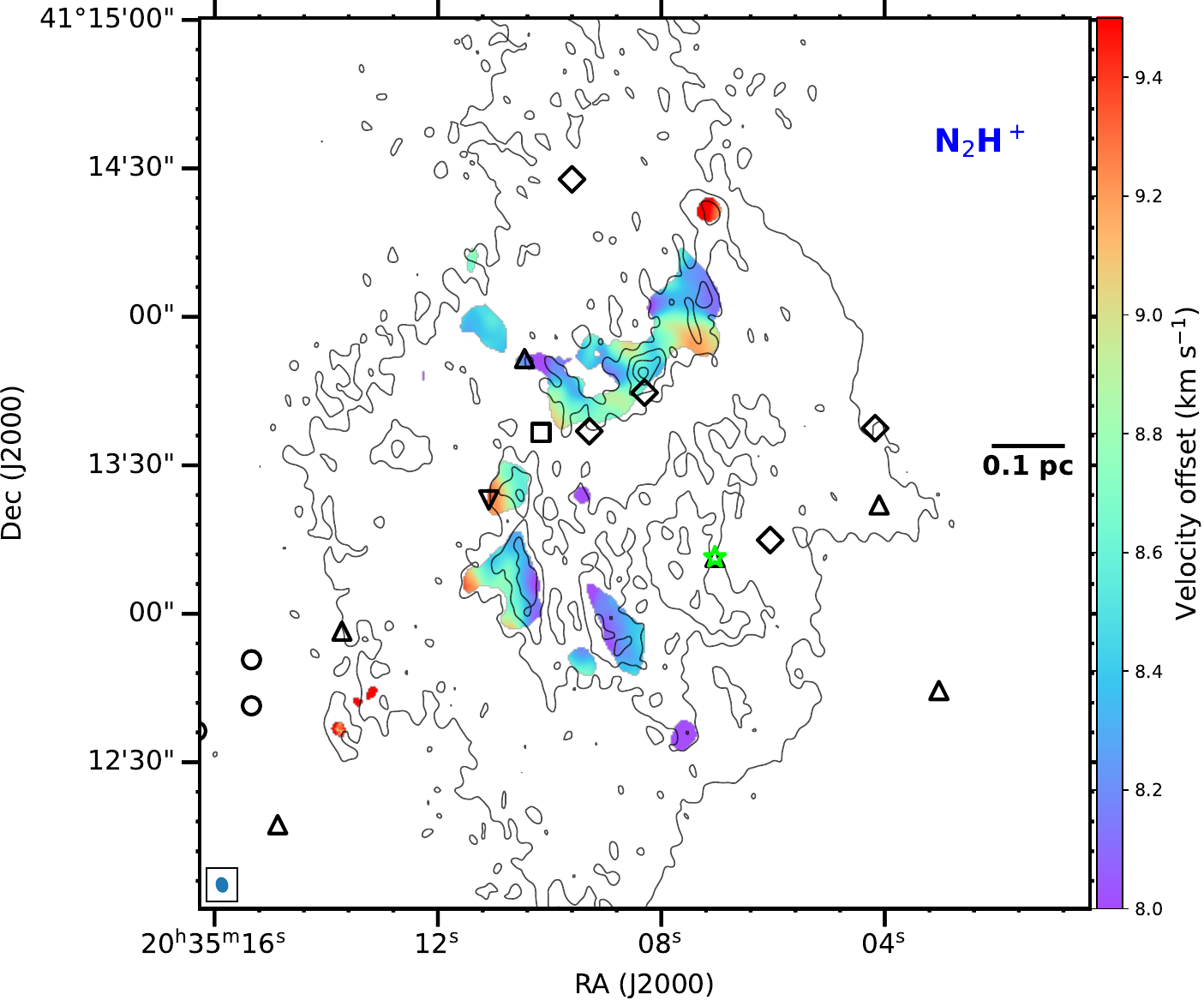} 
    \caption{Velocity field maps of SiO (top) and \ce{N2H+} (bottom). The black contours are the 3.6\,mm continuum emission. The \ce{N2H+} velocity field map is generated by fitting hyperfine structure lines of \ce{N2H+} per pixel using XCLASS while the SiO velocity map is the intensity-weighted velocity of the spectral line (i.e., moment 1 map). The black symbols are YSOs, and they are the same as presented in Fig.\,\ref{fig:rgb_image}. The velocity ranges in the color bars for both maps are fixed as the same. }
    \label{fig:sio_mom1_fwhm}
\end{figure}

\subsection{Association between SiO emission and the internal PDR}\label{sec:disscusion_sio}
As mentioned in Section \ref{sec:molecular_lines}, the SiO emission has a distinct spatial distribution compared with that of other species and is concentrated between the cold molecular gas and ionized gas regions (see Fig. \ref{fig:mom0_maps}). Figure\,\ref{fig:sio_spectra} shows the SiO spectral lines extracted from the NOEMA$+$30~m combined data (black curves) and NOEMA-only data (red curves). The NOEMA-only data has a better velocity resolution of 0.25\,\kms\ than the NOEMA$+$30~m combined data used for this study, but we only use this data for presenting the SiO spectra. These spectral lines are extremely narrow and weak ($\sim$\,0.8 -- 2.0\,\kms\ and a peak intensity $\sim$ 0.05\,Jy\,beam$^{-1}$). 

\begin{figure*}[!ht]
    \centering
    \includegraphics[width=0.46\textwidth]{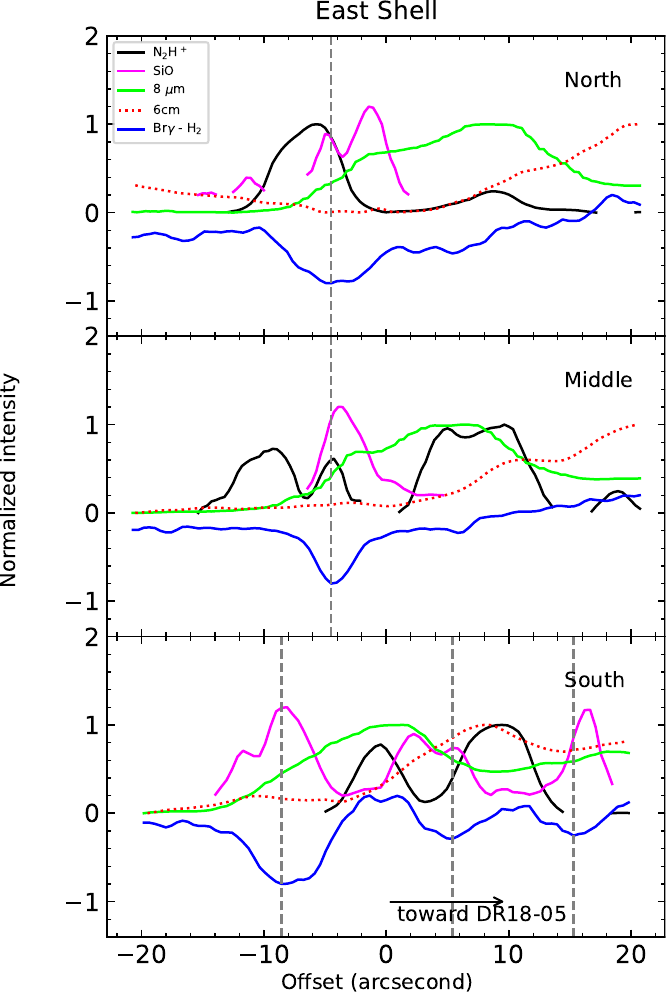}
    \includegraphics[width=0.415\textwidth]{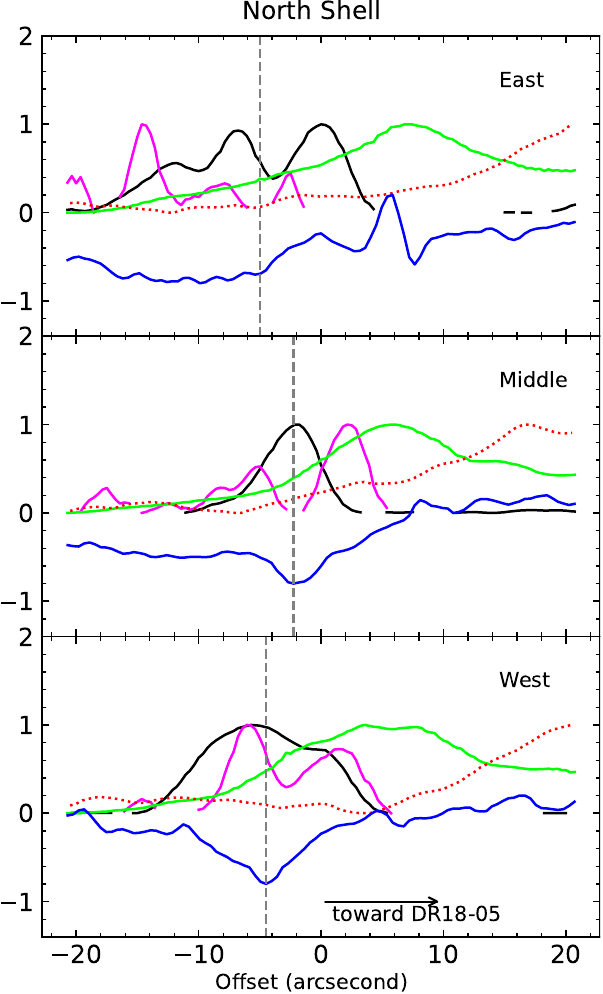}
    \caption{Intensity profiles of different tracers (see legend in the top left panel) along different cuts through the eastern shell (left column) and the northern shell (right column). All the molecular line intensities were extracted from their velocity-integrated intensity maps. The extents and directions of the cuts are shown in the left three-color image of Fig.\,\ref{fig:dr18_pdr_image}. For improved visibility, the Br$\gamma$-\ce{H2} intensity profiles are shifted by $-$0.8, while the SiO profiles are moved by $+$0.2 along the y-axis. The vertical dashed lines indicate the depths in the intensity profiles of Br$\gamma - $\ce{H2}.}
    \label{fig:intensity_profile}
\end{figure*}

In previous studies, SiO emission has been mainly associated with shocks in outflows of embedded YSOs (e.g., \citealt{Schilke1997,Gusdorf_SiO_models_2008,Sanchez-Monge2013,Duarte-Cabral2014_sio_outflows,Csengeri2016_atlasgal_sio}). Lines from 
SiO associated with outflows typically exhibit broad velocity components $>$ 8\,\kms\ and its abundance in outflows relative to \ce{H2}, $X$(SiO) is approximately $10^{-6}$ -- $10^{-9}$ (e.g., \citealt{Martin-Pintado1992_SiO_outflow,Sanchez-Monge2013,Duarte-Cabral2014_sio_outflows,Csengeri2016_atlasgal_sio}). On the other hand, SiO detected in this study shows extremely narrow line widths $\upsilon_{\rm FWHM} \sim 0.8 - 2.0$\,\kms, which are much narrower than the typical line widths originating from outflows. Nevertheless, if the axes of the outflows are on the plane of the sky, the Doppler width of outflow components can be quite modest. Thus, we compared the distributions of SiO emission in DR18 with that of the known YSOs identified by \cite{Comeron2022_DR18}, as displayed in the SiO velocity field map in the upper panel of Fig.\,\ref{fig:sio_mom1_fwhm}. However, there is only one Class I YSO (indicated by a black square marker in Fig.\,\ref{fig:sio_mom1_fwhm}), which is likely to drive an outflow (e.g., \citealt{Bontemps1996_outflow_class}), located at the gap between the east shell and the north shell of the \hii\ region. We can distinguish separate blue- and red-shifted emission regions, none of which appear to be associated with any YSO; this would contradict an outflow-based interpretation.  Toward the east shell region, the SiO emission is red-shifted toward the \hii\ region, whereas the SiO emission, toward the \ce{N2H+}, is blue-shifted. This gradient is the opposite of the velocity gradient seen in \ce{N2H+} (bottom panel of Fig.\,\ref{fig:sio_mom1_fwhm}), showing that the further components from the \hii\ region show red-shifted velocities. In the north shell region, \ce{N2H+} and SiO emission lines show more complicated velocity gradient features. In addition, we determined $N$(SiO) over areas twice bigger than the SiO beam size, using $1.8\times10^{12}~\int{T_{\rm mb}~d\upsilon}$ cm$^{-2}$, with a $T_{\rm ex}$ of 10\,K, taken from \cite{Csengeri2016_atlasgal_sio}. SiO column densities toward different positions along the east and north shells are $\sim 1.8 - 2.0\times10^{12}$\,cm$^{-2}$. Using \ce{H2} column densities ($1.8 - 3.8\times10^{22}$\,cm$^{-2}$) obtained from \cite{Bonne2023_dust_nh2_data}, the derived $X$(SiO) values are $\sim 5\times10^{-11} - 1.0\times10^{-10}$. The abundances are much lower than the typical abundances from outflows and closer to the values ($\sim 10^{-10} - 10^{-11}$) found toward low-velocity shocked gas regions experiencing global infall, converging flows, or cloud-cloud collisions toward star-forming molecular clouds (e.g., \citealt{Jimenez-Serra2010_sio_irdc, Nguyen2013, Duarte-Cabral2014_sio_outflows,Csengeri2016_atlasgal_sio,Cosentino2018_sio_irdc,Cosentino2019_sio_irdc_snr,Cosentino_SiO_IDRC_2020,Kim2023_sio_survey}), along with ($\sim$\,$10^{-11}$) determined toward the Orion bar region and the S 140 PDR (i.e., \citealt{Schilke2001_sio_pdr}). Thus, this SiO emission is unlikely related to any potential outflows associated with the Class I YSO source. 

Intriguingly, the SiO emission is significantly confined to narrow regions and spatially overlaps well with strong vibrational \ce{H2} line emission, evident in the negative values in the flux difference map in the right image of Fig.\,\ref{fig:dr18_pdr_image}. The \ce{H2} (1--0) $S$(1) emission line is enhanced in the vicinity of the \hii\ region where SiO emission is also detected. The excellent spatial agreement with the enhanced \ce{H2} $v=1-0$ $S$(1), narrow line widths, and lower abundances of SiO emission suggest that both trace PDR material around the \hii\ region. \ce{H2} lines can be excited by absorbing FUV radiation, followed by fluorescence \citep{Black1987_h2_pdr_excitation}. Finally, SiO molecules can enter the gas phase via the photo-desorption of the icy mantles of dust grains \citep{Walmsley1999_sio_pdr,Schilke2001_sio_pdr}. 

\begin{figure*} [!ht]
    \centering
    \includegraphics[width=0.33\textwidth]{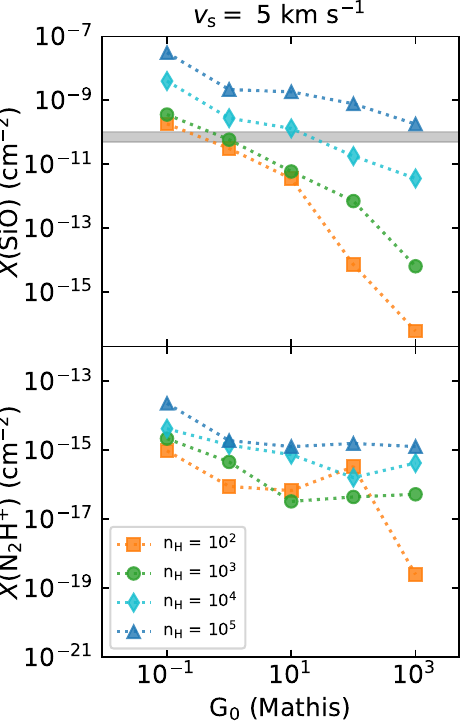}
    \includegraphics[width=0.33\textwidth]{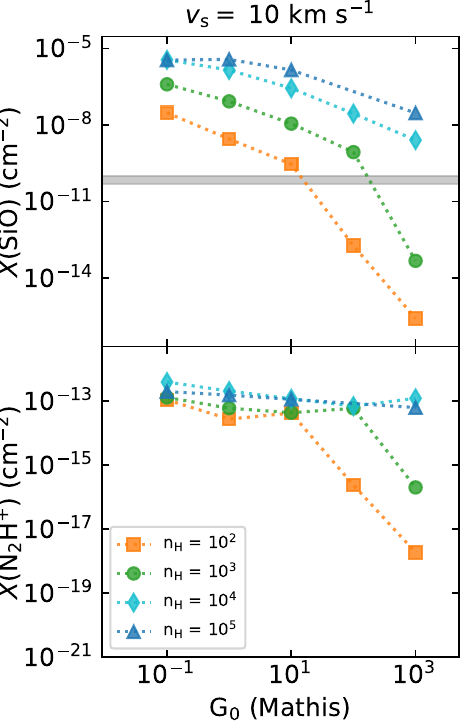}
    \includegraphics[width=0.33\textwidth]{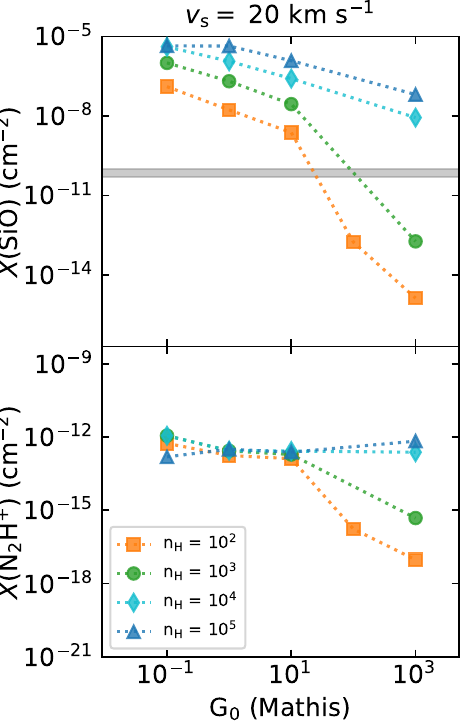}
    \caption{Fractional abundances, $N$(X)/$N$(\ce{H2}), of SiO and \ce{N2H+} relative to \ce{H2}, calculated across shocks propagating at 5 (left panels), 10 (middle panels), and 20 (right panels) \kms\ in the initial media density $n_{\rm H} = 10^2$ (red squares), $10^3$ (green circles), $10^4$ (cyan diamonds), and $10^5$ (blue triangles) cm$^{-3}$, as function of irradiating radiation field $G_{\rm 0}$ in Mathis unit. In the upper panels for SiO, the filled gray areas indicate the observed SiO abundances ($5\times10^{-11} - 1\times10^{-10}$) toward DR18.} 
    \label{fig:shock_model}
\end{figure*}

However, it is hard to discard a shock origin for SiO and \ce{H2} 1--0 $S$(1) emission because \ce{H2} can also be excited by collisions due to shock heating of \ce{H2} molecules \citep{Burton1989_h2_shock}. Thus, to investigate whether SiO and \ce{H2} are associated with PDRs or shocked regions, we compared their emission with infrared emission in the 4.5\,$\mu$m and 3.6\,$\mu$m \textit{Spitzer}/IRAC bands. Toward star-forming regions, especially infrared dark clouds, 4.5\,$\mu$m emission is found to be enhanced and extended, known as green 
fuzzies\footnote{The commonly employed three color code (red, green, and blue) used green for the 4.5\,$\mu$m band.}  \citep{Chambers2009_irdc_irac} or extended green objects (EGOs) \citep{Cyganowski2008_ego}. This enhancement is likely due to shock excited line emission, either from the \ce{H2} 0--0 $S$(9) line at 4.69\,$\mu$m or CO $v=1-0$ rovibrational bandhead at 4.3--5.2\,$\mu$m \citep{Noriega-Crespo2004_4p5_h2,Ray2023_jwst_outflow}. Consequently, green fuzzies mostly mark outflows (e.g., \citealt{Beuther2005_4p5_outflow,Cyganowski2011_ego_outflows}). In addition, unlike other IRAC bands, this band does not cover PAH lines and is considered a ``PAH-free'' band. On the other hand, the 3.6\,$\mu$m band includes 3.3\,$\mu$m features from neutral PAHs, while the 5.8\,$\mu$m and 8$\mu$m bands include emission from cationic PAHs (\ce{PAH+}) \citep{Bakes2001_PAH_I,Bakes2001_PAH_II,Benjamin2003_glimpse}. Comparing the emission in the 4.5\,$\mu$m band with that from these other bands can help to determine the contributions of \ce{H2} excited by shocks and by UV radiation, which is at the origin of the fluorescent IR emission from PAHs. It may also help elucidate the origin of the enhanced 2.12\,$\mu$m \ce{H2} emission and the narrow SiO emission. 
However, the [4.5\,$\mu$m]/[3.6\,$\mu$m] ratios toward the regions with the SiO detections and strong \ce{H2} emission are remarkably lower ([4.5\,$\mu$m]/[3.6\,$\mu$m]\,$<$\,0.65) than the color criteria value ([4.5\,$\mu$m]/[3.6\,$\mu$m]\,$\geq$\,1.8) for EGOs \citep{Chambers2009_irdc_irac} (see Fig.\,\ref{fig:ratio_4p5_3p6}).
Such low ratios might be due to either a lack of only a very small contribution from shock-excited \ce{H2} and/or CO lines to the 4.5\,$\mu$m band or strong PAH features at 3.6\,$\mu$m band, or both. The study of  \cite{van_den_Ancker2000_shock_pdr, van_den_Ancker2000_h2_pdr} reported that \ce{H2} lines of embedded sources are due to shocks, while toward more evolved objects (e.g., the shell of DR18), such lines mainly originate from a PDR as the envelopes disperse; meanwhile, the PDR components become dominant compared to the shock-excited emission.

Figure\,\ref{fig:intensity_profile} shows normalized intensity profiles of \ce{N2H+}, SiO, 8\,$\mu$m, and 6\,cm radio continuum emission along different cuts (black dashed lines in Fig.\,\ref{fig:dr18_pdr_image}) toward the east and north shells. In many cases, the SiO (indicated by pink curves) is enhanced at offset positions with enhanced 2.12\,$\mu$m \ce{H2} emission, evident by negative values on the Br$\gamma$ $-$ 2.12\,$\mu$m \ce{H2} map (blue curves), toward the east and north shell regions. Overall, these two tracers show substantial correlation over most of the six cuts of those shells. Interestingly, the north and middle cutouts of the east shell show that SiO emission is spatially detected between peaks of \ce{N2H+} emission (black curves), a dense gas tracer, and 8\,$\mu$m emission (bright green curves), which traces PDRs. Toward the middle and south cutouts in the east shell, we recognize that some peaks of SiO and \ce{N2H+} are situated close to 6\,cm continuum emission peaks tracing the position of DR18-05. These \ce{N2H+} peaks correspond to the gravitationally unbound cores (i.e., C9 and C4) and one bound core (i.e., C5). For the SiO peaks in the south cutout, the SiO emission is slightly red-shifted ($\upsilon_{\rm offset} > 8.8$\,\kms, while the \ce{N2H+} emission is relatively blue-shifted, $\sim \upsilon_{\rm offset} < 8.4$\,\kms. Thus, these two tracers do not trace the same molecular gas regions. All the spatial distributions and intensity profiles for different tracers indicate that the SiO emission toward DR18 seems conspicuously associated with the PDR surrounding the \hii\ region of DR18-05. 

\begin{table}
    \centering
    \small
    \caption{Input parameters of the Paris-Durham low-velocity shock models.}
    \label{tab:paris-durham}
    \begin{tabular}{l | c}
    \hline \hline
        \multicolumn{2}{c}{Fixed parameters}\\
        \hline 
        Magnetic parameter ($b$) & 1\\
        Cosmic ray ionization rate $\zeta$ (s$^{-1}$)  &  $1\times10^{-17}$\\
        PAHs abundance & $1\times10^{-8}$ \\
        Initial extinction & 0.1 \\
        \hline 
        \multicolumn{2}{c}{Free parameters}\\
        \hline
        Shock velocity ($\varv_{\rm s}$ in \kms) & 5, 10, 20 \\
        Radiation field ($G_{\rm 0}$ in Mathis unit) & 0.1, 10, $10^2$, $10^3$ \\
        Pre-shock proton density ($n_{\rm H}$ in cm$^{-3}$) & $10^{2}$, $10^{3}$, $10^{4}$, $10^{5}$ \\
    \hline
    \end{tabular}
    \tablefoot{The UV interstellar radiation field is in Mathis units; 1 Mathis unit corresponds to a flux of $1.92 \times 10^{-3}$ ergs cm$^{-2}$s$^{-1}$ integrated between 912 and 2400\,\AA, and the ratio relative to 1 Habing is 1.02.} 
\end{table}

To investigate the origin of SiO, we analyzed irradiated low-velocity shock models (project ID: shockgrid\_110\_q1m8\_2022) using the Paris-Durham shock code (version 1.1.0, revision 115) obtained from the InterStellar Medium DataBase (ISMDB)\footnote{https://app.ism.obspm.fr/ismdb/}. The detailed concept of irradiated shock models using the Paris-Durham shock code is described in \cite{Godard2019_irradiated_shock_models} and the descriptions for this grid of shock models used here are presented and explained in \cite{Kristensen2023_shock_model_ref}. The grid of shock models covers low-velocity shocks propagating through molecular gas environments with various external UV radiation intensities. Table\,\ref{tab:paris-durham} lists the fixed and free parameters used in the Paris-Durham low-velocity shock models we studied in this work. We examined shock velocities ($\varv_{\text{S}}$) ranging from 5 \kms\ to 20 \kms, initial density ($n_{\text{H}} = n(\text{H}) + 2n(\text{H}_2)$) in the range of $10^2$\,cm$^{-3}$ to $10^5$\,cm$^{-3}$, and UV radiation field intensities ($G_0$) from 0.1 to $10^3$ in Mathis unit \citep{Mathis1983_G0_unit}. However, the magnetic parameter, $b$ is fixed to 1, corresponding to a C-type shock. The initial transverse magnetic field and the magnetic parameter are expressed as $\frac{B_{\rm 0}}{1\,\mu{\rm G}} = b \times \left(\frac{n_{\rm H}}{1 {\rm cm}^{-3}} \right)^{0.5}$. In addition, the fractional abundance of PAHs relative to $n_{\rm H}$ is fixed to $10^{-8}$. In particular, the initial fractional elemental abundances ($n_{\rm x}/n_{\rm H}$) of silicon (Si) are $3\times10^{-6}$ in the gas-phase, as \ce{Si+}, and $3.37\times10^{-5}$ in grain cores. This grid only considers gas-phase chemistry and does not include grain-grain interactions or ice mantles on the grains (see \citealt{Kristensen2023_shock_model_ref} for details). However, grain erosion resulting in the release of elemental Si, Fe, and others into the gas phase is taken into account. Thus, in  cases with low-velocity shocks, the final SiO abundance greatly depends on the initial abundance of \ce{Si+} and formation routes of silicon-contained molecules in different physical conditions. 

Figure\,\ref{fig:shock_model} presents SiO and \ce{N2H+} abundances calculated across irradiated shocks propagating in different media, as a function of radiation field intensities. For all the models, $X$(SiO) and $X$(\ce{N2H+}) decrease as radiation field intensities increase. In these models, the maximum values of the fractional abundances relative to \ce{H2},  $N$(X)/$N$(\ce{H2}) are approximately close to the initial fractional abundance ($n_{\rm x}$/$n_{\rm H}$) of the gas-phase Si (i.e., \ce{Si+}). Overall, the computed SiO abundances for all the input shock velocities in high-density environments are several orders of magnitude greater than those in low-density cases. In the middle and right panels, for shock velocities of 10\,\kms\ and 20\,\kms, the abundances of $X$(SiO) can still be produced at higher levels in the UV-irradiated environments ($G_{\rm 0} \gtrsim 100$) than the observed values if the initial densities are $\sim 10^{3} - 10^{5}$\,cm$^{-3}$. According to \cite{Schneider2016_herschel_pillars_globules}, the measured UV radiation field in the region is about $G_{\rm 0}\sim 10^3$. However, the initial external UV radiation might be lower than $G_{\rm 0}\sim 10^3$ determined from the Herschel 70\,$\mu$m and 160\,$\mu$m fluxes because the DR18 region may be less exposed to the OB association and the external UV radiation field in the \hii\ region may be smaller. Considering that SiO emission is only detected toward the molecular shells where dense molecular cores are closely located, it is likely that the initial density is higher than $10^{3}$\,cm$^{-3}$. Therefore, the shock velocity creating the SiO emission in DR18 is likely slower than 10\,\kms.

In the left panels of Fig.\,\ref{fig:shock_model}, the 5\,\kms\ shock model irradiated by UV radiation of $G_{\rm 0} = 10^3$ and with an initial density of $10^{5}$\,cm$^{-3}$ demonstrates a strong agreement with the observed SiO abundances and the physical conditions of the DR18 region. \cite{Godard2019_irradiated_shock_models} found that only low-velocity C-type shocks ($\varv_{\rm s} \leq\,5$\,\kms) can exist  in typical PDRs ($n_{\rm H}\sim10^4$\,cm$^{-3}$ and $G_{\rm 0} \sim 100$). Low-velocity shocks in highly UV-illuminated regions ($G_{\rm 0} = 10^4$) only cause a slight compression of gas, which increases its thermal pressure and changes its interaction with the external UV field \citep{Godard2019_irradiated_shock_models}. The DR18 region likely falls between typical PDRs and the extremely irradiated regions. The SiO emission observed in DR18 is evidently associated with gas regions that have been compressed by low-velocity shocks ($\sim$\,5\,\kms) propagating within an initial medium density ranging between $10^4$ and $10^5$ cm$^{-3}$. These shocks are further influenced by external UV radiation ($G_{\rm 0}\sim 100 - 10^3$). In addition, the $X$(\ce{N2H+}) values for all the irradiated shock models are significantly lower than the expected values, $\sim 10^{-10} - 10^{-9}$ and undetectable. This possibly explains spatial and velocity discrepancies between SiO and \ce{N2H+} toward the east and north shells. By comparing the different shock models, we suggest that the SiO emission in the DR18 region originates from the compressed gas layers resulting from low-velocity shocks, which are likely produced by the initial expansion of the \hii\ region \citep[e.g.,][]{Liu2024_atoms_sio} and stellar winds, irradiated by external UV radiation.

\section{Summary}\label{sec:summary}
We carried out 3.6\,mm wavelength observations with the IRAM 30~m telescope and NOEMA toward DR18 to analyze the continuum radiation emission and selected spectral emission lines from various molecular species. This study addresses the complex molecular gas substructures inside DR18 for the first time. 

\begin{itemize}
\item[--] We detected ground- or near ground state emission lines from \ce{HCO+}, HCN, HNC, \ce{C2H}, \ce{H2CO}, SiO, \ce{N2H+}, \ce{and HC3N}, along with the rare isotopologues \ce{H^13CO+}, \ce{H^13CN}, and \ce{HN^13C}, and \ce{^13CS}. In addition, we also detected three deuterated species, DCN, DNC, and \ce{NH2D}. \ce{HCO+}, HCN, HNC, \ce{C2H}, and \ce{H2CO} show extended molecular gas distributions, whereas the emission from isotopologues and the three deuterated molecules arise from dense, compact regions. Interestingly, SiO emission shows a spatial distribution different from that of any other species.

\item[--]
The 3.6 mm continuum emission comprises a significant contribution from free-free radiation, as extrapolated from the 6 cm radio continuum emission. The range of spectral types (B0 – B0.5) of a star emitting the free-free radiation from the continuum emission approximately matches that of the B2 type star DR18-05. The cometary \hii\ region located in a cavity of the DR18 globule is likely formed by a champagne flow. Additionally, the shell structure seen in the cm continuum emission surrounding this \hii\ region is probably the result of expansion driven by the \hii\ region, as well as photoionization from DR18-05 and externally coming from stars in the OB2 association. The spatially extended ionized gas encompassing the entire DR18 globule is attributed to photoevaporation from the external OB stars in the Cyg OB2 association.

\item[--]
Using the velocity-integrated intensity map of \ce{N2H+}, we identified 18 compact cores that are all located around the \hii\ region of DR18-05. The average size determined via source extraction is 0.03\,pc. These cores are relatively colder ($T_{\rm HCN/HNC} < 30$\,K) than their surrounding regions ($T_{\rm HCN/HNC} > 30$\,K). Half of the cores (9 out of 18) are gravitationally bound and most of them are located behind PDRs surrounding the \hii\ region of DR18-05. This might suggest that these cores form as a result of stellar feedback that has triggered star formation.  

\item[--]
We found emission from SiO with narrow line widths ($\sim$0.8 -- 2.0\,\kms) and low abundance 
($\sim\,5\times10^{-11} - 1\times10^{-10}$). This SiO emission is not spatially coincident with that of other dense gas tracers, and its velocities show that it is not associated with the \ce{N2H+} cores that trace the densest gas. On the other hand, the SiO emission shows an excellent spatial agreement with PDRs located between the \ce{N2H+} cores and the \hii\ region of DR18-05. The PDRs also show stronger \ce{H2} emission. In addition, the ratios of emission of the [4.5]/[3.6] IRAC bands toward the SiO emitting regions do not show any evidence of shocks, as the values are much smaller ([4.5\,$\mu$m]/[3.6\,$\mu$m]\,$<$\,0.65) than typical ratios found toward shocked regions ([4.5\,$\mu$m]/[3.6\,$\mu$m]\,$\geq$\,1.8). In comparison to the irradiated shock models, we suggest that the SiO emission encompassing the \hii\ region indicates that the molecular gas regions are slightly compressed by low-velocity shocks (with $\sim$ 5\,\kms) irradiated by external UV radiation (with intensities ranging from 100 to 1000), as they propagate through the medium with $n_{\rm H} \sim 10^4$ to $10^5$\,cm$^{-3}$. These shocks are presumably generated by the initial \hii\ region expansion and possibly by stellar winds as well.

\end{itemize}

\begin{acknowledgements}
 This work is based on observations made with the Institut de Radioastronomie Millim\'etrique (IRAM) 30~m telescope and the Northern Extended Millimeter Array (NOEMA). W.-J. K. was supported by DLR/Verbundforschung Astronomie und Astrophysik Grant 50 OR 2007. A.S.-M.\ acknowledges support from the RyC2021-032892-I grant funded by MCIN/AEI/10.13039/501100011033 and by the European Union `Next GenerationEU’/PRTR, as well as the program Unidad de Excelencia María de Maeztu CEX2020-001058-M, and support from the PID2020-117710GB-I00 (MCI-AEI-FEDER, UE). S.A.D. acknowledges the M2FINDERS project from the European Research Council (ERC) under the European Union's Horizon 2020 research and innovation programme (grant No 101018682). D.~S. acknowledges support from the European Research Council under the Horizon 2020 Framework Program via the ERC Advanced Grant Origins 83 24 28 (PI: Th. Henning). W.-J.~K., N.~S, and P.~S. acknowledge support by the Deutsche Forschungsgemeinschaft via the collaborative research center SFB 1601 (project ID 500700252), subprojects A2 and B2.

\end{acknowledgements}

%
\bibliographystyle{aa} 
\bibliography{aa51998-24corr} 
%
\newpage
\onecolumn
\begin{appendix}
\section{Velocity-integrated intensity maps}
\begin{figure*}[!ht]
    \centering
    \includegraphics[width=0.40\textwidth]
    {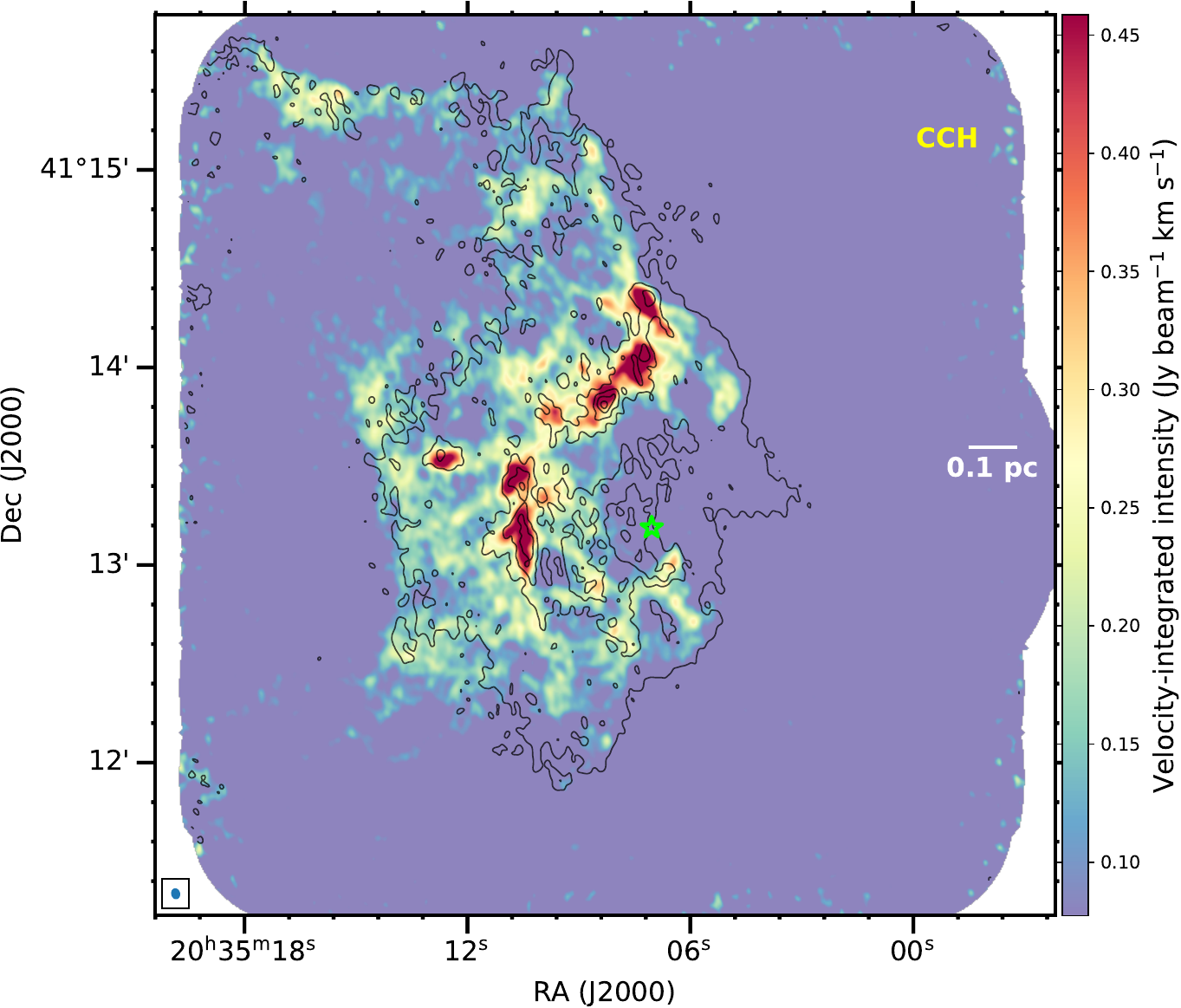}
    \includegraphics[width=0.40\textwidth]{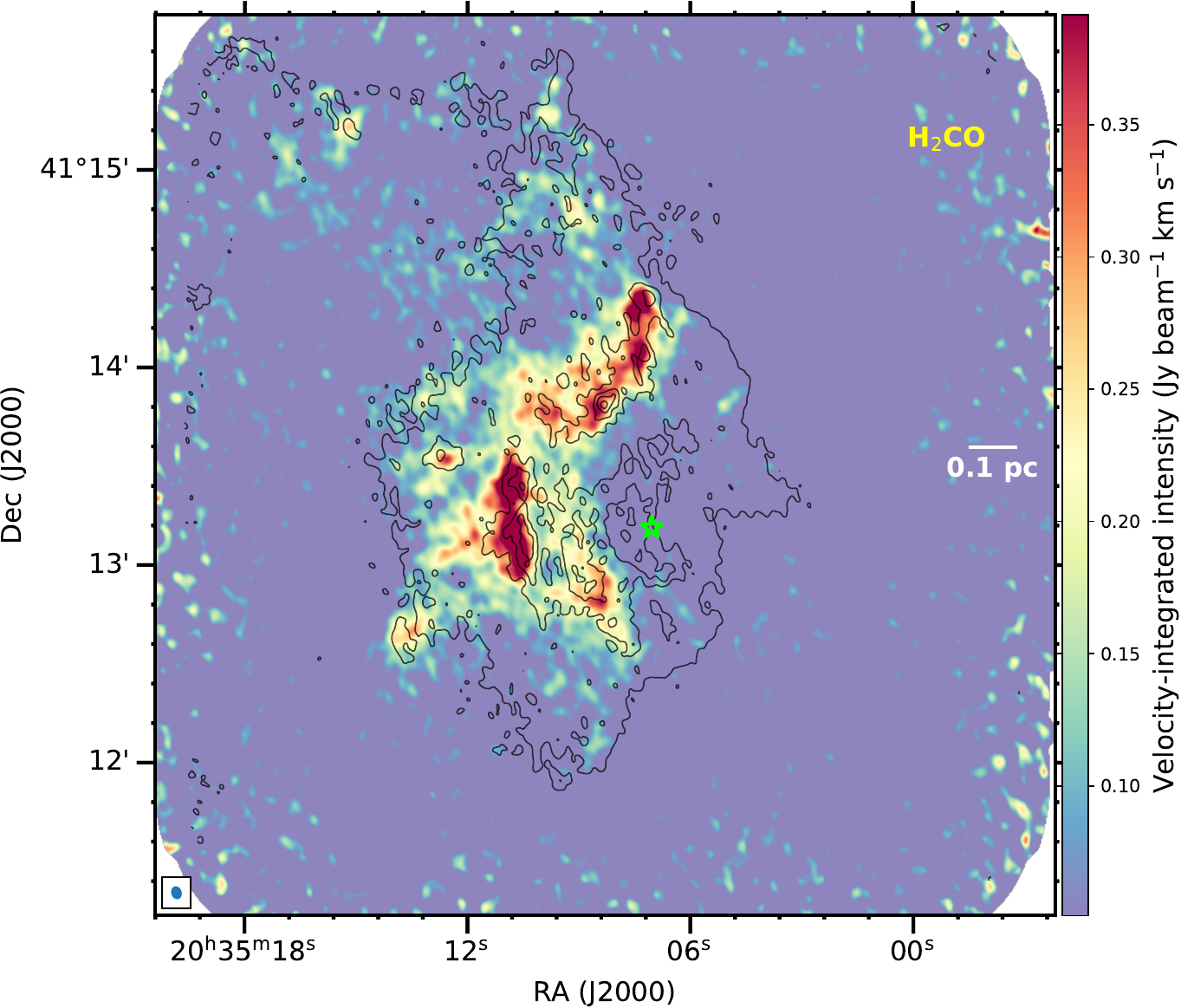}
    \includegraphics[width=0.40\textwidth]{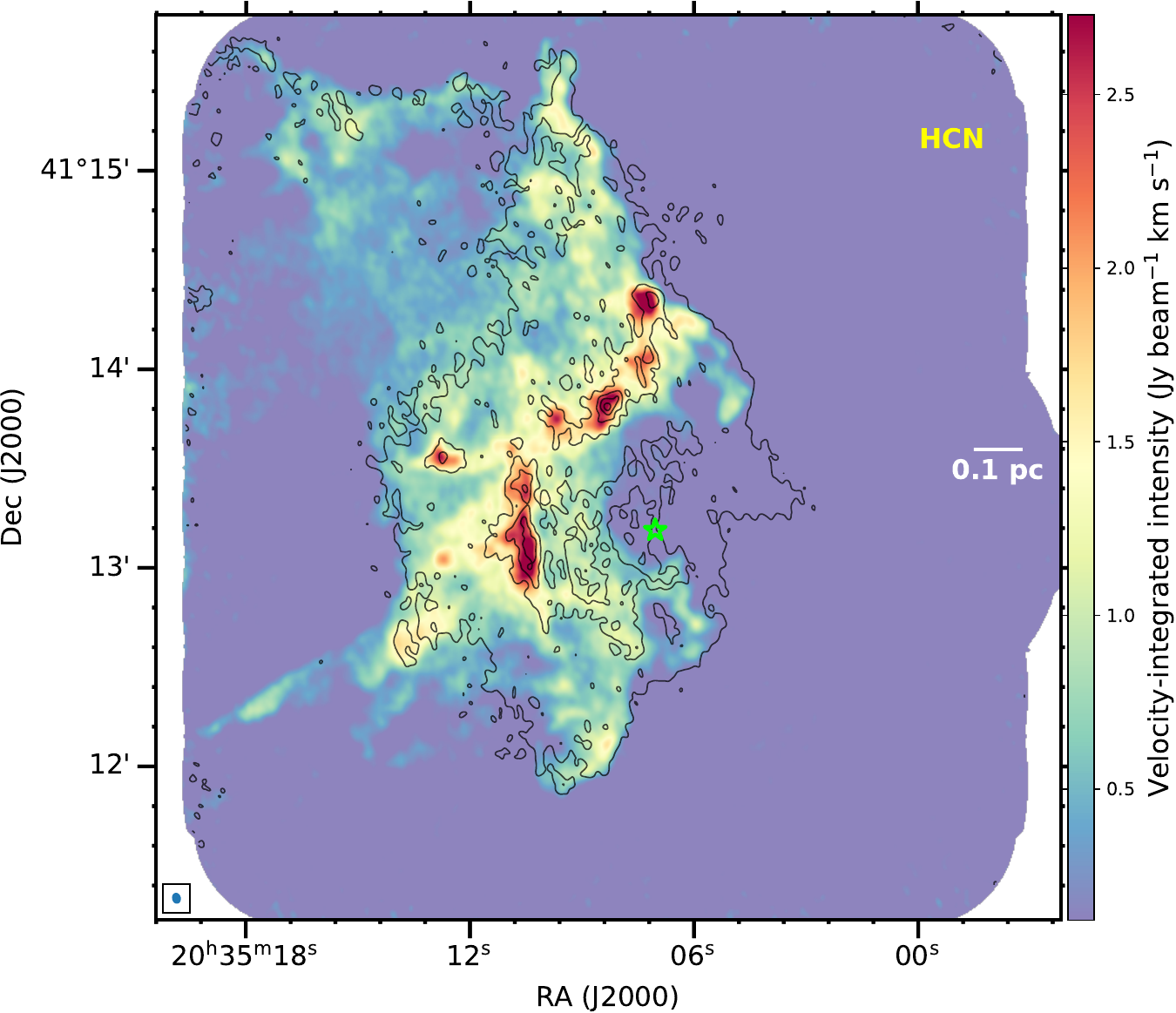}
    \includegraphics[width=0.40\textwidth]{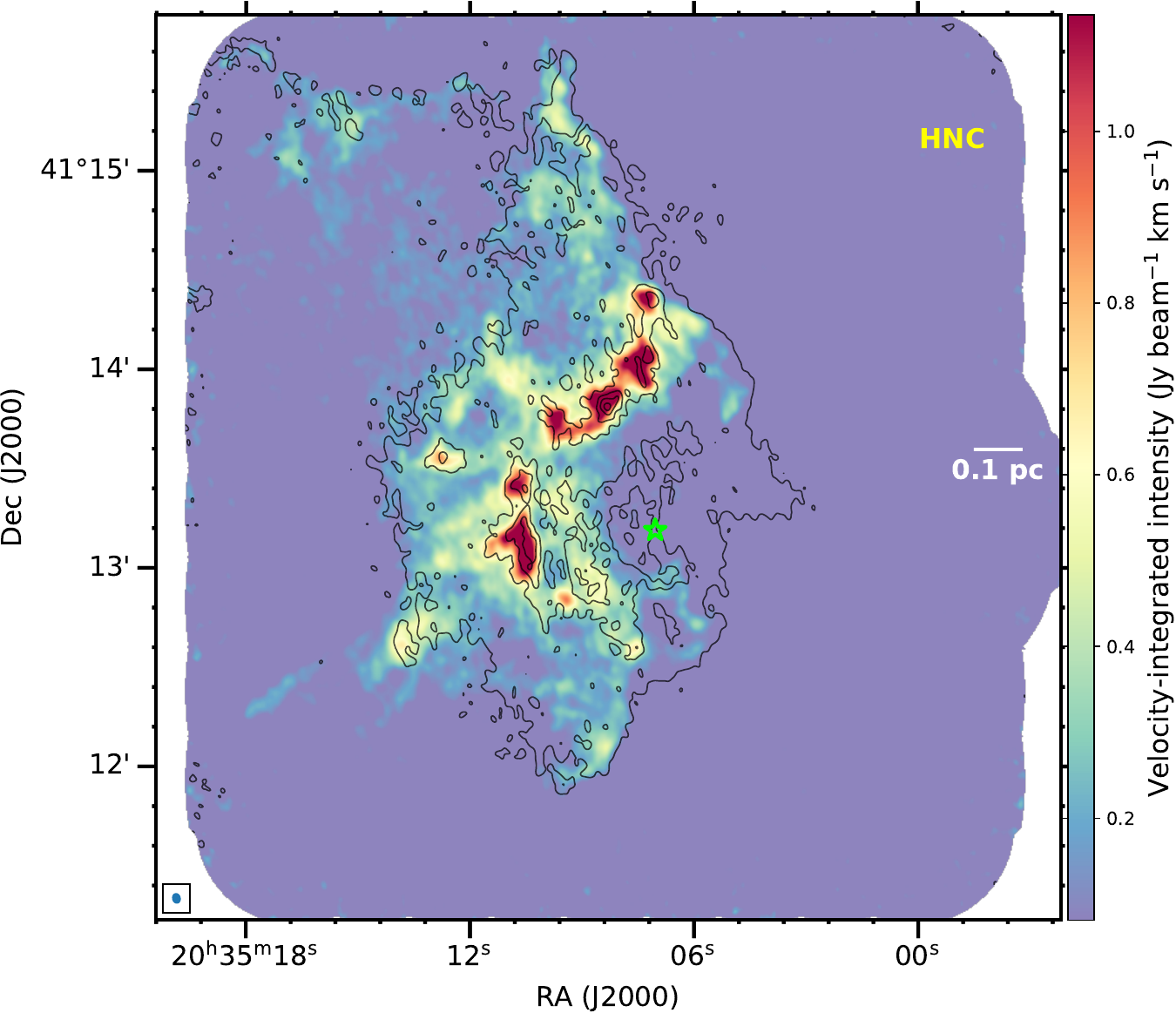}
    \includegraphics[width=0.40\textwidth]{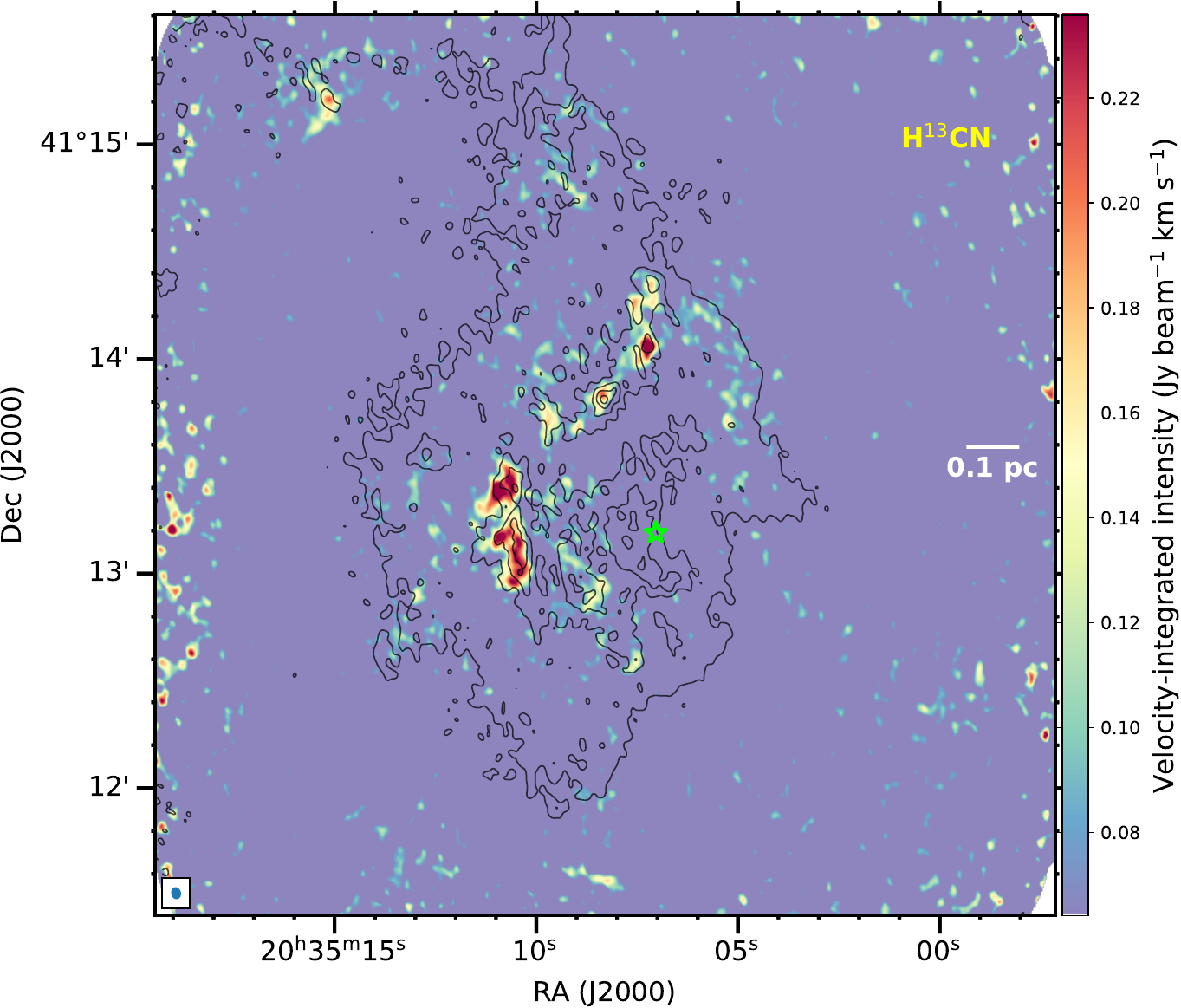}
    \includegraphics[width=0.40\textwidth]{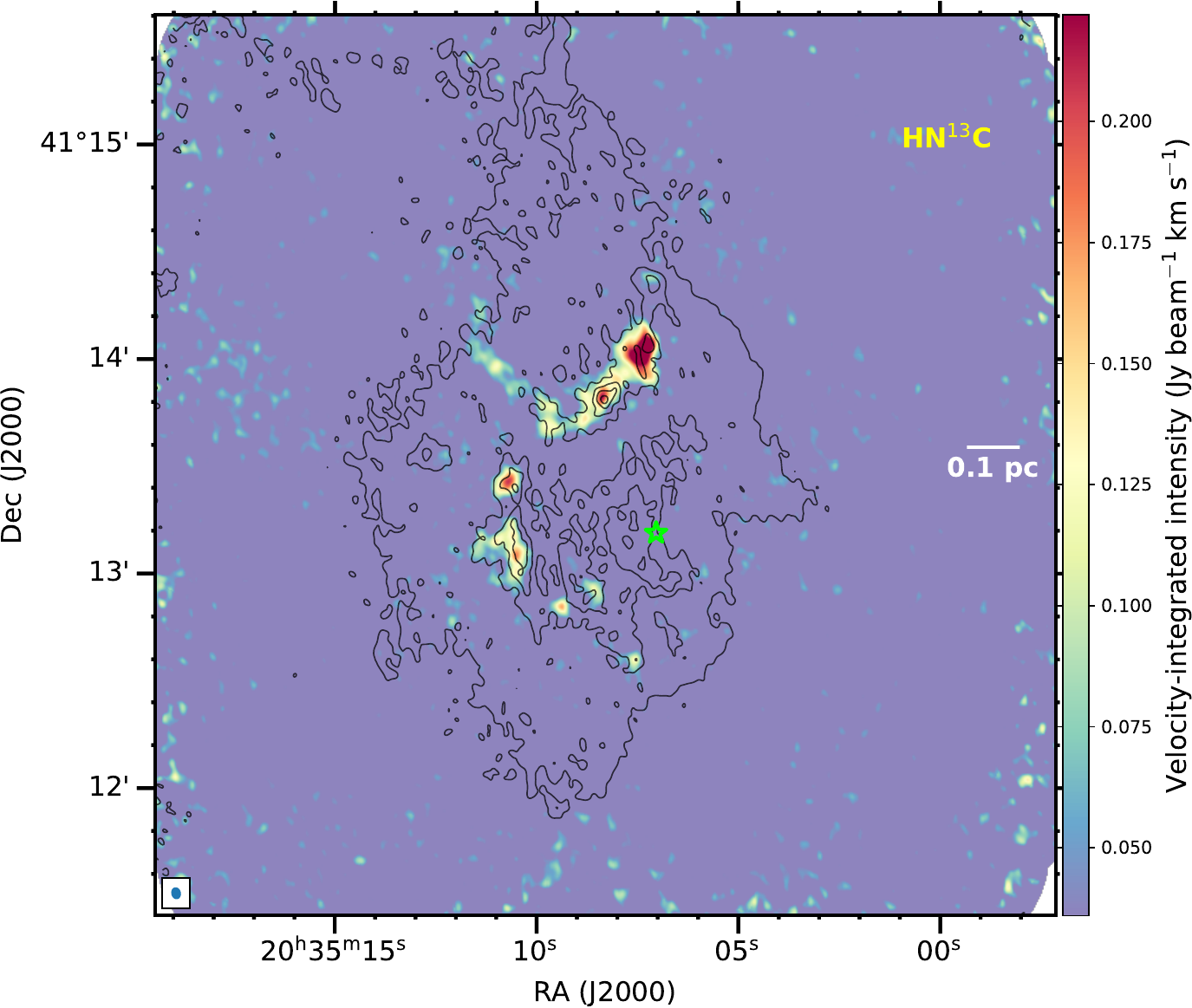}   
    \caption{Velocity-integrated intensity maps of CCH (integrated from 6\,\kms\ to 14\,\kms), \ce{H2CO} (6\,\kms\ to 11.6\,\kms), HCN ($-$0.4\,\kms to 18.8\,\kms, including all the hfs transitions), HNC (7\,\kms\ to 13.5\,\kms), \ce{H^13CN} (0\,\kms\ to 16.4\,\kms\ including all the hfs transitions), and \ce{HN^13C} (6.8\,\kms\ to 10.8\,\kms) from top to bottom. The star symbol and contours are the same as in Fig.\,\ref{fig:mom0_maps}.}
    \label{appendix:mom0_extend_gas}
\end{figure*}

\begin{figure*}[!ht]
    \centering
    \includegraphics[width=0.40\textwidth]{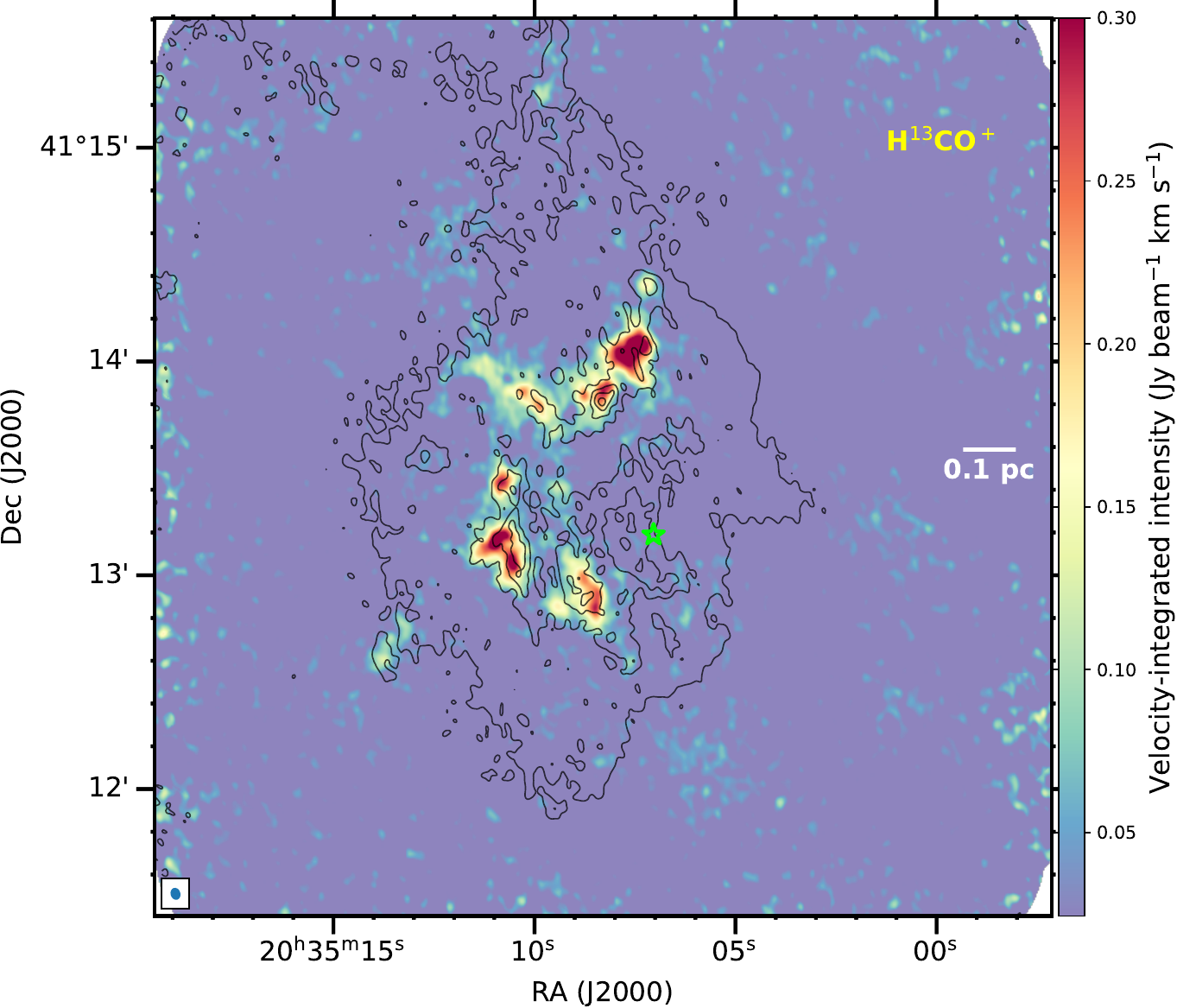}
    \includegraphics[width=0.40\textwidth]{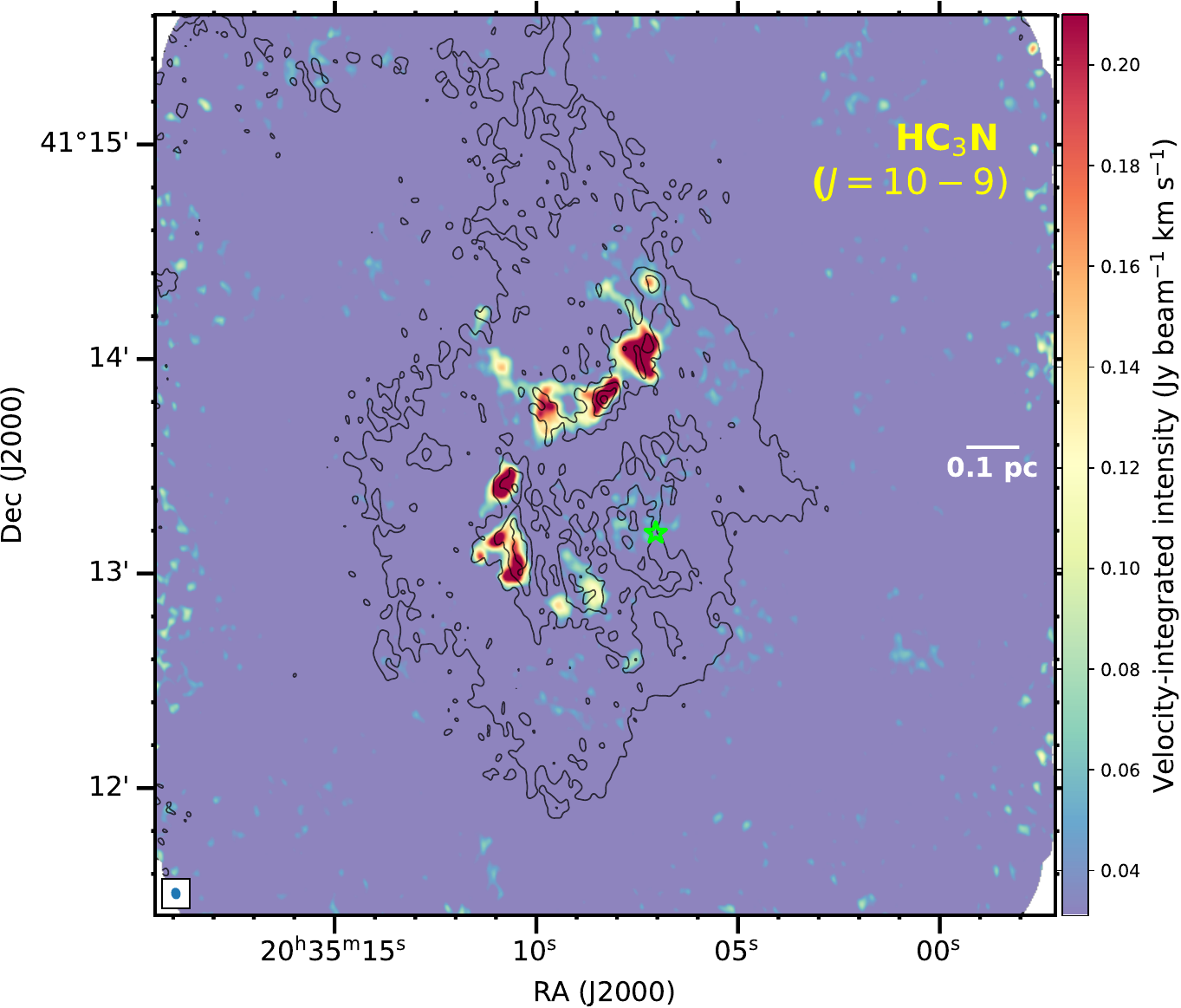}
    \includegraphics[width=0.40\textwidth]{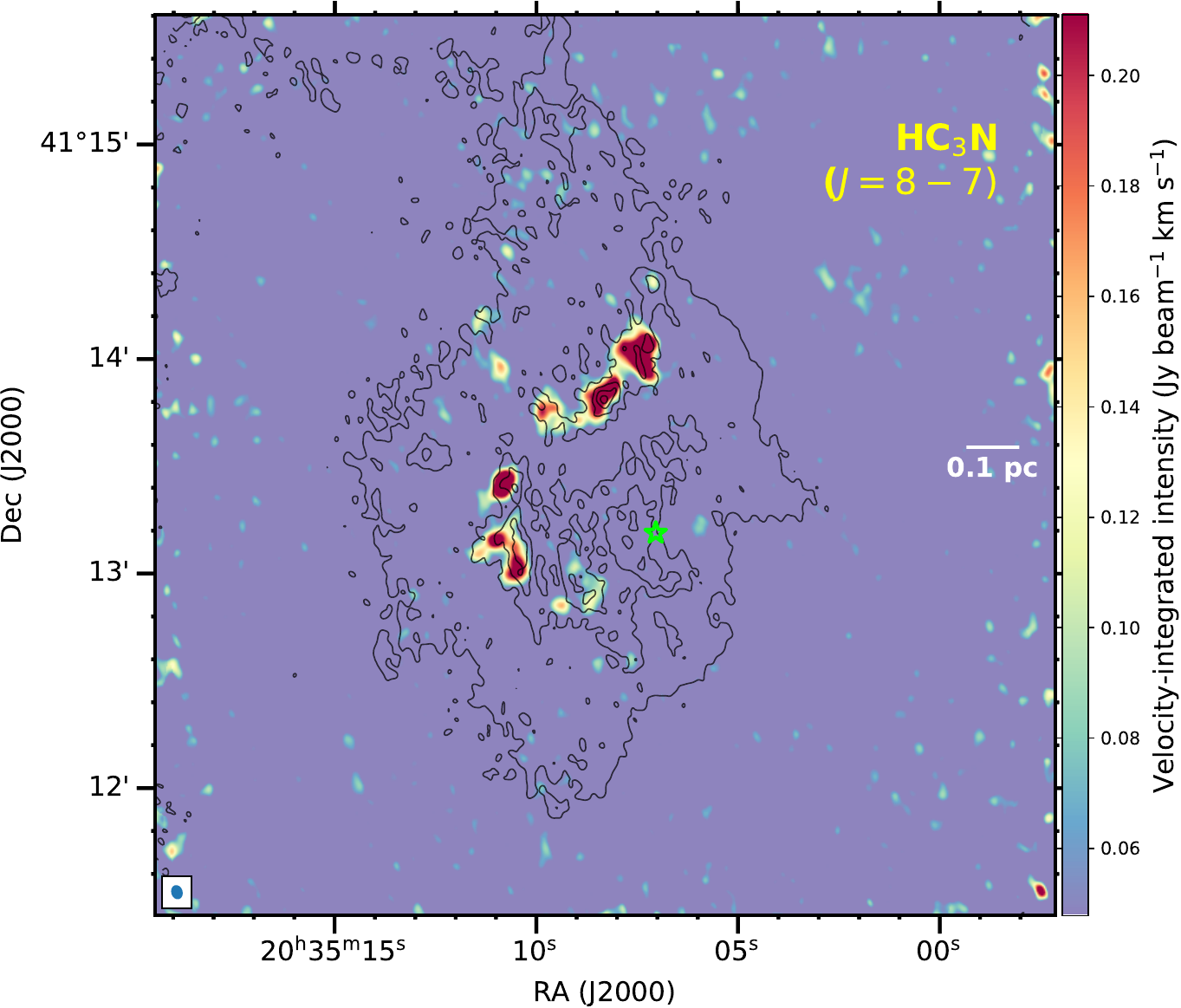}
    \includegraphics[width=0.40\textwidth]{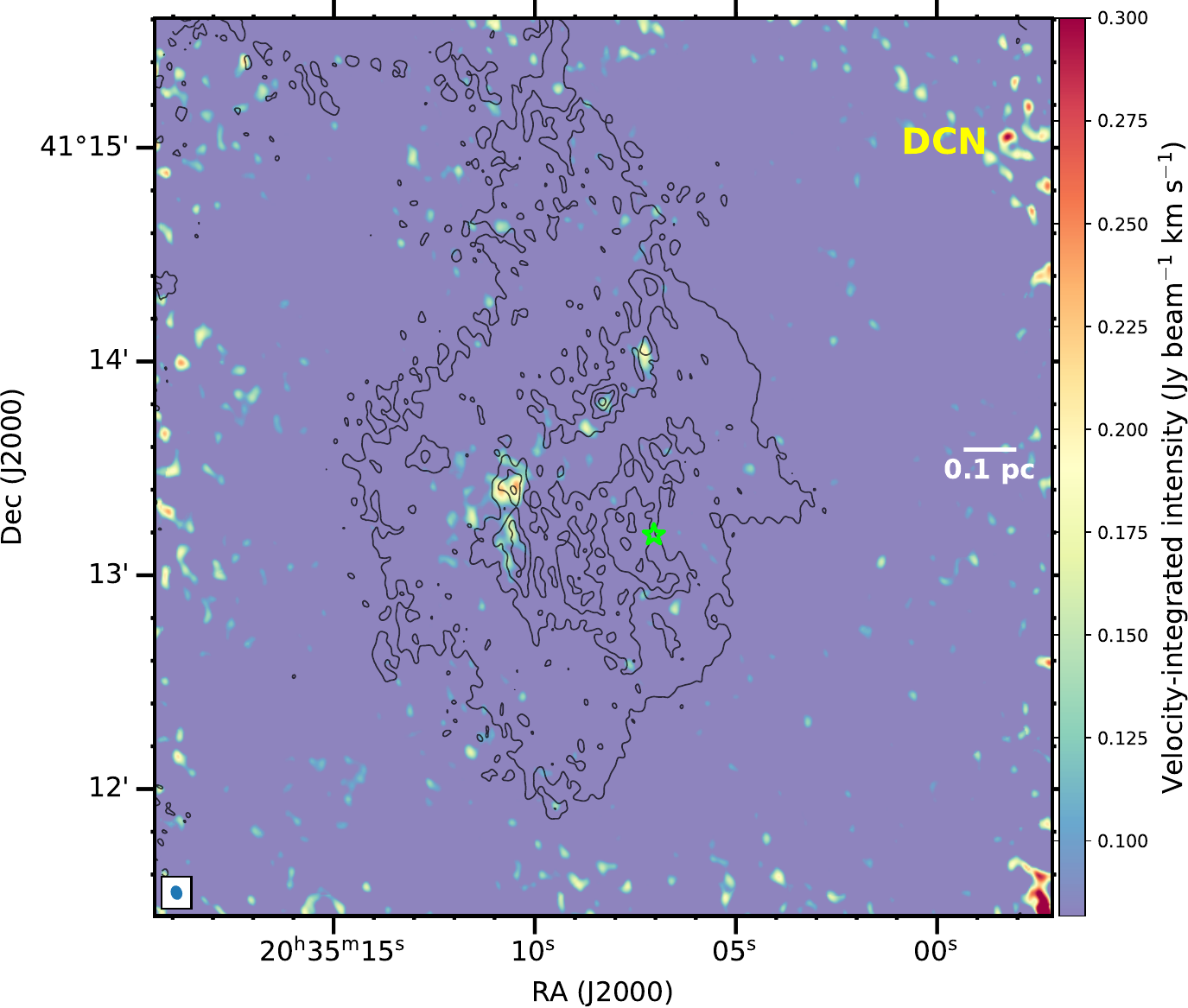}
    \includegraphics[width=0.40\textwidth]{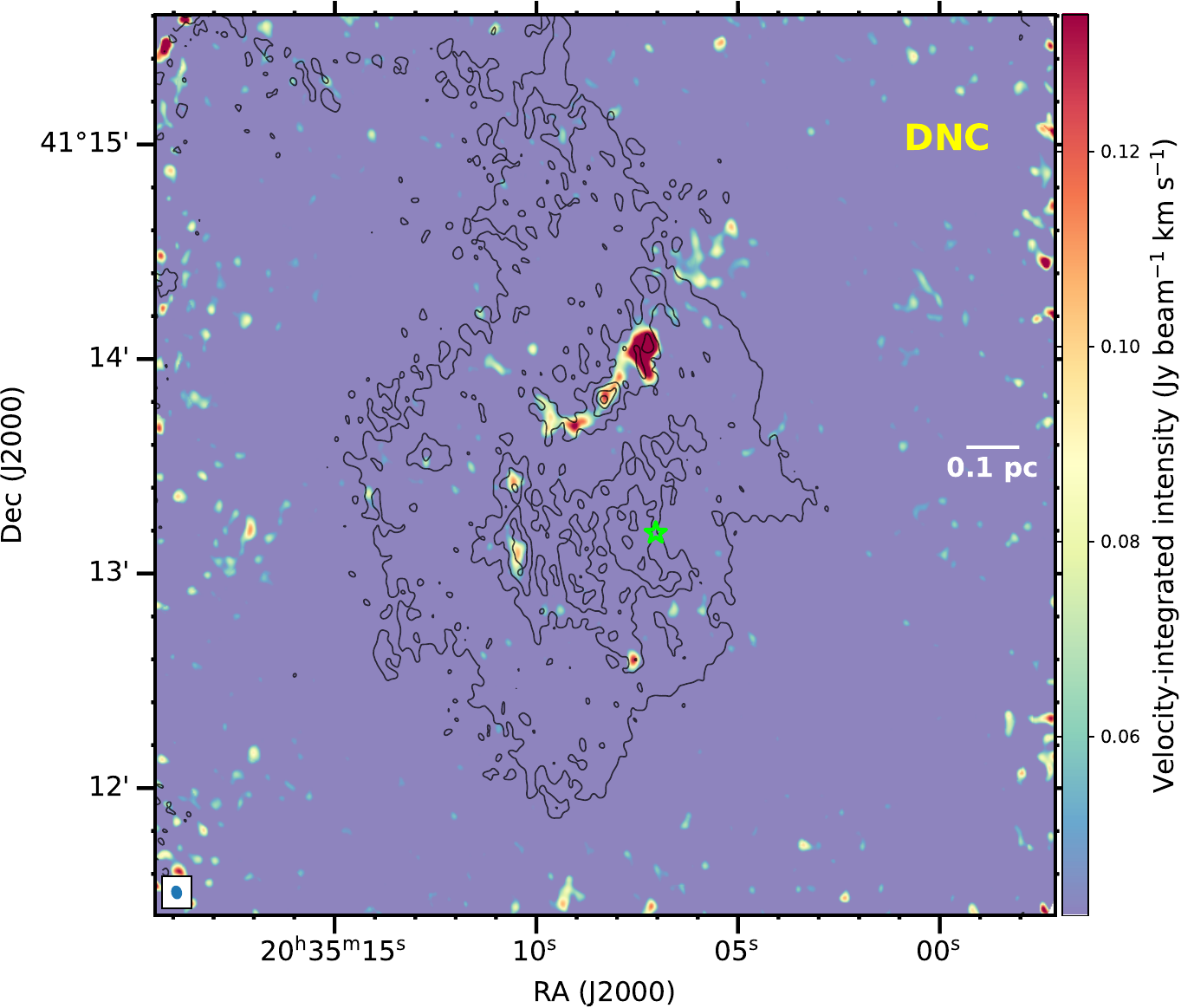}
    \includegraphics[width=0.40\textwidth]{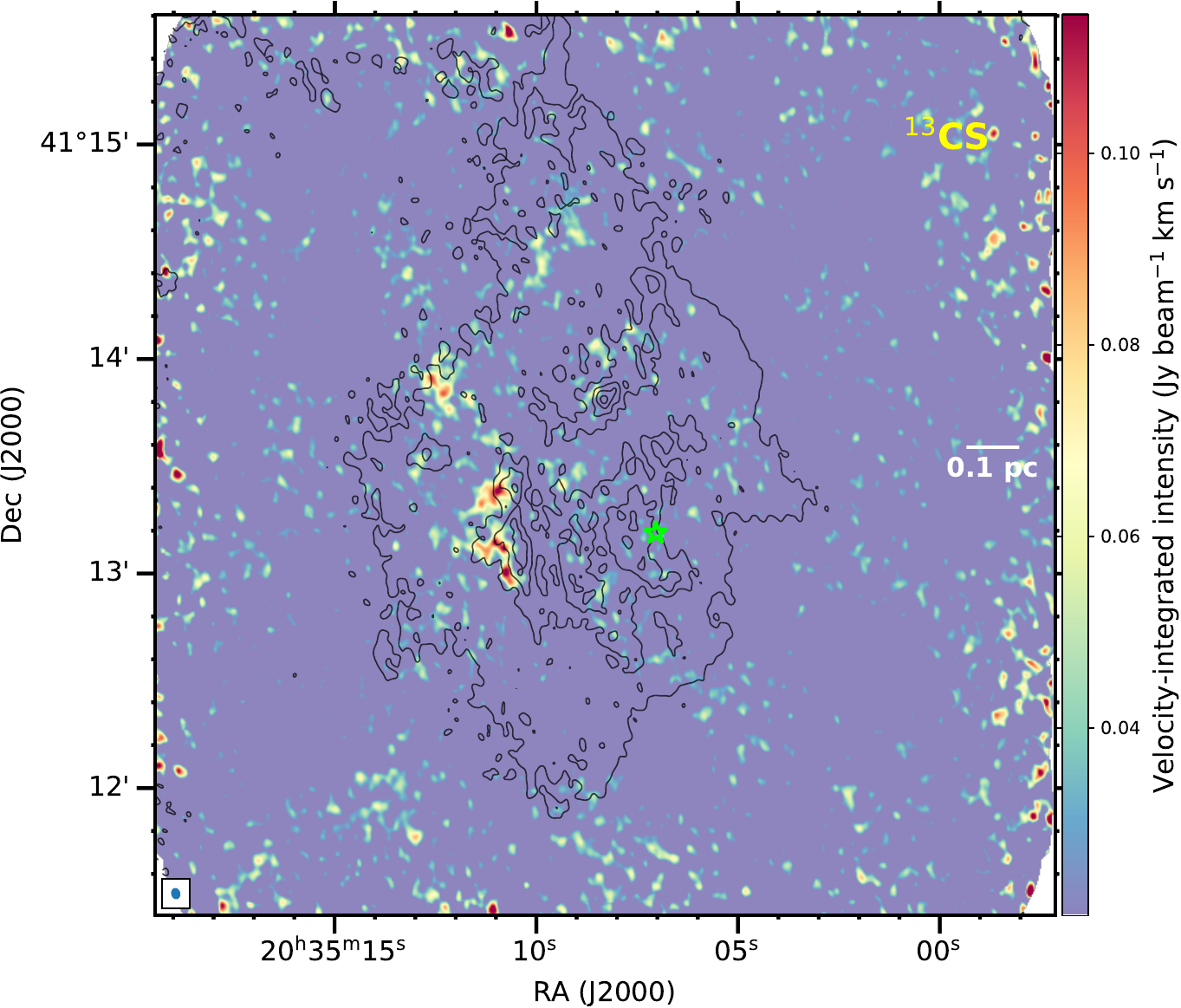} 
    \caption{Velocity-integrated intensity maps of \ce{H^13CO+} (6.8\,\kms\ to 11.6\,\kms), \ce{HC3N} ($J=10-9$) (7.6\,\kms\ to 10.8\,\kms), \ce{HC3N} ($J=8-7$) (7.6\,\kms\ to 10.8\,\kms), DCN (5\,\kms\ to 17\,\kms\ spanning all the hfs transitions), DNC (7\,\kms\ to 10\,\kms), and \ce{^13CS} (7\,\kms\ to 10\,\kms) from top to bottom. The star symbol and contours are the same as in Fig.\,\ref{fig:mom0_maps}.}
    \label{appendix:mom0_dense_gas}
\end{figure*}

\clearpage
\newpage
\section{$T_{\rm HCN/HNC}$ and $T_{\rm \ce{H^13CN}/\ce{HN13C}}$ maps}
\begin{figure}[h!]
    \centering
    \includegraphics[width=0.40\textwidth]{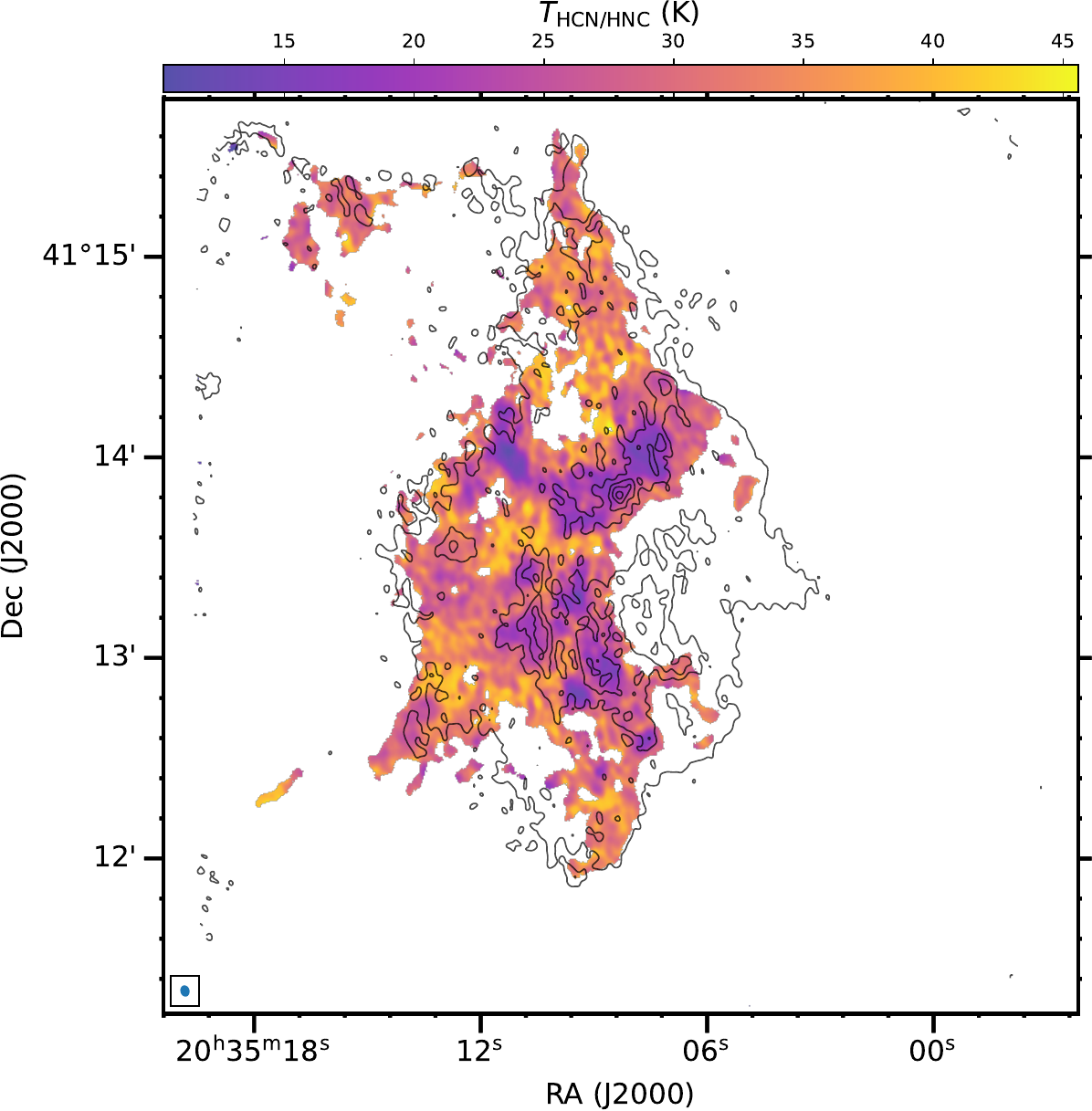}
\includegraphics[width=0.40\textwidth]{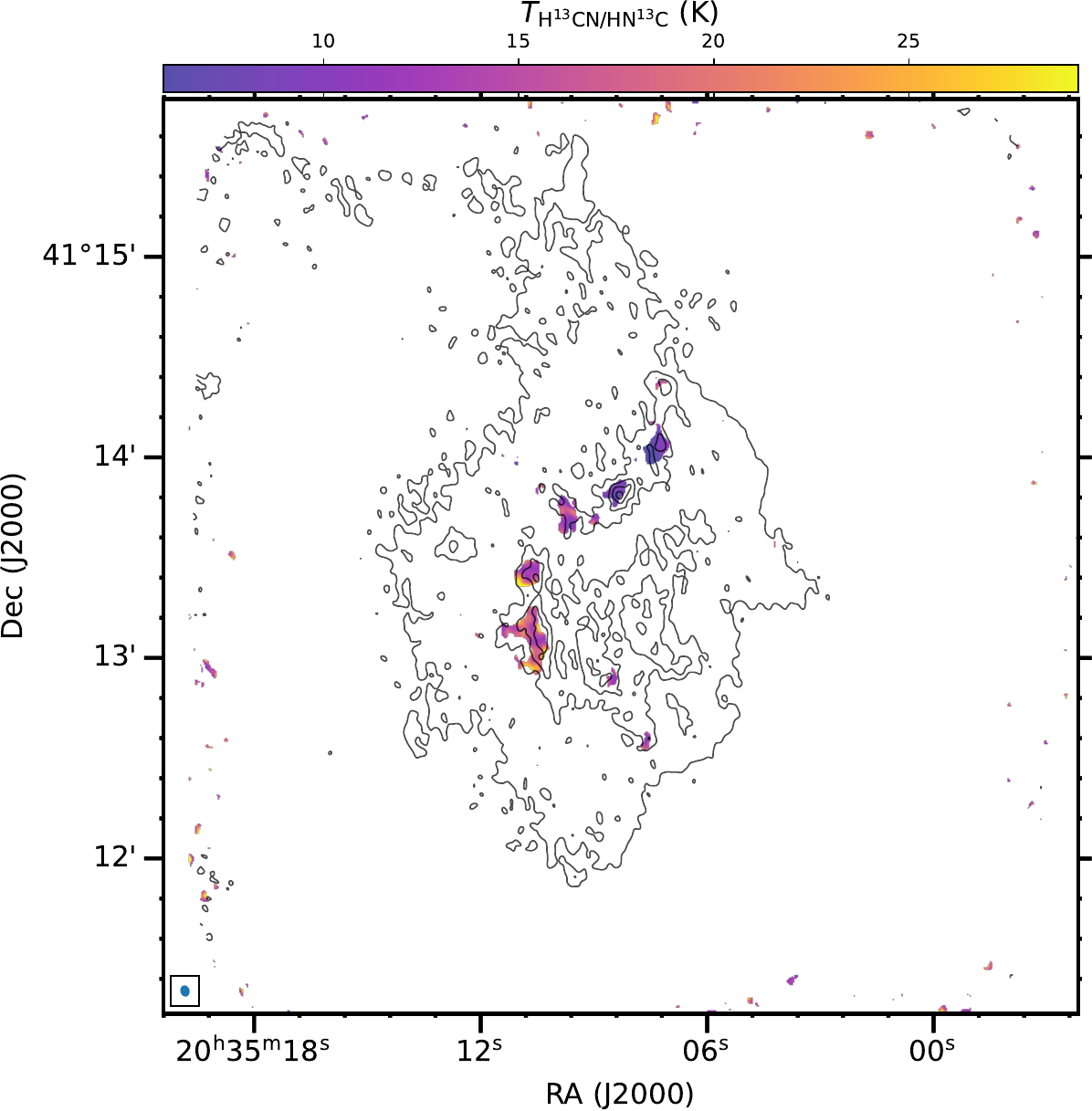}
    \caption{Gas kinetic temperature map separately derived by using the ratios of the HCN/HNC and \ce{H^13CN}/\ce{HN^13C} emission lines. The black contours represent the 3.6\,mm continuum emission. The beams for the $T_{\rm HCN/HNC}$ and $T_{\rm \ce{H^13CN}/\ce{HN^13C}}$ maps are displayed in the lower left corner. }
    \label{fig:appendix_gas_temp}
\end{figure}
\section{\ce{N2H+} spectral lines extracted toward the identified compact cores.}
\begin{figure}[h!]
    \centering
    \includegraphics[width=0.28\textwidth]{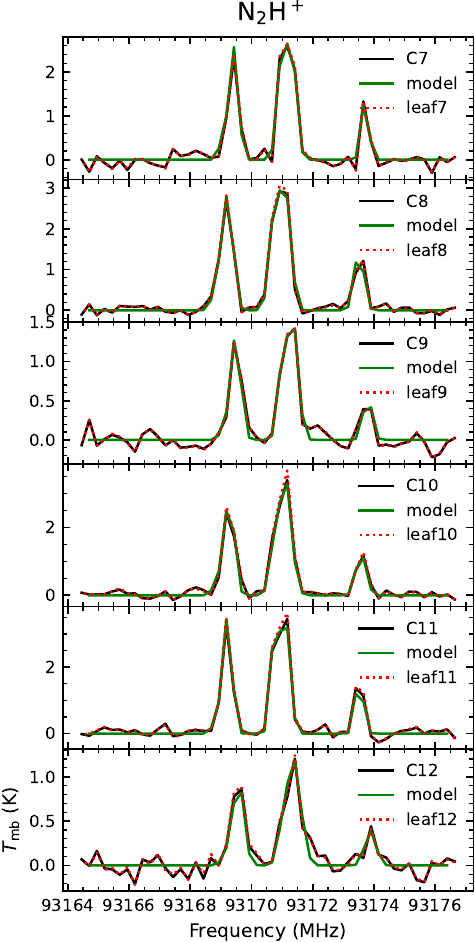}
        \includegraphics[width=0.27\textwidth]{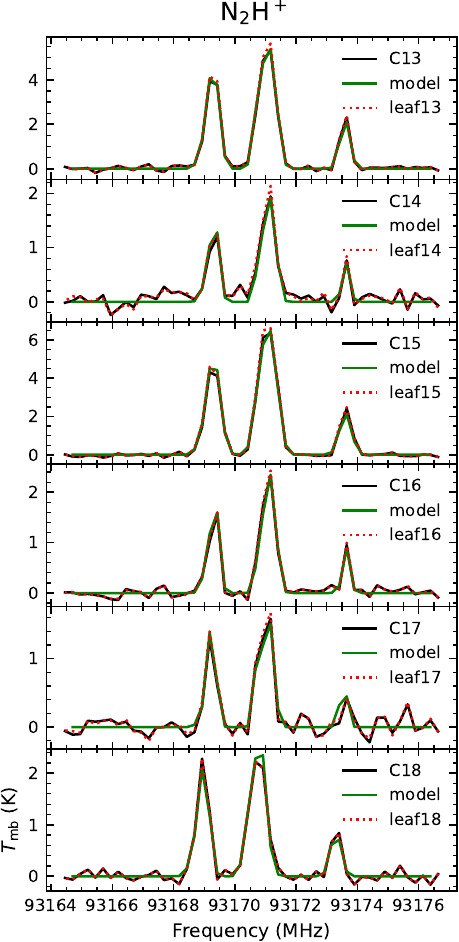}
    \caption{\ce{N2H+} spectral lines (black curves) extracted over the circular regions of the compact cores, overlapped with \ce{N2H+} spectral lines (red dotted curves) extracted over the leaf structures from the \texttt{astrodendro} and the XCLASS modeled spectra (green curves) in a $T_{\rm mb}$ scale}.
    \label{appendix:n2hp_xclass}
\end{figure}

\clearpage
\newpage
\section{Flux ratio map of \textit{Spitzer}/IRAC 4.5\,$\mu$m and 3.6\,$\mu$m bands.}
\begin{figure}[h!]
    \centering
    \includegraphics[width=0.50\textwidth]{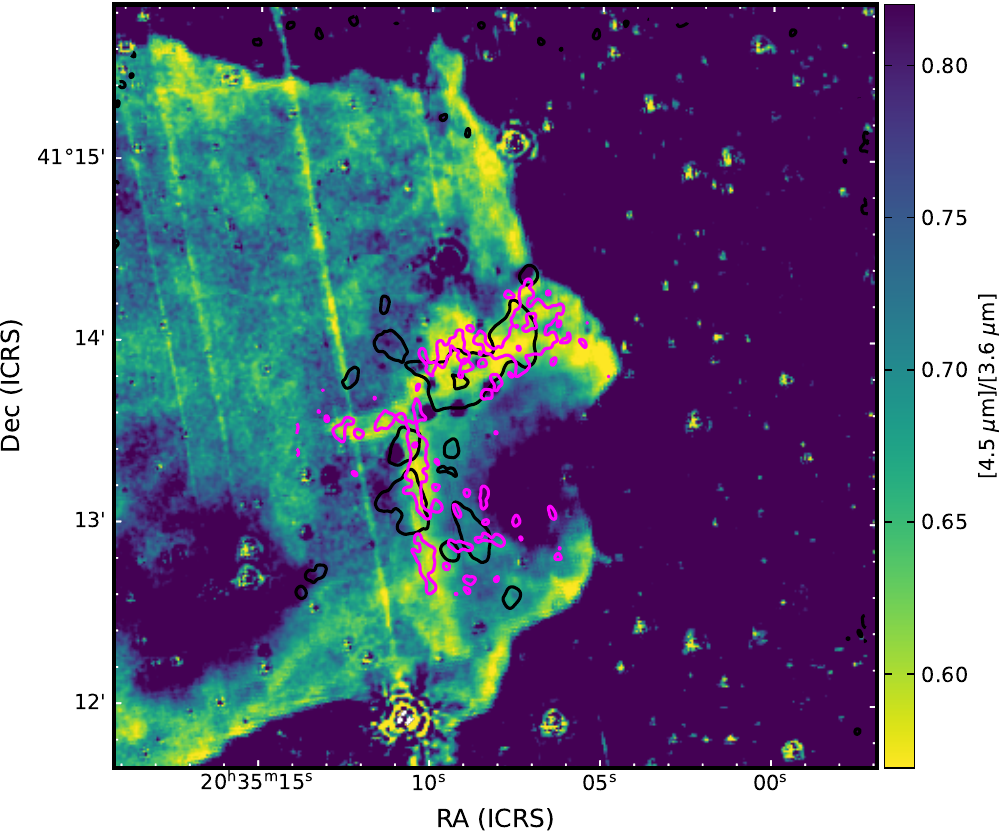}
    \caption{Ratio map of 4.5\,$\mu$m over 3.6\,$\mu$m fluxes. The magenta and black contours for SiO and \ce{N2H+}, respectively, are the same as the right image of Fig.\,\ref{fig:dr18_pdr_image}.}
    \label{fig:ratio_4p5_3p6}
\end{figure}

\end{appendix}

\end{document}